 \documentclass[final,3p,times]{elsarticle}

\usepackage{graphicx}
\usepackage{amssymb}
\usepackage{setspace}
\usepackage{lineno}
\usepackage{booktabs}  
\usepackage{threeparttable}
\usepackage{subfigure}
\usepackage{amsmath}
\usepackage{color}
\usepackage{multirow}
\usepackage{hyperref} 
\hypersetup{hidelinks}
\usepackage{float}

\biboptions{numbers,sort&compress}

\journal{Elsevier}

\begin{document}

\begin{frontmatter}


\title{Revisit to the WGVC schemes: a nonlinear order-preserving and spectral-property-optimized methodology and its enhancement}



\author{Kang He\fnref{1,2}}

\author{Hongwei Liu\fnref{1}}

\author{Tongbiao Guo\fnref{1}}

\author{Xinliang Li\corref{cor1}\fnref{1,2}}
\ead{lixl@imech.ac.cn}

\author{Zhiwei He\corref{cor2}\fnref{3,4}}
\ead{he_zhiwei@iapcm.ac.cn}

\address[1]{State Key Laboratory of High Temperature Gas Dynamics, Institute of Mechanics, Chinese Academy of Sciences, Beijing 100190, China}
\address[2]{School of Engineering Science, University of Chinese Academy of Sciences, Beijing 100049, China}
\address[3]{Institute of Applied Physics and Computational Mathematics, Beijing 100088, China}
\address[4]{National Key Laboratory of Computational Physics, Beijing 100088, China}

\cortext[cor1]{Corresponding author}
\cortext[cor2]{Principal corresponding author}

\begin{abstract}
The numerical simulation of supersonic complex flow problems demands capabilities in identifying multiscale structures and capturing shocks, imposing stringent requirements on the numerical scheme. The capability to identify multiscale structures is closely related to the spectral properties of the numerical scheme. Currently, existing methods to improve the spectral properties of finite difference schemes face shortcomings such as parallel difficulties (compact methods) or introducing unnecessary dispersion errors at low wavenumbers due to accuracy loss (spectral-like optimization methods). In this paper, we proposed an order-preserving spectral properties optimization method based on the group velocity control theory: the weighted group velocity control (WGVC) scheme. This method, centered around the concept of group velocity, achieves low-wavenumber accuracy control and mid-wavenumber group velocity control by designing smoothness indicators and nonlinear weighting approach for wave packets. Furthermore, by embedding the WGVC scheme into shock-capturing schemes such as WENO/TENO scheme, we not only preserve the spectral properties of the WGVC scheme at medium to low wavenumbers but also enhance the shock-capturing capability of the scheme. Theoretical and numerical experiments verify that the new method has advantages such as order-preserving, small dispersion and dissipation errors, and is very suitable for numerical simulation of complex flow problems such as turbulence-shock boundary layer interactions.
\end{abstract}




\end{frontmatter}


\section{Introduction}

The supersonic complex flow problem remains a persistent challenge and focal point in computational fluid dynamics. Of particular interest is the construction of shock-capturing schemes with high order and resolution. When addressing such issues, it is crucial to consider the existence of multiscale structures, such as turbulence, and discontinuous structures, like shock waves \cite{he2015}. For accurate simulation of flow structures across different scales, numerical schemes need to exhibit favorable spectral properties-keeping dispersion and dissipation errors minimal across a wide range of wavenumbers. However, to ensure computational stability, especially near discontinuities, it becomes necessary to introduce appropriate numerical dissipation for suppressing non-physical oscillations \cite{pirozzoli2002}. The inherent contradiction between these two requirements poses a significant challenge in the research and development of numerical schemes.

Finite difference (FD) methods employ difference quotients at discrete grid points to construct difference equations for approximating solutions to differential equations, and accuracy and spectral properties (dispersion error, dissipation error) \cite{vichnevetsky1987,trefethen1982} are two main characteristics. Regarding finite difference methods, accuracy refers to the order of truncation error when the numerical scheme expands Taylor series around finite grid points. The significance of accuracy becomes significant only when the grid is sufficiently dense. When the number of grid points is limited, resolution is commonly used to describe the numerical scheme's ability to effectively identify the range of wavenumbers. For problems involving multiscale structures such as turbulence, spectral properties can reflect the ability of the numerical scheme to identify large-scale structures (low wavenumber components) and small-scale structures (medium to high wavenumber components). For a single wave equation, the exact solution propagates harmonics of different wavenumbers at the same velocity. However, in discrete finite difference methods, due to the presence of dispersion errors, harmonics with different wavenumbers have inconsistent phase velocities. Low-wavenumber components propagate at speeds close to the physical solution, while high-wavenumber components are truncated, causing changes in the speed and direction of wave propagation.

To enhance the spectral properties of finite difference schemes, many scholars have developed compact schemes \cite{lele1992,ma2001,dexun2001,deng2000} and spectral-like optimization methods \cite{tam1993,bogey2004,sun2011,li2022}. Compact schemes utilize derivative information at finite stencil points to enhance the spectral properties of numerical methods, such as central compact schemes \cite{lele1992}, upwind compact \cite{ma2001} and super-compact schemes \cite{dexun2001}, and weighted compact nonlinear schemes \cite{deng2000}. However, compact schemes, which necessitate solving linear systems of equations, encounter challenges in large-scale parallel applications and involve high computational costs \cite{kim2013}, especially in direct numerical simulations (DNS). On the other hand, spectral-like optimization methods typically trade accuracy for improved spectral properties. Instances of such methods include Dispersion Relation-Preserving (DRP) schemes \cite{tam1993}, Minimal Dispersion Controllable Dissipation (MDCD) schemes \cite{sun2011}, and Minimal Dispersion Adaptive Dissipation (MDAD) schemes \cite{li2022}. The fundamental idea behind these methods involves introducing free parameters by relaxing certain coefficients of the numerical scheme. These free parameters are then determined through specific criteria to enhance spectral properties. However, Cunha \cite{cunha2014} highlighted that while such numerical methods aim for intermediate wavenumber resolution, they come at the cost of reduced accuracy and introduce more errors in the low wavenumber range. When the grid is sufficiently resolved, their computational results may be worse than those of corresponding standard difference schemes.

It can be seen that how to enhance the spectral properties in finite difference schemes is still to be well solved although there has been great progress in this area.  In such area, the most important concept is the group velocity in time-dependent partial differential equations within finite difference schemes which has been widely discussed \cite{vichnevetsky1987,trefethen1982,cunha2014}. Here, group velocity is a physical quantity describing the speed of wave packet propagation. It is the ratio of the derivative of the wave packet's phase velocity to its frequency and is associated with the dispersion relationship of the numerical scheme. Trefethen \cite{trefethen1982} studied the significance of group velocity in wave propagation, numerical dispersion, stability analysis, parasitic waves, and the impact of grid refinement on group velocity. Building on wave propagation theory, Vichnevetsky \cite{vichnevetsky1987} analyzed the spurious errors in finite difference methods for hyperbolic equations. He used phase velocity and group velocity to analyze phenomena such as parasitic reflection or scattering occurring at boundaries and non-uniform grids. Cunha \cite{cunha2014} utilized relative dispersion and group velocity to assess the errors in spectral-like optimization methods and their corresponding standard difference schemes at different wavenumber ranges. 

In order to not only enhance the spectral properties but also preserve order of  finite difference schemes, we propose a new nonlinear optimized methodology, i.e., the Weighted Group Velocity Control (WGVC) scheme in this paper. This approach, based on the concept of group velocity, achieves accuracy control in the low-wavenumber range, group velocity control in the mid-wavenumber range, and significantly improves the dispersion relationship of the finite difference scheme by designing smoothness indicators and nonlinear weighting strategies for wave packets. Furthermore, by embedding WGVC scheme into the ENO shock-capturing schemes \cite{jiang1996,henrick2005,borges2008,fu2016,fu2018}, we have developed the WGVC-WENO and WGVC-TENO schemes, enhancing the robustness and shock-capturing capabilities of WGVC scheme. Numerical results demonstrate that this novel method effectively improves the spectral properties of the finite difference scheme while maintaining accuracy. It is particularly suitable for numerical simulations of multiscale problems such as turbulence and aerodynamic noise and has been successfully applied in direct numerical simulations of shock-turbulence boundary layer interaction. 

The structure of the remaining sections is as follows: the second section introduces some background knowledges related to group velocity, the third section presents the construction of WGVC scheme, providing a detailed overview of the theory and construction process, and discusses how to embed WGVC scheme into ENO-like shock-capturing schemes to enhance shock-capturing capabilities, the fourth section analyzes spectral properties of the proposed schemes using Fourier method, the fifth section validates the new method's accuracy, dissipation, and other numerical characteristics using various numerical examples, and the final section provides a summary.

\section{Background knowledges}

Group velocity in finite difference schemes serves as a crucial physical quantity for assessing the resolution and stability of computational results. For multiscale complex flow problems, differential schemes need to possess two characteristics: broad-scale resolution capability and the ability to handle discontinuities. The resolution capability depends on the level of dispersion error and dissipation error, closely related to the group velocity. Moreover, parasitic waves generated due to discontinuities often lead to numerical oscillations, and the group velocity is the only meaningful speed for these parasitic waves \cite{trefethen1982}. This section introduce concepts related to group velocity.

\subsection{Group velocity in numerical solutions}

First, consider a one-dimensional linear single-wave equation in the following form:
\begin{equation} \label{eq1}
   \frac{\partial u}{\partial t}+\frac{\partial f}{\partial x}=0,\quad f=au,\quad a=\mathrm{const}. 
\end{equation}

The exact solution corresponding to Eq.(\ref{eq1}) is:
\begin{equation} \label{eq2}
   u(x,t)=e^{\mathrm{i}(\kappa x-\omega t)}.
\end{equation}

For each real wavenumber $\kappa$, there is a corresponding real frequency $\omega$ that satisfies Eq.(\ref{eq2}), and the relationship between wavenumber $\kappa$ and frequency $\omega$, denoted by  $\omega=\omega(\kappa)$, is called the dispersion relation. For the exact solution, from Eq.(\ref{eq2}), it can be observed that waves propagate with a constant velocity $a=\omega/\kappa$, where $a$ is known as the phase velocity. When waves appear in a group, the situation becomes more complex. Assuming an initial moment with $u(x,0)$, the corresponding Fourier transform is:
\begin{equation} \label{eq3}
   u(x,0)=\int_{-\infty}^{+\infty}f(\kappa)e^{\mathrm{i}\kappa x}\mathrm{d}\kappa,
\end{equation}
when $t\geq0$, according to Eq.(\ref{eq2}):
\begin{equation} \label{eq4}
   u(x,t)=\int_{-\infty}^{+\infty}f(\kappa)\exp\left\{\mathrm{i}t\bigg[\frac{\kappa x}t-\omega(\kappa)\bigg]\right\}\mathrm{d}\kappa.
\end{equation}

Assuming $x/t$ is a constant, i.e., $x/t=a=\mathrm{const}$, from Eq.(\ref{eq4}), it can be observed that as $t$ approaches infinity, along $x/t=a$, the oscillation frequency of the wave increases with the increase of the wave number $\kappa$. What can be observed are only the waves that satisfy the following formula:
\begin{equation} \label{eq5}
   \frac{\mathrm{d}}{\mathrm{d}\kappa}\bigg[\omega(\kappa)-\frac{\kappa x}{t}\bigg]=0,
\end{equation}
which are actually waves with frequencies reaching extreme values as the wave number varies, satisfying $\mathrm{d}\omega/\mathrm{d}\kappa=x/t$, define:
\begin{equation} \label{eq6}
   D(\kappa)=\mathrm{d}\omega(\kappa)/\mathrm{d}\kappa,
\end{equation}
where $D(\kappa)$ represents the group velocity. During the propagation of a wave group, waves with similar amplitudes and periods superimpose to form a wave train, with the envelope of the wave train being the wave packet. The phase velocity reflects the propagation speed of an individual wave, while the group velocity reflects the propagation speed of the wave packet. If the dispersion relationship can be consistently maintained, it can be seen from Eq.(\ref{eq6}) that the group velocity is a fixed value consistent with the phase velocity. Conversely, inconsistency in group velocity indicates the occurrence of dispersion. Further studies on group velocity can be found in \cite{vichnevetsky1987,trefethen1982}.

\subsection{Classification of difference schemes}

Discretizing the spatial derivative term in Eq.(\ref{eq1}), we obtain the semi-discrete equation as follows:
\begin{equation} \label{eq7}
\frac{\partial u_j}{\partial t}+\frac{F_j}{\Delta x}=0,\quad j=0,...,N,
\end{equation}
where $\Delta x$ is grid size, and for uniform grids, it satisfies $x_j=j\Delta x$. $F_j/\Delta x$ is the difference approximation of $\partial f/\partial x$ and takes the following form:
\begin{equation} \label{eq8}
\frac{F_j}{\Delta x}=\frac1{\Delta x}\sum_{m=-l}^rb_mf_{j+m}.
\end{equation}

The coefficient $b_m$ in Eq.(\ref{eq8}) is a constant. According to order requirements, it is obtained by Taylor expansion of the flux $f_{j+m}$ at the corresponding stencil points. In addition, Eq.(\ref{eq7}) is generally expressed in the following conservative form:
\begin{equation} \label{eq9}
   \frac{\partial u_j}{\partial t}+\frac{\hat{F}_{j+1/2}-\hat{F}_{j-1/2}}{\Delta x}=0,
\end{equation}
where $\hat{F}_{j+1/2}$ is the numerical flux which has an approximation of the flux $F_j$ at the boundary $x_{j+1/2}$. The subsequent discussion on the construction of numerical schemes will primarily focus on $\hat{F}_{j+1/2}$.

For spatial discretization, according to Eq.(\ref{eq2}), let's assume that the solution takes the following form:
\begin{equation} \label{eq10}
   u(x_j,t)=\hat{u}(t)e^{\mathrm{i}\kappa x_j},\quad\hat{u}(0)=\hat{u}_0,
\end{equation}
substituting Eq.(\ref{eq10}) into Eq.(\ref{eq7}), we can obtain:
\begin{equation} \label{eq11}
   \frac{d\hat{u}(t)}{dt}+\mathrm{i}a\tilde{\kappa}\hat{u}(t)=0,\quad\hat{u}(0)=\hat{u}_0,
\end{equation}
\begin{equation} \label{eq12}
   \tilde{\kappa}=\frac{1}{\mathrm{i}\Delta x}\sum_{m=-l}^rb_me^{\mathrm{i}\kappa(m\Delta x)},
\end{equation}
where $\tilde{\kappa}$ is called the modified wavenumber. Solving Eq.(\ref{eq11}), we can obtain:
\begin{equation} \label{eq13}
   u(x_j,t)=\hat{u}_0e^{\mathrm{i}(\kappa x_j-\tilde{\kappa}at)}.
\end{equation}

From Eq.(\ref{eq12}), it can be observed that the modified wavenumber $\tilde{\kappa}$ varies for different  schemes. For convenience of discussion, introducing the effective wavenumber $\alpha=\kappa\Delta x$ with $\alpha\in(0,\pi]$ and the modified effective wavenumber $\Xi(\alpha)=\tilde{\kappa}\Delta x$. Substituting $\Xi=\mathrm{Re}(\Xi)+\mathrm{i}\mathrm{Im}(\Xi)$ into Eq.(\ref{eq13}), we can obtain:
\begin{equation} \label{eq14}
   u(x_j,t)=\hat{u}_0e^{\frac{\mathrm{Im}(\Xi)}{\Delta x}at}e^{\mathrm{i}\kappa(x_j-\frac{\mathrm{Re}(\Xi)}\alpha at)}.
\end{equation}

From Eq.(\ref{eq14}), it is evident that $\mathrm{Im}(\Xi)$ affects the amplitude of the computational results and is related to the dissipation error of the difference scheme, while $\mathrm{Re}(\Xi)$ influences the phase of the computational results and is associated with the dispersion error of the difference scheme. When $\mathrm{Im}(\Xi)=0$ and $\mathrm{Re}(\Xi)=\alpha$, the numerical solution is equal to the exact solution. For a given linear scheme, the modified effective wavenumber $\Xi(\alpha)$ can be obtained through theoretical analysis. For nonlinear schemes, $\Xi(\alpha)$ can also be approximated through numerical methods \cite{li2005,pirozzoli2006}. 

Moreover, from Eq.(\ref{eq14}), it can be observed that for components with different effective wavenumbers, the phase velocity of the numerical solution varies. This causes the waves of different wavenumbers to continually disperse and broaden over time, especially for high wavenumber components. To investigate the propagation characteristics of wave packets in the numerical solution, Fu \textit{et al}. \cite{fu1997} defined the numerical group velocity based on Eq. (\ref{eq14}):
\begin{equation} \label{eq15}
   D^{\circ}(\alpha)=\frac{\mathrm{d}}{\mathrm{d}\alpha}[\mathrm{Re}(\Xi)],
\end{equation}
where $D^{\circ}(\alpha)$ is the gradient of $\mathrm{Re}(\Xi)$ with respect to $\alpha$, and for the exact solution, that is Eq.(\ref{eq2}), $D^{\circ}(\alpha)=1$. Fu \textit{et al}. classified the scheme into three types based on the magnitude of $D^{\circ}(\alpha)$: fast scheme, slow scheme, and mixed scheme. It should be noted that the mixed scheme here does not refer to a hybrid of two schemes but rather indicates that the numerical scheme behaves as a slow scheme in certain wavenumber ranges and as a fast scheme in others. Specifically:

\begin{itemize}
    \item Fast Schemes (abbreviated as $FST$):
    \begin{equation} \label{eq16}
         D^{\circ}(\alpha){>}1,\quad0{<}\alpha{\leqslant}\pi.
     \end{equation}
\end{itemize}

\begin{itemize}
    \item Slow Schemes (abbreviated as $SLW$):
    \begin{equation} \label{eq17}
         D^{\circ}(\alpha)<1,\quad0<\alpha\leqslant\pi.
     \end{equation}
\end{itemize}

\begin{itemize}
    \item Mixed Schemes (abbreviated as $MXD$):
    \begin{equation} \label{eq18}
       \begin{cases}
       D^{\circ}(\alpha)>1,\quad0<\alpha\leqslant\alpha_0<\pi,\\
       D^{\circ}(\alpha)<1,\quad\alpha_0<\alpha\leqslant\pi.
       \end{cases}
     \end{equation}
\end{itemize}

From Eqs.(\ref{eq16}-\ref{eq18}), it can be observed that the numerical solution corresponding to the $FST$ scheme exhibits a wave group movement speed faster than the physical solution, while the numerical solution of the $SLW$ scheme has a wave group movement speed slower than the physical solution. The $MXD$ scheme typically exhibits $FST$ characteristics in the mid-low wavenumber range and $SLW$ characteristics in the high wavenumber range. Due to the strong dissipative effects in the high wavenumber range and the difficulty in constructing a purely $FST$-type scheme, $MXD$ scheme is considered to represent $FST$ scheme. For simplicity, the remainder of this paper will no longer distinguish between $FST$ and $MXD$ schemes and will uniformly use the term $MXD$ scheme.

\subsection{The group velocity control (GVC) theory}
As discussed previously, the concept of group velocity is often used to analyze the spectral properties of difference schemes. However, this concept is more utilized to explain numerical oscillations near shock waves \cite{zhang1991,fu1997} in the Chinese academic community. The group velocity control (GVC) theory proposed by  Fu \textit{et al}. \cite{fu1997} is a typical one. The GVC theory proposed by Fu \textit{et al}. \cite{fu1997} posits that, based on Fourier decomposition, a shock wave can be decomposed into the superposition of harmonics with different wavenumbers. Inconsistencies in the group velocities of these harmonics after numerical discretization cause a misalignment between waves of different scales, leading to numerical oscillations. By controlling the group velocities of the numerical scheme before and after the discontinuity, the propagation of this misalignment can be suppressed, thereby inhibiting numerical oscillations. To be more specific, when using the $MXD$ scheme before the shock, the numerical solution's velocity is faster than the physical solution's velocity. Conversely, when using the $SLW$ scheme after the shock, the numerical solution's velocity is slower than the physical solution's velocity. Both cases lead to the dispersion and broadening of perturbation waves, causing numerical oscillations. By employing a strategy of utilizing a $SLW$ scheme before a discontinuity and a $MXD$ scheme after it, non-physical oscillations can be effectively suppressed, thus preventing the propagation of errors. Several GVC schemes \cite{ma2001,fu1997,li2005} have been developed based on the GVC theory, and the classical $2nd$-order GVC scheme \cite{fu1997} is as follows. 

For the flux $\hat{F}_{j+1/2}$, the numerical flux of the $2nd$-order accuracy GVC scheme is given by:
\begin{equation} \label{eq19}
\hat{F}_{j+1/2}=\frac{1+\operatorname{SS}\left(u_{j+1/2}\right)}2\hat{F}_{j+1/2}^{MXD}+\frac{1-\operatorname{SS}\left(u_{j+1/2}\right)}2\hat{F}_{j+1/2}^{SLW},
\end{equation}

\begin{equation} \label{eq20}
   \begin{aligned}
       &\hat{F}_{j+1/2}^{MXD}=(3f_j-f_{j-1})/2,\\
       &\hat{F}_{j+1/2}^{SLW}=(f_{j+1}+f_j)/2.
   \end{aligned}
\end{equation}

In which, $\hat{F}_{j+1/2}^{MXD}$ is the $2nd$-order upwind scheme, categorized as a $MXD$ scheme, and $\hat{F}_{j+1/2}^{SLW}$ is the $2nd$-order central scheme, categorized as a $SLW$ scheme. $\operatorname{SS}$ denotes the shock-structure function, utilized for determining the positions before and after a shock wave, and is defined as:
\begin{equation} \label{eq21}
   \mathrm{SS}(u)=\mathrm{sign}(\frac{\partial u}{\partial x}.\frac{\partial^2u}{\partial x^2}),
\end{equation}
for a right-propagating shock wave, $\mathrm{SS}(u)=-1$ represents the wavefront, whereas  $\mathrm{SS}(u)=1$ signifies the wave rear.

The GVC theory aims to control the group velocities of schemes before and after shock waves, with the hope of directing waves of different scales towards the shock wave to suppress numerical oscillations. It is noteworthy that, unlike common shock-capturing schemes \cite{jiang1996,henrick2005,borges2008,fu2016,fu2018}, which often emphasize the introduction and control of numerical viscosity to enhance scheme characteristics, the GVC theory places more emphasis on manipulating the group velocity of the numerical scheme while introducing a certain level of numerical viscosity. This approach is more conducive to improving the resolution of the numerical scheme. However, the GVC theory still has some limitations. As seen from Eqs.(\ref{eq19}-\ref{eq20}), GVC schemes rely on the $\mathrm{SS}$ (or shock indicator), and frequent changes in sign are detrimental to computational stability. Additionally, the $SLW$ and $MXD$ schemes used before and after the shock wave are both linear schemes. Relying solely on linear schemes makes it challenging to achieve non-oscillatory properties.

\section{The proposed method}
According to the GVC theory, He \textit{et al}. \cite{he2014} introduced a kind of weighted GVC schemes. The core of WGVC schemes lies in the design of a smoothness indicator (for shocks or small-scale wave packets which are usually treated as shocks numerically) and a nonlinear weighting approach. The numerical results has confirmed that the WGVC schemes cannot achieve the essentially non-oscillatory shock-capturing property \cite{he2014}. However, the WGVC schemes shows spectral properties while do not have the order-reduced problem comparing to the previous spectral-like schemes (mainly linear). Such numerical experiments \cite{he2014} indicate that the GVC theory is more suitable as an optimization principle for wave packets and can be employed to enhance the spectral properties of finite difference schemes. The corresponding WGVC schemes may be view as a kind of nonlinear spectral-property optimized methodology. For such reasons, the WGVC schemes is called nonlinear spectral-like schemes \cite{he2014} while the smoothness indicator designed for WENO schemes is found to not work too well. In this section, we revisit clearly the WGVC schemes as an optimization principle by introduce a new smooth indicator. Moreover, a new method to enhance the shock-capturing ability of WGVC schemes further is proposed.

\subsection{A new WGVC scheme}

\subsubsection{Construction of WGVC schemes}
\begin{figure}[ht]
	\centering
	\includegraphics[scale=1.]{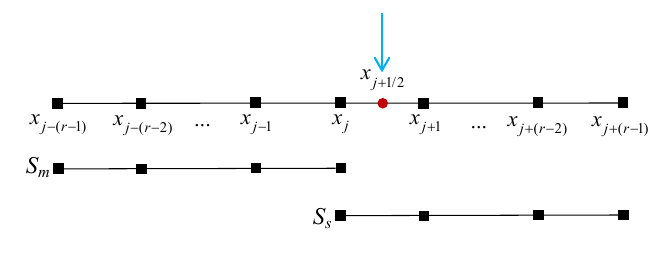}
	\caption{Diagram of WGVC scheme stencils.}
	\label{fig.1}
\end{figure}

The WGVC scheme initially divides the stencil points into $S_m$ and $S_s$ subsets, as shown in Fig.\ref{fig.1}. It is important to note that here, $S_m$ and $S_s$ are just used for determining the shock position and the weights for the $SLW$ and $MXD$ schemes. The stencil points employed for both $SLW$ and $MXD$ schemes are $\left\{{{x_{j-(r-1)}},...,{x_{j+(r-1)}}} \right\}$. The specific forms of the $SLW$ and $MXD$ schemes will be discussed later in the paper. If the sub-stencil $S_m$ is not smooth, it indicates that $x_{j+1/2}$ is located ahead of the shock wave, resulting in a weight of $1$ for $SLW$ scheme and a weight of $0$ for $MXD$ scheme. Conversely, if $S_s$ is not smooth, indicating the wave rear, $SLW$ scheme is assigned a weight of $0$, and $MXD$ scheme is assigned a weight of $1$.

Now, let's consider how to design the $SLW$ and $MXD$ schemes used in WGVC scheme. As we know that a stencil with $2r$ points $\left\{{{x_{j-(r-1)}},...,{x_{j+(r-1)}}} \right\}$ can yield a linear upwind scheme of up to $2r-1$ order. By designating one of the stencil points' corresponding coefficients as a free parameter, a linear scheme of $2r-2$ order can be obtained:
\begin{equation} \label{eq22}
   L(f_j;\sigma)=\sigma\cdot f_{j-r}+\sum_{l=-(r-1)}^{r-1}a_l(\sigma)\cdotp f_{j+l},
\end{equation}
the conservation form satisfies: 
\begin{equation} \label{eq23}
   \hat{F}_{j+1/2}\left(\sigma\right)=\sum_{l=-(r-1)}^{r-1}b_{l}(\sigma)\cdot f_{j+l}.
\end{equation}

For $5th$ and $7th$-order linear schemes, where $r=3$ and $r=4$ respectively, the parameters in Eqs.(\ref{eq22}-\ref{eq23}) are as shown in Tables.\ref{tab.1}-\ref{tab.2}.
  
\begin{table}[ht]  
\centering  
\begin{threeparttable}  
\caption{The values of $a_l$}  
\label{tab.1}  
\begin{tabular}{ccccccccc}  
\toprule
  & $a_{_{-4}}(\sigma)$ & $a_{_{-3}}(\sigma)$ & $a_{_{-2}}(\sigma)$ & $a_{_{-1}}(\sigma)$ & $a_{_{0}}(\sigma)$ & $a_{_{1}}(\sigma)$ & $a_{_{2}}(\sigma)$ & $a_{_{3}}(\sigma)$ \\
\midrule
$r=3$ & - & $\sigma$ & $1/12-5\sigma$ & $-2/3+10\sigma$ & $-10\sigma$ & $2/3+5\sigma$ & $-1/12-\sigma$ & - \\
$r=4$ & $\sigma$ & $-1/60-7\sigma$ & $3/20+21\sigma$ & $-3/4-35\sigma$ & $35\sigma$ & $3/4-21\sigma$ & $-3/20+7\sigma$ & $1/60-\sigma$ \\
\bottomrule  
\end{tabular}  
\end{threeparttable}  
\end{table}

\begin{table}[ht]  
\centering  
\begin{threeparttable}  
\caption{The values of $b_l$}  
\label{tab.2}
\begin{tabular}{cccccccc}  
\toprule
  & $b_{_{-3}}(\sigma)$ & $b_{_{-2}}(\sigma)$ & $b_{_{-1}}(\sigma)$ & $b_{_{0}}(\sigma)$ & $b_{_{1}}(\sigma)$ & $b_{_{2}}(\sigma)$ & $b_{_{3}}(\sigma)$ \\
\midrule
$r=3$ & - & $-\sigma$ & $-1/12+4\sigma$ & $7/12-6\sigma$ & $7/12+4\sigma$ & $-1/12-\sigma$ & - \\
$r=4$ & $-\sigma$ & $1/60+6\sigma$ & $-2/15-15\sigma$ & $37/60+20\sigma$ & $37/60-15\sigma$ & $-2/15+6\sigma$ & $1/60-\sigma$ \\
\bottomrule  
\end{tabular}  
\end{threeparttable}  
\end{table}

\begin{figure}[ht]
\centering
\subfigure[dispersion]{
\includegraphics[width=8cm]{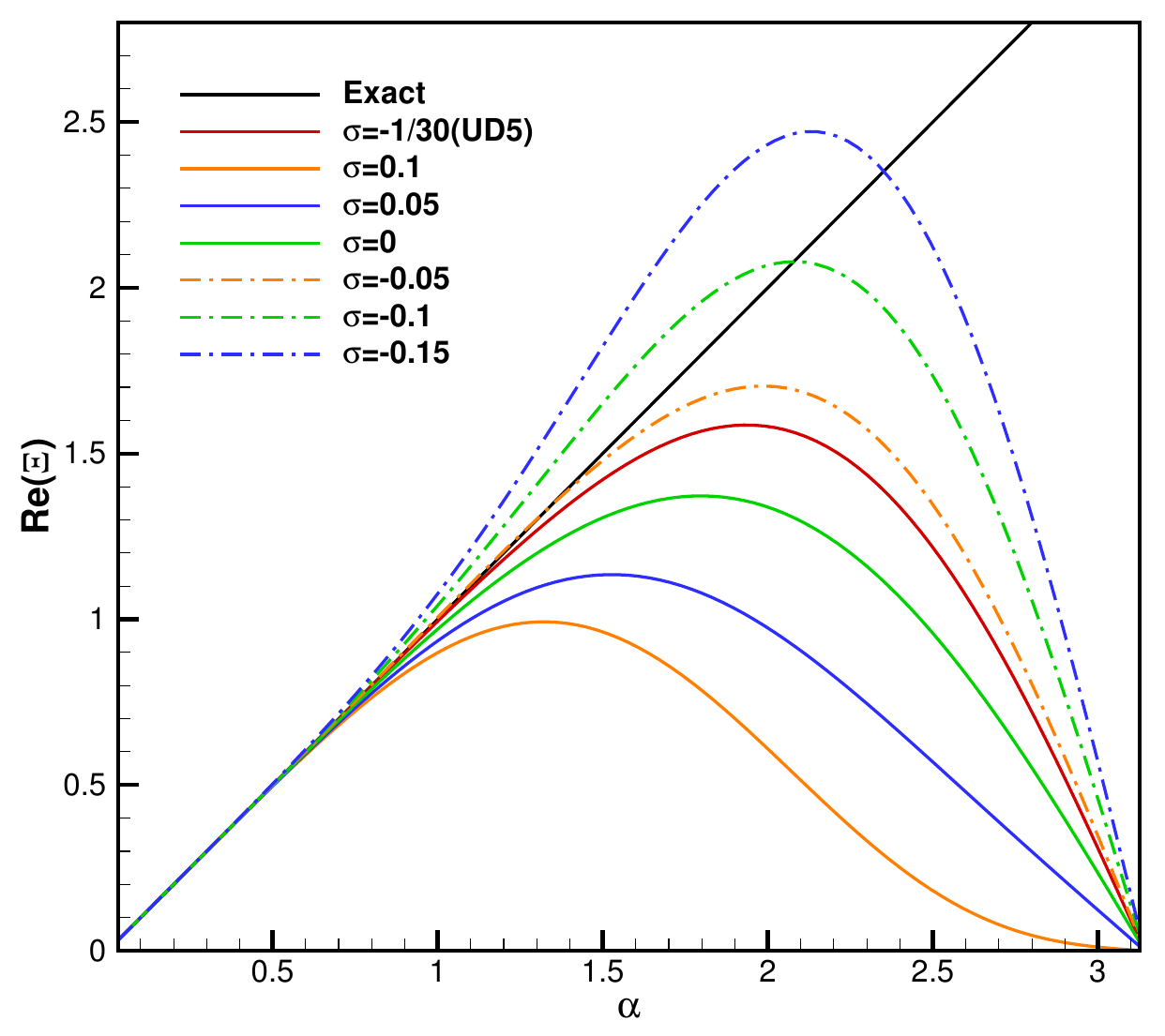}}
\subfigure[dissipation]{
\includegraphics[width=8cm]{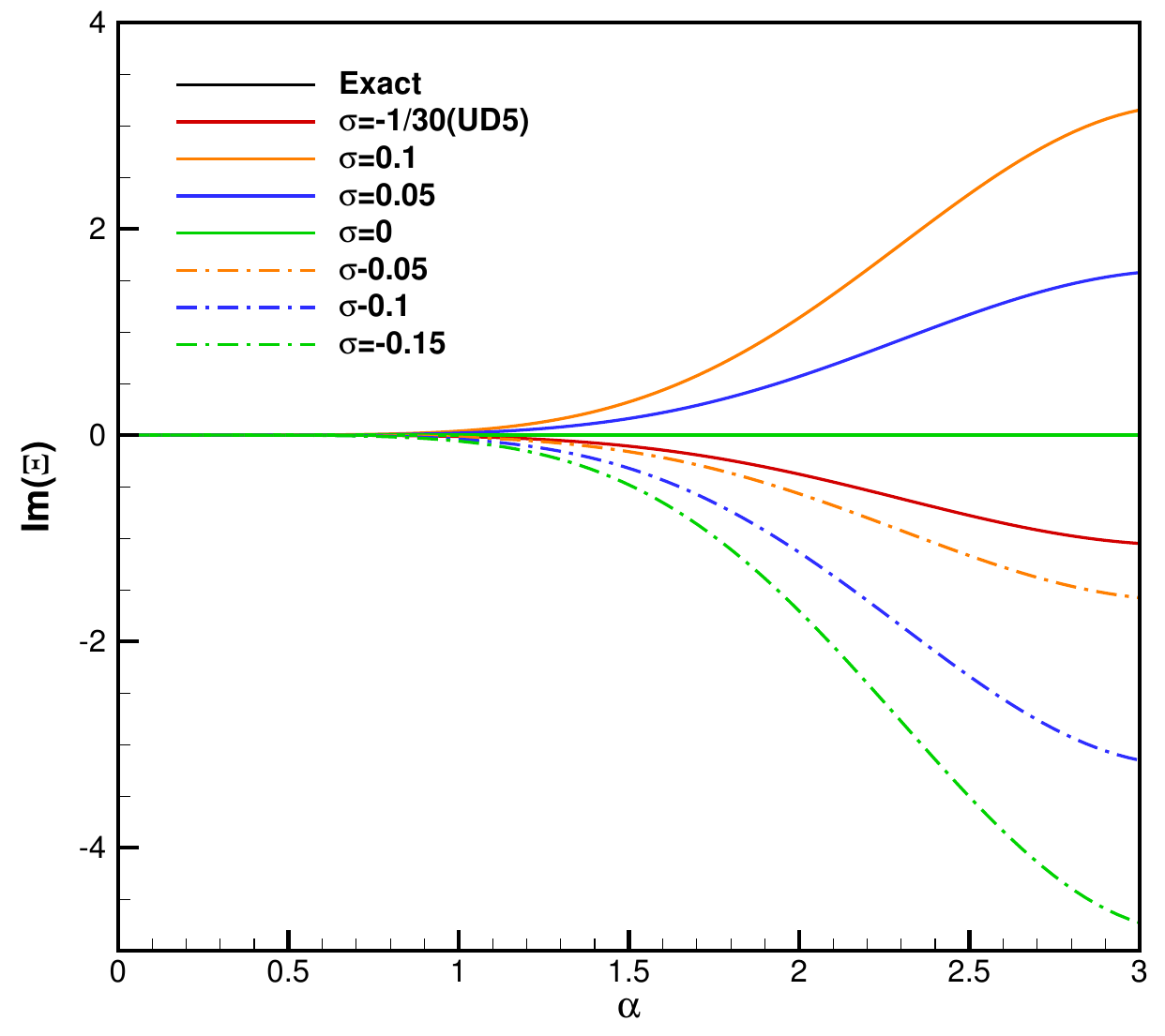}}
\caption{ Changes in dispersion and dissipation characteristics of $\hat{F}_{j+1/2}(\sigma)$, with free parameter $\sigma$ varies from -0.15 to 0.1.}
\label{fig.2}
\end{figure}

The free parameters $\sigma$ in Eqs.(\ref{eq22}-\ref{eq23}) have an impact on the numerical scheme's dispersion and dissipation characteristics. Fig.\ref{fig.2} illustrates the variations in dispersion and dissipation characteristics as the free parameter $\sigma$ varies from $-0.15$ to $0.1$ for $r=3$. It can be observed that, with different values of $\sigma$, the difference scheme corresponding to Eq.(\ref{eq23}) gradually transitions from a $MXD$ scheme to a $SLW$ scheme. Exploiting this property allows the identification of a $SLW$ scheme and a $MXD$ scheme, denoted by the respective free parameters $\sigma_s$ and $\sigma_m$. When the free parameter is set to $\sigma_0=-1/30$, it corresponds to a $5th$-order upwind scheme. The selection and optimization of parameters  $\sigma_s$ and $\sigma_m$ will be discussed in detail in the subsection $4.1$. Thus, we can determine the $SLW$ scheme $\hat{F}_{j+1/2}(\sigma_s)$ and the $MXD$ scheme $\hat{F}_{j+1/2}(\sigma_m)$ required by WGVC scheme.

In addition to the condition mentioned earlier, which ensures the use of a $SLW$ scheme before a discontinuity and a $MXD$ scheme after it, we also aim to ensure that within smooth regions, a weighted combination of the $MXD$ and $SLW$ schemes can recover a $2r-1$ order. In other words, we seek to satisfy:
\begin{equation} \label{eq24}
\hat{F}_{j+1/2}\left(\sigma_{0}\right)=D_{m}\hat{F}_{j+1/2}\left(\sigma_{m}\right)+D_{s}\hat{F}_{j+1/2}\left(\sigma_{s}\right),
\end{equation}
where $D_m$ and $D_s$ are ideal weight coefficients for $\sigma_m$ and $\sigma_s$, and their specific values for $r=3$ and $r=4$ can be found in Table.\ref{tab.3}. The weighting of the $MXD$ and $SLW$ schemes can refer to WENO or Targeted ENO (TENO) schemes \cite{jiang1996,henrick2005,borges2008,fu2016,fu2018}. In \cite{he2014}, the weighting method proposed by Jiang \textit{et al}. \cite{jiang1996} was employed: 
\begin{equation} \label{eq25}
\alpha_{k}=\frac{D_{k}}{\left(\beta_{k}+\varepsilon\right)^{2}},\quad \tilde{\omega}_k=\frac{\alpha_{k}}{\alpha_{m}+\alpha_{s}},\quad k=m,s,
\end{equation} 
where $\varepsilon$ is a small quantity to avoid having a denominator of $0$, usually taken as $1\times 10^{-6}$,  $\beta_k$ represents the smoothness indicator, which satisfies:
\begin{equation} \label{eq26}
\beta_{k}=\sum_{l=1}^{r-1}\int_{x_{j-1/2}}^{x_{j+1/2}}\Delta x^{2l-1}(q_{k}^{(l)})^{2}\mathrm{d}x,\quad k=m,s,
\end{equation}
for $r=3$, we have:
\begin{equation} \label{eq27}
  \begin{aligned}
    &\beta_{m}= \frac{13}{12}\Big(f_{j-2}-2f_{j-1}+f_{j}\Big)^{2}+\frac{1}{4}\Big(f_{j-2}-4f_{j-1}+3f_{j}\Big)^{2}, \\
    &\beta_{s}= \frac{13}{12}\Big(f_j-2f_{j+1}+f_{j+2}\Big)^2+\frac14\Big(3f_j-4f_{j+1}+f_{j+2}\Big)^2,  
  \end{aligned}
\end{equation}
by utilizing Eq.(\ref{eq23})and Eq.(\ref{eq25}), we can derive the expression for WGVC scheme:
\begin{equation} \label{eq28}
\hat{F}_{j+1/2}=\tilde{\omega}_{m}\hat{F}_{j+1/2}\left(\sigma_{m}\right)+\tilde{\omega}_{s}\hat{F}_{j+1/2}\left(\sigma_{s}\right).
\end{equation}

WGVC scheme combines the $SLW$ and $MXD$ schemes by designing smoothness indicators for wave packets and implementing a nonlinear weighting approach. This method, while satisfying the GVC theory, yields some beneficial numerical characteristics: 
\begin{itemize}
    \item The introduction of the nonlinear weighting mechanism reduces the risk of frequent switching between the $SLW$ and $MXD$ schemes. Moreover, the nonlinear mechanism is better suited for handling discontinuities compared to the linear $SLW$/$MXD$ scheme, further enhancing the stability of the scheme.
    
    \item From the discussion in this subsection regarding the $SLW$/$MXD$ scheme, it is evident that optimizing linear difference schemes for better group velocity characteristics often comes at the cost of order. But the use of $SLW$ and $MXD$ schemes expands the space for resolution optimization. It is worth noting that WGVC scheme also possesses the capability to maintain accuracy, as indicated by Eq.(\ref{eq24}). Theoretical and numerical analyses will further validate this point.
\end{itemize}

\subsubsection{A new smoothness indicator for WGVC schemes}
\begin{figure}[ht]
	\centering
	\includegraphics[scale=1.]{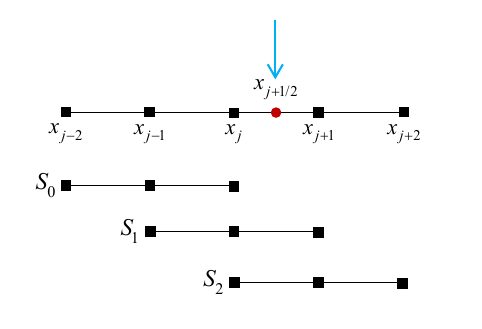}
	\caption{Diagram of WENO scheme stencils.}
	\label{fig.3}
\end{figure}

Furthermore, we aim to enhance the WGVC scheme from three aspects: order-preserving, spectral characteristic optimization, and improved shock-capturing capability. First, let's discuss the ability to preserve accuracy. Subsection $3.1.1$ outlined how to construct the WGVC scheme using the weighted approach of the WENO-JS scheme. In addition to this, other weighting mechanisms can be employed \cite{henrick2005,borges2008}. Both the smoothness indicator and weighting approach in the weighting mechanism can impact the performance of the scheme. In this subsection, we introduce a smoothness indicator suitable for the WGVC scheme and analyze its theoretical accuracy. To begin, let's review the weighted mechanism and theoretical accuracy analysis of the classical WENO scheme. For the sake of discussion, we will focus solely on the $5th$-order, i.e., $r=3$. Fig.\ref{fig.3} illustrates the stencil points for the WENO scheme, which includes three sub-stencils. A convex combination of these three sub-stencils form the WENO scheme: 
\begin{equation} \label{eq29}
   \hat{F}^{JS}_{j+1/2}=\sum_{k=0}^2\omega_k^{JS}h_k,
\end{equation}

\begin{equation} \label{eq30}
   \alpha_{k}=\frac{d_{k}}{\left(\beta_{k}+\varepsilon\right)^{2}}, \quad \omega_{k}^{JS}=\frac{\alpha_{k}}{\sum_{k=0}^2\alpha_k},
\end{equation}
where $d_{k}$ are the ideal weight coefficients, which generate the $5th$-order upwind scheme, $h_{k}$ represent a set linear $3rd$-order linear scheme:
\begin{equation} \label{eq31}
   \begin{aligned}
   &h_0=\frac{1}{3}f_{j-2}-\frac{7}{6}f_{j-1}+\frac{11}{6}f_j, \\
   &h_1=-\frac{1}{6}f_{j-1}+\frac{5}{6}f_j+\frac{1}{3}f_{j+1}, \\
   &h_2=\frac{1}{3}f_j+\frac{5}{6}f_{j+1}-\frac{1}{6}f_{j+2}. 
   \end{aligned}
\end{equation}

The expression for the smoothness indicator $\beta_k$ corresponding to the sub-stencils are:
\begin{equation} \label{eq32}
   \begin{aligned}
   &\beta_{0}= \frac{13}{12}(f_{j-2}-2f_{j-1}+f_{j})^{2}+\frac{1}{4}(f_{j-2}-4f_{j-1}+3f_{j})^{2},  \\
   &\beta_{1}= \frac{13}{12}(f_{j-1}-2f_{j}+f_{j+1})^{2}+\frac{1}{4}(f_{j-1}-f_{j+1})^{2},  \\
   &\beta_{2}= {\frac{13}{12}}(f_{j}-2f_{j+1}+f_{j+2})^{2}+{\frac{1}{4}}(3f_{j}-4f_{j+1}+f_{j+2})^{2}. 
   \end{aligned}
\end{equation}

Expanding Eq.(\ref{eq32}) at $x_j$, we obtain:
\begin{equation} \label{eq33}
   \begin{aligned}
   &\beta_0=f_j^{'2}\Delta x^2+(\frac{13}{12}f_j^{''2}-\frac{2}{3}f_j^{'}f_j^{'''})\Delta x^4+(-\frac{13}{6}f_j^{''}f_j^{'''}+\frac{1}{2}f_j^{'}f_j^{(4)})\Delta x^5+O(\Delta x^6),\\
   &\beta_1=f_j^{'2}\Delta x^2+(\frac{13}{12}f_j^{''2}+\frac{1}{3}f_jf_j^{'''})\Delta x^4+O(\Delta x^6),\\
   &\beta_2=f_j^{'2}\Delta x^2+(\frac{13}{12}f_j^{''2}-\frac{2}{3}f_j^{'''})\Delta x^4+(\frac{13}{6}f_j^{''}f_j^{'''}-\frac{1}{2}f_j^{'}f_j^{(4)})\Delta x^5+O(\Delta x^6).
   \end{aligned}
\end{equation}

Combining Eq.(\ref{eq30}), we can derive the weight coefficients as follows:
\begin{equation} \label{eq34}
   \omega_k^{JS}=\begin{cases}d_k+O(\Delta x^2),&f_j^{'}\ne0,\\d_k+O(\Delta x),&f_j^{'}=0.\end{cases}
\end{equation}

The specific derivation process can refer to \cite{henrick2005,borges2008}. Henrick \textit{et al}. provided sufficient conditions to guarantee the convergence accuracy of the $5th$-order WENO scheme \cite{henrick2005}:
\begin{equation} \label{eq35}
   \omega_{k}-d_{k}=O(\Delta x^{3}).
\end{equation}

From Eq.(\ref{eq34}), it can be observed that the weighting method in Eq.(\ref{eq30}) evidently does not satisfy Eq.(\ref{eq35}). Borges \textit{et al}. introduced an efficient weighting method, commonly referred to as WENO-Z \cite{borges2008}:
\begin{equation} \label{eq36}
   \alpha_{k}=d_{k}\Bigg(1+(\frac{\tau_{5}}{\beta_{k}+\varepsilon})^{q}\Bigg),\quad \omega_{k}^{z}=\frac{\alpha_{k}}{\sum_{k=0}^2\alpha_k},
\end{equation}
where $\tau_5$ is the global smoothness indicator, defined as follows in the $5th$-order case:
\begin{equation} \label{eq37}
   \tau_{5}=\mid \beta_{2}-\beta_{0}\mid.
\end{equation}

The convergence accuracy corresponding to the WENO-Z weighting satisfies:
\begin{equation} \label{eq38}
   \omega_k^z=\begin{cases}d_k+O(\Delta x^{3q}),&f_j^{'}\ne0,\\d_k+O(\Delta x^q),&f_j^{'}=0.\end{cases}
\end{equation}

The weighting method for Eq.(\ref{eq36}) is also based on the smoothness indicator from Eq.(\ref{eq26}). When $q$ is set to $1$, the critical point attains $4th$-order accuracy, and when $q$ is set to $2$, the critical point can be restored to 5th-order accuracy, albeit with an increase in dissipation. In the discussion above regarding the convergence accuracy of the WENO scheme, the most fundamental criterion is whether Eq.(\ref{eq35}) can be satisfied. Unlike the WENO scheme, the sub-stencil for WGVC scheme consists of two $4th$-order accuracy linear schemes, as shown in Eq.(\ref{eq28}). Therefore, the sufficient condition for WGVC scheme to achieve convergence accuracy is:
\begin{equation} \label{eq39}
   \tilde{\omega}_{k}-D_{k}=O(\Delta x^{2}).
\end{equation}

Compared to the WENO scheme, the sufficient condition for critical accuracy convergence is relaxed by one order, making it easier to achieve $5th$-order accuracy. Both of the two weighting methods mentioned above have the same smoothness indicator as in Eq.(\ref{eq26}). Additionally, there are other forms of smoothness indicators \cite{fan2014,ha2013}. In this paper, we have designed a new smooth indicator based on Legendre polynomials for WGVC scheme. Firstly, we assume that each region has a set of local coordinates that satisfy $(x,y,z)\in[-1/2,1/2]\times[-1/2,1/2]\times[-1/2,1/2]$. The Legendre polynomial in $[-1/2,1/2]$ is as follows \cite{balsara2009}:
\begin{equation} \label{eq40}
   q_{0}(\xi)=1,\quad q_{1}(\xi)=\xi,\quad q_{2}(\xi)=\xi^{2}-\frac{1}{12},\quad q_{3}(\xi)=\xi^{3}-\frac{3}{20}\xi,\quad\ldots 
\end{equation}
subsequently, in the sub-stencil, two types of interpolation polynomials were reconstructed:
\begin{equation} \label{eq41}
   q_{k,0}=\begin{cases}f_j+\sum_{l=1}^{(r-1)/2}f_j^{(2l)}q_{2l}(x),&\quad\mathrm{mod}(r,2)=1,\\\\f_j+\sum_{l=1}^{(r-2)/2}f_j^{(2l)}q_{2l}(x),&\quad\mathrm{mod}(r,2)=0.\end{cases}
\end{equation}

\begin{equation} \label{eq42}
   q_{k,1}=\begin{cases}f_j+\sum_{l=1}^{(r-1)/2}f_j^{(2l-1)}q_{2l-1}(x),&\quad\mathrm{mod}(r,2)=1,\\\\f_j+\sum_{l=1}^{r/2}f_j^{(2l-1)}q_{2l-1}(x),&\quad\mathrm{mod}(r,2)=0.\end{cases}
\end{equation}

After obtaining the interpolation polynomial, the smooth indicator can be calculated:
\begin{equation} \label{eq43}
   \beta_{k}=\frac{\sum_{l=1}^{r-1}\int_{x_{j-1/2}}^{x_{j+1/2}}\Delta x^{2l-1}(q_{k,1}^{(l)})^{2}\mathrm{d}x+\varepsilon}{\sum_{l=1}^{r-1}\int_{x_{j-1/2}}^{x_{j+1/2}}\Delta x^{2l-1}(q_{k,0}^{(l)})^{2}\mathrm{d}x+\varepsilon}, \quad k=m,s,
\end{equation}
where $q_k^{(l)}$ is the $l-th$ order approximate partial derivative of $q_k$, $\varepsilon$ is an infinitesimal quantity, and in this paper, we set $\varepsilon=1\times10^{-40}$. For $r=3$, these smoothness indicators can be expressed explicitly as:
\begin{equation} \label{eq44}
   \beta_{k}=\frac{\frac{13}{3}(f_{k}^{''})^{2}+\varepsilon}{(f_{k}^{'})^{2}+\varepsilon},\quad k=m,s,
\end{equation}
the stencil $S_m$ gives:
\begin{equation} \label{eq45}
   \begin{aligned}
     &f_m'=f_{j-2}-4f_{j-1}+3f_j,\\
     &f_m''=f_{j-2}-2f_{j-1}+f_j,
   \end{aligned}
\end{equation}
the stencil $S_s$ gives:
\begin{equation} \label{eq46}
   \begin{aligned}
     &f_s'=3f_j-4f_{j+1}+f_{j+2},\\
     &f_s''=f_j-2f_{j+1}+f_{j+2},
   \end{aligned}
\end{equation}
inspired by \cite{lis2022}, we have adopted the following weighting method:
\begin{equation} \label{eq47}
   \alpha_{k}=D_{k}\Bigg(1+(\frac{\tau_{8}}{\beta_{k}+\varepsilon})^{q}\Bigg),\quad \tilde{\omega}_{k}=\frac{\alpha_{k}}{\alpha_{m}+\alpha_{s}},\quad k=m,s,
\end{equation}
where $q$ is set to $1$, $\tau_8$ is the global smoothness indicator, and it is defined as follows:
\begin{equation} \label{eq48}
   \tau_{8}=\left[D^{(4)}f_{j}\right]^{2}=\left(f_{j-2}-4f_{j-1}+6f_{j}-4f_{j+1}+f_{j+2}\right)^{2}.
\end{equation}

Expanding Eq.(\ref{eq44}-\ref{eq48}) into Taylor series, we derive the weighting coefficients to satisfy: $\tilde{\omega}_{k}-D_{k}=O(\Delta x^{6})$, which can achieve sufficient condition Eq.(\ref{eq39}) for accuracy convergence. The calculation results of the numerical example in section $5$ are also verified this.

\subsection{Enhancement of shock-capturing capability}
Numerical analysis shows that the developed WGVC scheme in this study demonstrates superior scale resolution in the crucial medium to low wavenumber range compared to the upwind scheme. However, as discussed above, such schemes can not achieve the essential non-oscillatory property for shock capturing due to the intrinsic linear nature of the GVC theory. In our previous work \cite{he2014}, a switch function is used to toggle between WGVC schemes and WENO schemes \cite {jiang1996} which are widely used for shock-capturing. In this paper, we further proposed novel WGVC-WENO and WGVC-TENO schemes by embedding the WGVC scheme into WENO/TENO schemes \cite{jiang1996,henrick2005,borges2008,fu2016,fu2018} in order to achieve this complementarity. It is important to note that "\textbf{\textit{embedded}}" is distinct from "\textbf{\textit{hybrid}}" \cite{he2014}. The embedded method developed in this study directly restores WGVC scheme from WENO/TENO schemes in smooth regions, rather than reverting to the linear upwind scheme.

\subsubsection{WGVC-WENO scheme}
To embed the WGVC scheme into WENO scheme, it is necessary to decompose WGVC scheme (Eqs.(\ref{eq28},\ref{eq47})) according to WENO's sub-stencils (Eq.(\ref{eq31})), resulting in:
\begin{equation} \label{eq50}
   \hat{F}_{j+1/2}^{WGVC}=\sum_{k=0}^{2}g_{k}h_{k},
\end{equation}
where $h_k$ represent a set of linear $3rd$-order linear scheme as Eq.(\ref{eq31}), $g_k$ is the corresponding weights of WGVC scheme in WENO's sub-stencils.
\begin{equation} \label{eq51}
   g_{k}=\tilde{\omega}_{m}d_{k}\left(\sigma_{m}\right)+\tilde{\omega}_{s}d_{k}\left(\sigma_{s}\right),
\end{equation}
 $\tilde{\omega}_{m}$ and $\tilde{\omega}_{s}$ in Eq.(\ref{eq51}) can be obtained by Eq.(\ref{eq47}), and $d_{k}(\sigma)$ has the following form:
\begin{equation} \label{eq52}
   \begin{aligned}
   &d_0(\sigma)=-3\sigma,\\
   &d_1(\sigma)=0.5-3\sigma,\\
   &d_2(\sigma)=0.5+6\sigma,
   \end{aligned}
\end{equation}
subsequently, using $g_k$ to replace the ideal weighting coefficient $d_k$ in the WENO-Z scheme (Eq.(\ref{eq36}), we have:
\begin{equation} \label{eq53}
   \alpha_{k}=g_{k}\Bigg(1+(\frac{\tau_{5}}{\beta_{k}+\varepsilon})^{q}\Bigg),\quad \tilde{\omega}_{k}^{z}=\frac{\alpha_{k}}{\sum_{k=0}^2\alpha_k},
\end{equation}
finally, the embedding between the WGVC scheme and WENO scheme can be obtained, which is denoted as WGVC-WENO scheme for simplicity.
\begin{equation} \label{eq54}
    \hat{F}_{j+1/2}=\sum_{k=0}^{2}\tilde{\omega}_{k}^{z}h_{k}.
\end{equation}

The embedded method preserves both the dispersion and dissipation advantages of WGVC scheme while retaining shock-capturing capabilities. As demonstrated in the spectral analysis in Fig.\ref{fig.5}, WGVC scheme and WGVC-WENO scheme exhibit low dispersion and dissipation characteristics in the mid-to-low wavenumber ranges, surpassing even the performance of the $5th$-order upwind scheme.

\subsubsection{WGVC-TENO scheme}
\begin{figure}[ht]
	\centering
	\includegraphics[scale=0.75]{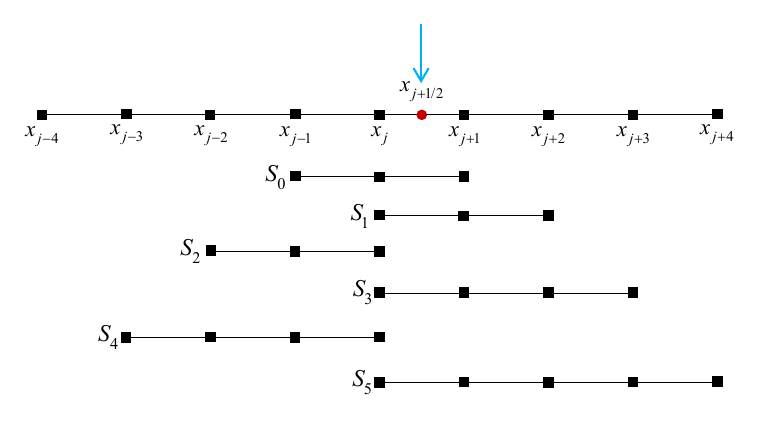}
	\caption{Diagram of TENO scheme stencils.}
	\label{fig.4}
\end{figure}

The TENO scheme proposed by Fu \textit{et al}. \cite{fu2016,fu2018} employs a set of low-order candidate stencils with increasing width and has garnered considerable attention in recent years, particularly in numerical simulations of shock-turbulence interactions. In this subsection, we further embed WGVC scheme into the TENO scheme. Fig.\ref{fig.4} provides an illustrative diagram of the stencil points for the TENO scheme, and as shown in the Fig.\ref{fig.4}, the sub-stencils for the $5th$-order TENO scheme are the same as those in the WENO scheme.

\begin{equation} \label{eq55}
   \hat{F}_{j+1/2}^{T}=\sum_{k=0}^{2}\omega_{k}^{T}h_{k},
\end{equation}
where $h_k$ can be obtained using Eq.(\ref{eq31}), and $\omega_{k}^{T}$ satisfies:
\begin{equation} \label{eq56}
   \omega_{k}^{T}=\frac{d_{k}\delta_{k}}{\sum_{k=0}^{2}d_{k}\delta_{k}},
\end{equation}
$\delta_k$ is a cutoff function with the following form:
\begin{equation} \label{eq57}
   \delta_k=\begin{cases}0,&\text{if}\chi_k<C_T,\\1,&\text{otherwise},\end{cases}
\end{equation}
where $C_T$ is a cutoff threshold that affects the robustness and spectral properties of the scheme. It can be a constant \cite{fu2016} or an adaptive selection \cite{fu2018}. The $C_T$ value in this paper is $1\times 10^{-5}$, and $\chi_k$ in Eq.(\ref{eq57}) satisfies:
\begin{equation} \label{eq58}
   \gamma_{k}=\left(1+\frac{\tau_{5}}{\beta_{k}+\varepsilon}\right)^{6}, \quad \chi_{k}=\frac{\gamma_{k}}{\sum_{k=0}^{2}\gamma_{k}}.
\end{equation}
the smoothness indicators in Eq.(\ref{eq58}) can be obtained by Eq.(\ref{eq32}). Subsequently, replace the ideal weighting coefficients $d_k$ in Eq.(\ref{eq56}) with $g_k$ (Eq.(\ref{eq51})), as follows:
\begin{equation} \label{eq59}
   \tilde{\omega}_k^T=\frac{g_k\delta_k}{\sum_{k=0}^2g_k\delta_k},
\end{equation}
the embedded scheme, denoted as WGVC-TENO scheme, is as follows:
\begin{equation} \label{eq60}
   \hat{F}_{j+1/2}=\sum_{k=0}^{2}\tilde{\omega}_k^{T}h_{k}.
\end{equation}

When comparing Eq.(\ref{eq36}) with Eq.(\ref{eq53}) and Eq.(\ref{eq56}) with Eq.(\ref{eq59}), it is evident that the implementation of the embedded method is straightforward. It only requires replacing the ideal weighting coefficients $d_k$ in the nonlinear weighting with the weights $g_k$, which is the corresponding weights of WGVC scheme in WENO/TENO's sub-stencils. Besides, the WGVC scheme can be embedded into other shock-capturing schemes such as the monotonicity-preserving (MP) scheme \cite{suresh1997,he2016}. Numerical experiments validate the effectiveness of this approach. Additionally, the strategy to enhance the shock-capturing capability of spectral-like optimized schemes includes hybrid methods. In comparison to hybrid methods, the reasons for adopting the embedded method in this paper can be summarized as follows:
\begin{itemize}
    \item The embedded method exhibits superior spectral properties at the crucial medium to low wavenumbers. As demonstrated in Fig.\ref{fig.5} and the preceding discussions, we observe that the embedded method can maintain the spectral properties of the WGVC scheme at medium to low wavenumbers, while the hybrid method may lead to a deterioration in spectral properties.
    \item The embedded method, being independent of shock indicators, possesses greater integrity. It reduces the risk of divergence associated with inaccurate shock recognition and, at the same time, minimizes the intervention of artificial parameters in shock indicators.
\end{itemize}

\section{Spectral properties analysis}
In this section, we discuss the optimization of spectral properties for WGVC schemes. Then, we conduct some quantitative analysis of the spectral properties of the proposed schemes.

\subsection{Determination of free parameters in the WGVC schemes}
 Factors influencing spectral properties not only include the nonlinear weighting mechanism introduced in subsection $3.1.2$ but also the two parameters, $\sigma_m$ and $\sigma_s$, as included in Eq.(\ref{eq28}). Fig.\ref{fig.2} shows that the free parameters $\sigma_m$ and $\sigma_s$ directly affect the numerical scheme's dispersion and dissipation. It is crucial to establish principles for determining these parameters, and optimization principles from spectral-like optimization methods \cite{tam1993,bogey2004,sun2011,li2022} can be employed. One design principle is to ensure that the error functions corresponding to the free parameters achieve a minimum within a certain range of wavenumbers. An example of such an error function, as used by Tam \textit{et al}. in the DRP scheme \cite{tam1993} , is:
\begin{equation} \label{eq49}
   E=\int_{(\kappa\Delta x)_{l}}^{(\kappa\Delta x)_{h}}\left|e_{d}(\kappa\Delta x)\right|^{2}d(\kappa\Delta x)=\int_{(\kappa\Delta x)_{l}}^{(\kappa\Delta x)_{h}}\left|\kappa\Delta x-\tilde{\kappa}\Delta x\right|^{2}d(\kappa\Delta x),
\end{equation}
where $\tilde{\kappa}\Delta x$ represents the imaginary part of the corrected wave number associated with the exact solution, $(\kappa\Delta x)_{l}$ represents the lower bound of the wave number interval, and $(\kappa\Delta x)_{h}$ represents the upper bound of the wave number interval.

The fundamental idea behind such optimization principles is to minimize the error functions associated with dispersion or dissipation, ensuring that the numerical scheme exhibits optimal dispersion or dissipation characteristics within a specified wavenumber range. Additionally, Li \textit{et al}. \cite{li2005} proposed a robustness-based optimization method, referred to as the robustness optimization principle. This approach does not pursue optimal spectral properties but rather aims to enhance the robustness of the numerical scheme. The robustness optimization principle involves determining free parameters by solving a shock tube problem. For the one-dimensional Sod shock-tube problem, under the initial conditions: when $x<1/2$: $u=0,\rho=\rho_t,p=p_t$, when $x{\geqslant}1/2$: $u=0,\rho=0.125,p=0.1$. That is, the left side has high pressure and density, while the right side has low pressure and density. Assuming $p_t=\rho_t$, if $p_s$ increases, it indicates a stronger shock intensity. By solving this problem with different sets of free parameters $\sigma_m$ and $\sigma_s$, the maximum pressure $p_t$ achievable under each set is denoted as $p_{max}$. The value of $p_{max}$ can reflect WGVC scheme's robustness, through numerical experiments, the determined free parameters $\sigma_m$ and $\sigma_s$, as well as the ideal weight coefficients $D_m$ and $D_s$, are shown in Table.\ref{tab.3}.

\begin{table}[ht]  
\centering  
\begin{threeparttable}  
\caption{The values of free parameters and ideal weight coefficients.}  
\label{tab.3}  
\begin{tabular}{cccccc}  
\toprule
  & $\sigma_m$ & $\sigma_s$ & $\sigma_0$ & $D_m$ & $D_s$  \\
\midrule
$r=3$ & -0.07773 & 0 & -1/30 & 0.42883 & 0.57117  \\
$r=4$ & 0.02205 & 0 & 1/40 & 0.32394 & 0.67606 \\
\bottomrule  
\end{tabular}  
\end{threeparttable}  
\end{table}

Numerical analysis results (Fig.\ref{fig.5}) indicate that the spectral properties of the WGVC scheme are superior to those of upwind scheme at medium to low wavenumbers. This endows WGVC scheme with a broader scale recognition capability across various wavenumbers.

\subsection{The effect of shock-capturing mechanism on the final schemes}
In this subsection, we discuss the effect of shock-capturing mechanism on the final schemes. For linear schemes, analytical results can be obtained through Fourier analysis. Spectral analysis of nonlinear numerical methods requires numerical discretization methods, such as the ADR method proposed by Pirozzoli \textit{et al}. \cite{pirozzoli2006} and the simplified ADR method proposed by Li \textit{et al}. \cite{li2005}. We employ Li \textit{et al}.'s ADR analysis method to investigate the proposed schemes.

\begin{figure}[H]
\centering
\subfigure[comparison of dispersion characteristics]{
\includegraphics[width=8cm]{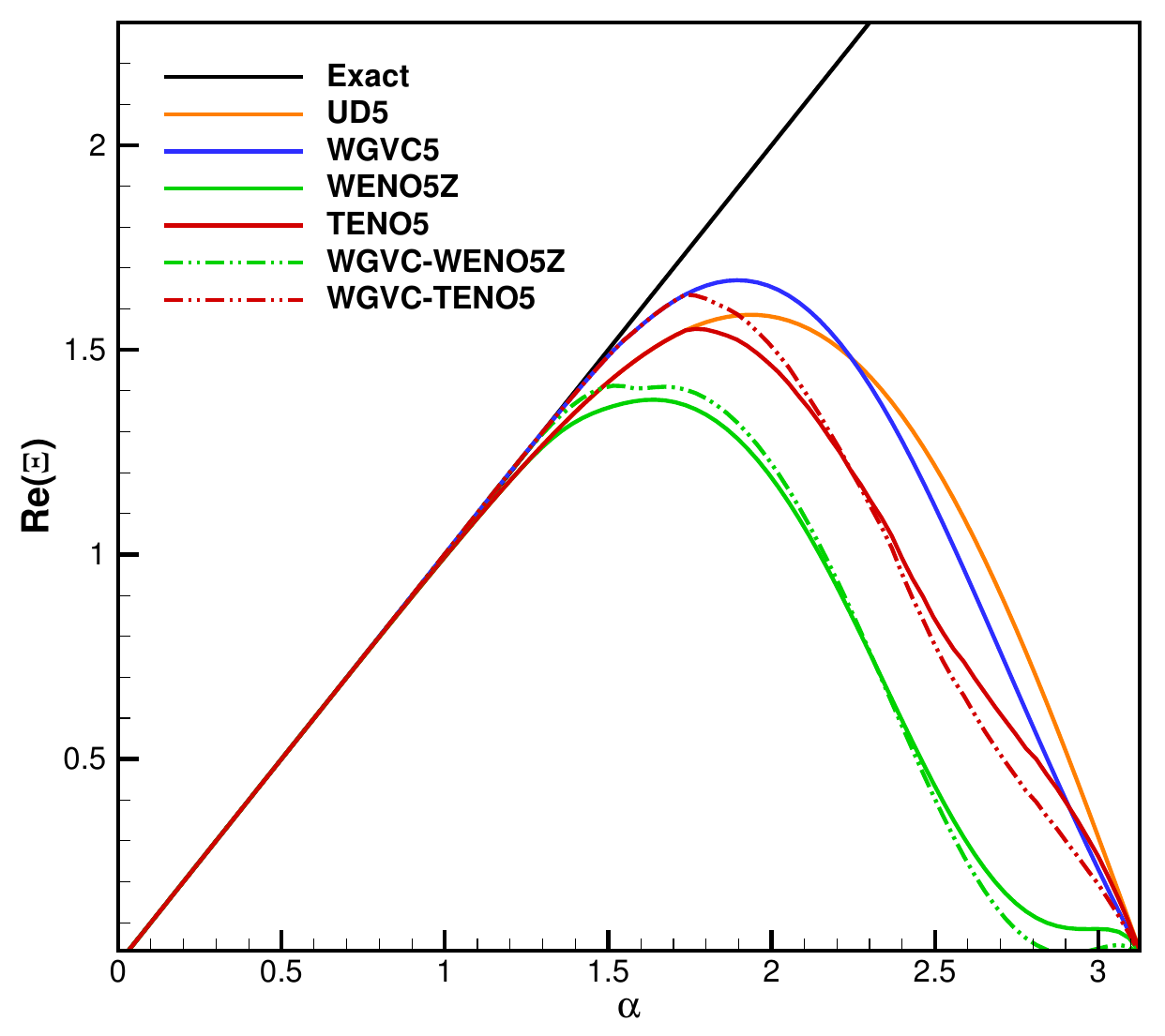}}
\subfigure[comparison of dissipation characteristics]{
\includegraphics[width=8cm]{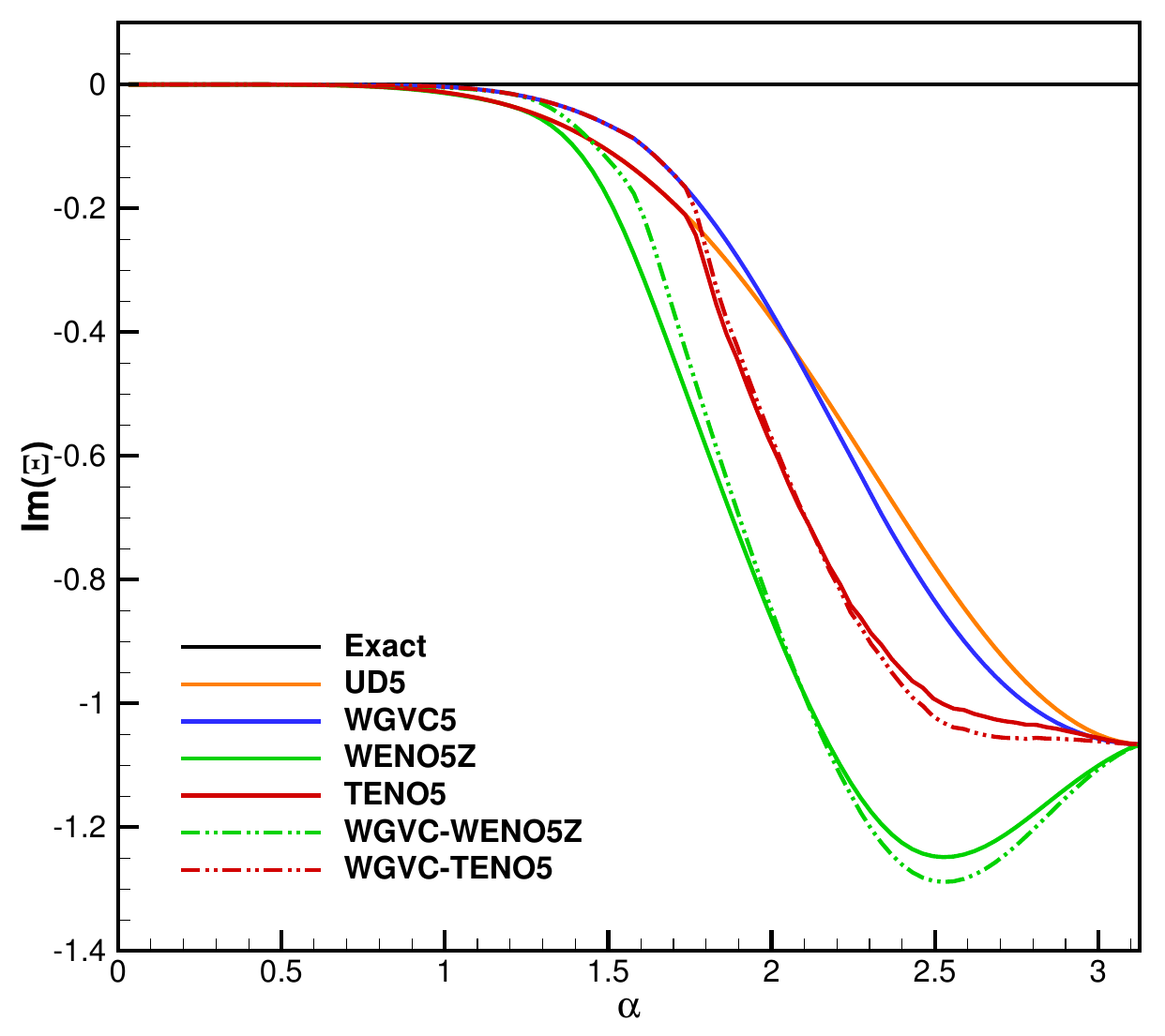}}
\subfigure[comparison of relative  dispersion characteristics]{
\includegraphics[width=8cm]{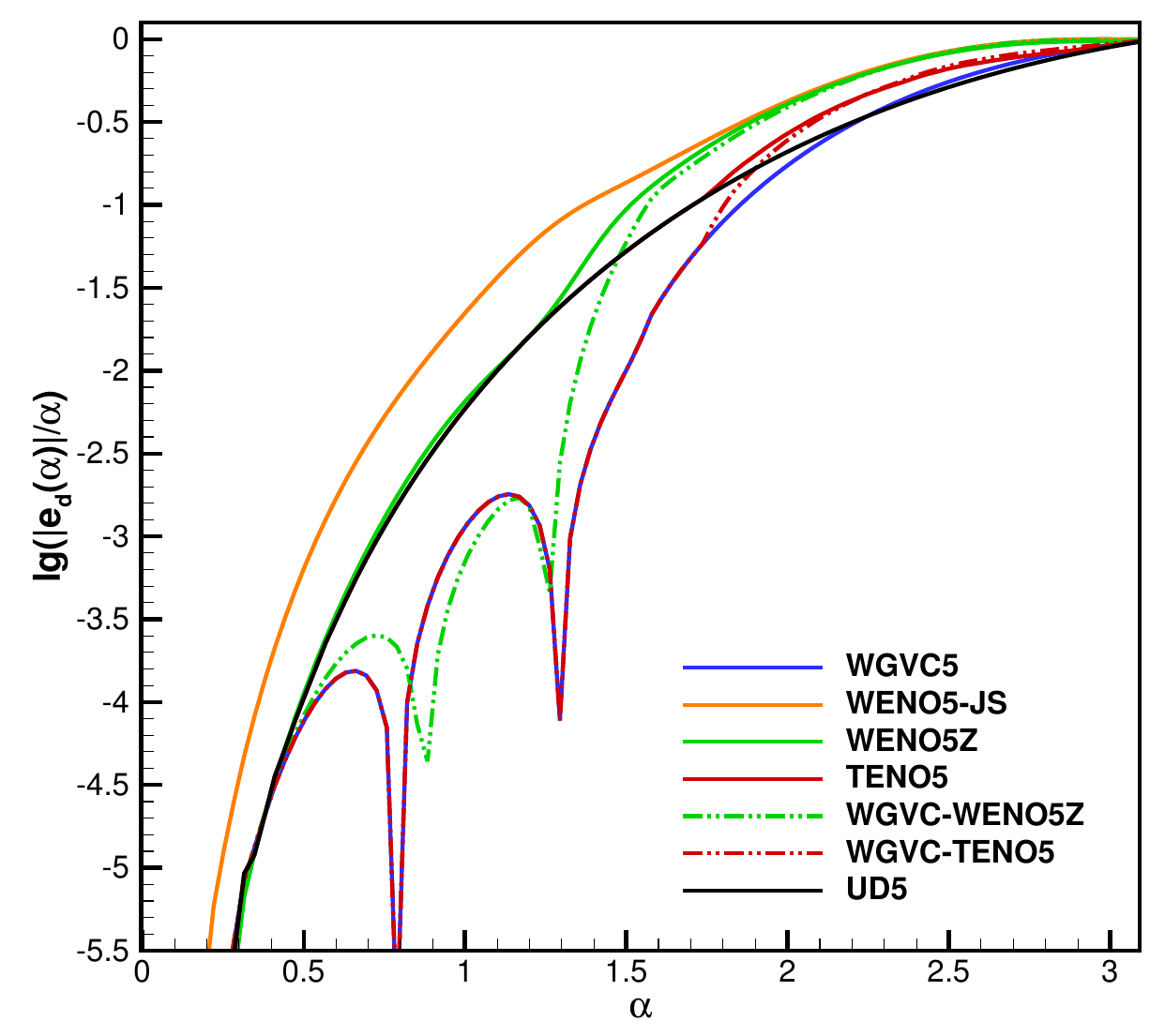}}
\caption{Comparison of spectral properties of different numerical schemes.}
\label{fig.5}
\end{figure}

Fig.\ref{fig.5}(a) compares the dispersion characteristics of different schemes in the wavenumber domain. Fig.\ref{fig.5}(b) compares the dissipation characteristics of different schemes. From Fig.\ref{fig.5}(a) and Fig.\ref{fig.5}(b), it can be observed that within the wavenumber range below $2.2$, the dispersion and dissipation errors of WGVC5 scheme are both lower than those of the $5th$-order upwind scheme. Additionally, the embedded WGVC-WENO5Z and WGVC-TENO5 schemes exhibit lower dispersion and dissipation errors compared to the WENO5Z and TENO5 schemes within the same wavenumber range, thereby preserving the numerical characteristics of WGVC5 scheme. Fig.\ref{fig.5}(c) compares the relative dispersion characteristics of different schemes, where the relative dispersion $e_d$ is defined as \cite{cunha2014}:
\begin{equation} \label{eq61}
   e_{d}(\alpha)=\kappa\Delta x-\tilde{\kappa}\Delta x.
\end{equation}

Cunha \textit{et al}. \cite{cunha2014} pointed out that this relative dispersion error is a direct consequence similar to the accuracy loss observed in spectral-like optimized schemes, primarily present in the low wavenumber range. Once accumulated over longer periods of time and distances, it can significantly impact the accuracy of computational results. From Fig.\ref{fig.5}(c), it can be observed that the WGVC5 scheme, as well as the WGVC-WENO5Z and WGVC-TENO5 schemes, all maintain consistency with the $5th$-order upwind scheme in the low wavenumber range, without incurring additional dispersion error loss.

\section{Numerical experiments}
In this section, we use different numerical cases to assess the new schemes, including 1D linear advection problem, 1D Euler cases, and 2D Euler cases. The uniform mesh is used for both 1D and 2D problems and the Local Lax–Friedrichs (LLF) \cite{lax1954} is utilized for flux splitting. All cases are advanced in time using the $3rd$-order Runge-Kutta method \cite{shu1988}:
\begin{equation} \label{eq62}
   \begin{aligned}
     &u^{(1)}=u^{n}+\Delta t{\mathcal R}(u^{n}), \\
     &\begin{aligned}u^{(2)}=\frac{3}{4}u^{n}++\frac{1}{4}u^{(1)}+\frac{1}{4}\Delta t{\mathcal R}(u^{(1)}),\end{aligned} \\
     &u^{n+1}=\frac{1}{3}u^{n}+\frac{2}{3}u^{(2)}+\frac{2}{3}\Delta t{\mathcal R}(u^{(2)}),
    \end{aligned}
\end{equation}
where ${\mathcal R}$ is the spatial operator used to calculate partial derivative of spatial terms, such as Eq.(\ref{eq8}). For all numerical examples except for accuracy analysis, the time step $\Delta t$ for 1D cases is:
\begin{equation} \label{eq63}
   \Delta t=\left|\frac{\eta\Delta x}{(\mid u\mid+c)}\right|_{\min},
\end{equation}
the time step for 2D cases is:
\begin{equation} \label{eq64}
  \Delta t=\eta\frac{\Delta t_x\Delta t_y}{\Delta t_x+\Delta t_y}, \quad 
  \Delta t_x=\left|\frac{\Delta x}{(\mid u\mid+c)}\right|_{\min}, \quad 
  \Delta t_y=\left|\frac{\Delta y}{(\mid v \mid+c)}\right|_{\min},
\end{equation}
where the CFL number $\eta$ is set to be $0.6$. $c$ is the speed of sound defined by $c=\sqrt{\gamma p/\rho}$. 

\subsection{Accuracy Analysis}
To test the critical accuracy of the proposed schemes, we solve the linear advection equation with the following initial condition (Eq.(\ref{eq1})) \cite{henrick2005}, where $a=1$.
\begin{equation} \label{eq65}
   u(x,t=0)=\sin\biggl(\pi x-\frac{\sin(\pi x)}{\pi}\biggr).
\end{equation}

The computational domain for this case is $[-1,1]$, and the computation time is set to $t=2$. Periodic boundary conditions are applied on both the left and right boundaries. This case has two critical points, satisfying $f^{'}=0$ and $f^{'''}\neq0$. The exact solution corresponding to Eq.(\ref{eq65}) is:
\begin{equation} \label{eq66}
   u_{{e}}(x,t)=\sin\Bigg(\pi(x-t)-\frac{\sin(\pi(x-t))}{\pi}\Bigg).
\end{equation}

The time step is $\Delta t=8\Delta x^{5/3}$ to ensure that the temporal discretization error is sufficiently small, with the spatial discretization error being the dominant factor. Different numerical methods are evaluated using the $L_1$, $L_2$ norms to assess numerical accuracy, and the results are shown in Tables.\ref{tab.4}-\ref{tab.6}.
 
\begin{table}[ht]  
\centering    
\begin{threeparttable}  
\caption{Convergence properties of WGVC5,UD5 for the 1D linear advection equation.}  
\label{tab.4}  
\begin{tabular}{ccccccccc}  
\toprule
\multirow{2}{*}{N} &
\multicolumn{4}{c}{WGVC5}&\multicolumn{4}{c}{UD5}\\
\cmidrule(l){2-5} \cmidrule(l){6-9}  
    & ${L_1}$ & Order & ${L_2}$ & Order & ${L_1}$ & Order & ${L_2}$ & Order\\  
\midrule  
   50  &1.419E-03  &-      &1.274E-03  &-       &1.421E-03 &-      &1.275E-03  &	-\\
   100	&4.415E-05	&5.006	&3.980E-05	&5.001	 &4.421E-05	&5.006	&3.980E-05	&5.001\\
   200	&1.379E-06	&5.001	&1.243E-06	&5.001	 &1.379E-06	&5.003	&1.243E-06	&5.001\\
   400	&4.307E-08	&5.001	&3.881E-08	&5.001	 &4.307E-08	&5.001	&3.881E-08	&5.001\\
   800	&1.346E-09	&5.000	&1.213E-09	&5.000	 &1.346E-09	&5.000	&1.213E-09	&5.000\\
\bottomrule  
\end{tabular}  
\end{threeparttable}  
\end{table} 

\begin{table}[ht]  
\centering    
\begin{threeparttable}  
\caption{Convergence properties of WENO5Z,WGVC-WENO5Z for the 1D linear advection equation.}  
\label{tab.5}  
\begin{tabular}{ccccccccc}  
\toprule
\multirow{2}{*}{N} &
\multicolumn{4}{c}{WENO5Z}&\multicolumn{4}{c}{WGVC-WENO5Z}\\
\cmidrule(l){2-5} \cmidrule(l){6-9}  
    & ${L_1}$ & Order & ${L_2}$ & Order & ${L_1}$ & Order & ${L_2}$ & Order\\  
\midrule  
   50	&1.420E-03  &-	    &1.273E-03  &-		&1.417E-03	&-	    &1.272E-03	&-\\
   100	&4.422E-05	&5.005	&3.979E-05	&4.999	&4.416E-05	&5.004	&3.979E-05	&4.999\\
   200	&1.379E-06	&5.003	&1.243E-06	&5.001	&1.379E-06	&5.001	&1.243E-06	&5.001\\
   400	&4.307E-08	&5.001	&3.881E-08	&5.001	&4.307E-08	&5.001	&3.881E-08	&5.001\\
   800	&1.346E-09	&5.000	&1.213E-09	&5.000	&1.346E-09	&5.000	&1.213E-09	&5.000\\
\bottomrule  
\end{tabular}  
\end{threeparttable}  
\end{table}

\begin{table}[ht]  
\centering    
\begin{threeparttable}  
\caption{Convergence properties of TENO5,WGVC-TENO5 for the 1D linear advection equation.}  
\label{tab.6}  
\begin{tabular}{ccccccccc}  
\toprule
\multirow{2}{*}{N} &
\multicolumn{4}{c}{TENO5}&\multicolumn{4}{c}{WGVC-TENO5}\\
\cmidrule(l){2-5} \cmidrule(l){6-9}  
    & ${L_1}$ & Order & ${L_2}$ & Order & ${L_1}$ & Order & ${L_2}$ & Order\\  
\midrule  
   50	&1.421E-03	&-	    &1.274E-03	&-		&1.419E-03	&-	    &1.274E-03	&-\\
   100	&4.421E-05	&5.006	&3.980E-05	&5.001	&4.415E-05	&5.006	&3.980E-05	&5.001\\
   200	&1.379E-06	&5.003	&1.243E-06	&5.001	&1.379E-06	&5.001	&1.243E-06	&5.001\\
   400	&4.307E-08	&5.001	&3.881E-08	&5.001	&4.307E-08	&5.001	&3.881E-08	&5.001\\
   800	&1.346E-09	&5.000	&1.213E-09	&5.000	&1.346E-09	&5.000	&1.213E-09	&5.000\\
\bottomrule  
\end{tabular}  
\end{threeparttable}  
\end{table}

From the results in Tables.\ref{tab.4}-\ref{tab.6}, it can be observed that all of the methods achieve $5th$-order convergence accuracy. Comparing WGVC5 with the UD5 scheme, WENO5Z with WGVC-WENO5Z, and TENO5 with WGVC-TENO5, it is evident that at low grid resolutions, WGVC5, WGVC-WENO5Z, and WGVC-TENO5 exhibit smaller norm errors and lower numerical dissipation, consistent with the spectral analysis results shown in Fig.\ref{fig.5}.

\subsection{One-dimensional cases}
In this subsection, we solve the 1D Euler equations to assess the characteristics of different numerical schemes. Benchmark cases include shock-tube problems, blast wave problem, and shock–density wave interaction problems. The 1D Euler equations can be written as:
\begin{equation} \label{eq67}
   \frac{\partial\mathbf{U}}{\partial t}+\frac{\partial\mathbf{F}(\mathbf{U})}{\partial x}=0,
\end{equation}
where $\mathbf{U}=(\rho,\rho u,E)^T$, $\mathbf{F}(\mathbf{U})=(\rho u,\rho u^2+p,u(E+p))^T$, $\rho$ is the density, $u$ is the velocity, $p$ is the pressure, $E=\rho(e+u^{2}/2)$ is the total energy. For an ideal gas, the thermal energy $e$ can be obtained through $p=(\gamma-1)\rho e$, and here $\gamma$ is the ratio of the specific heats.

\subsubsection{Shock-tube problems}
The classic shock-tube problems include the Sod shock-tube problem and the Lax shock-tube problem, with the following initial conditions:
\begin{itemize}
    \item Case 1: Sod shock-tube problem \cite{sod1978}
    \begin{equation} \label{eq68}
       (\rho,u,p)=\begin{cases}(1.0,0.0,1.0),&0.0\leq x<0.5,\\
                               (0.125,0.0,0.1),&0.5\leq x\leq1.0.\end{cases}
    \end{equation}
    
    \item Case 2: Lax shock-tube problem \cite{lax1954}
    \begin{equation} \label{eq69}
       (\rho,u,p)=\begin{cases}(0.445,0.698,3.528),&0.0\le x<0.5,\\
                            (0.500,0.000,0.571),&0.5\le x\le1.0.\end{cases}
     \end{equation}
\end{itemize}

\begin{figure}[H]
\centering
\subfigure[Density profiles of Sod shock-tube problem]{
\includegraphics[width=8cm]{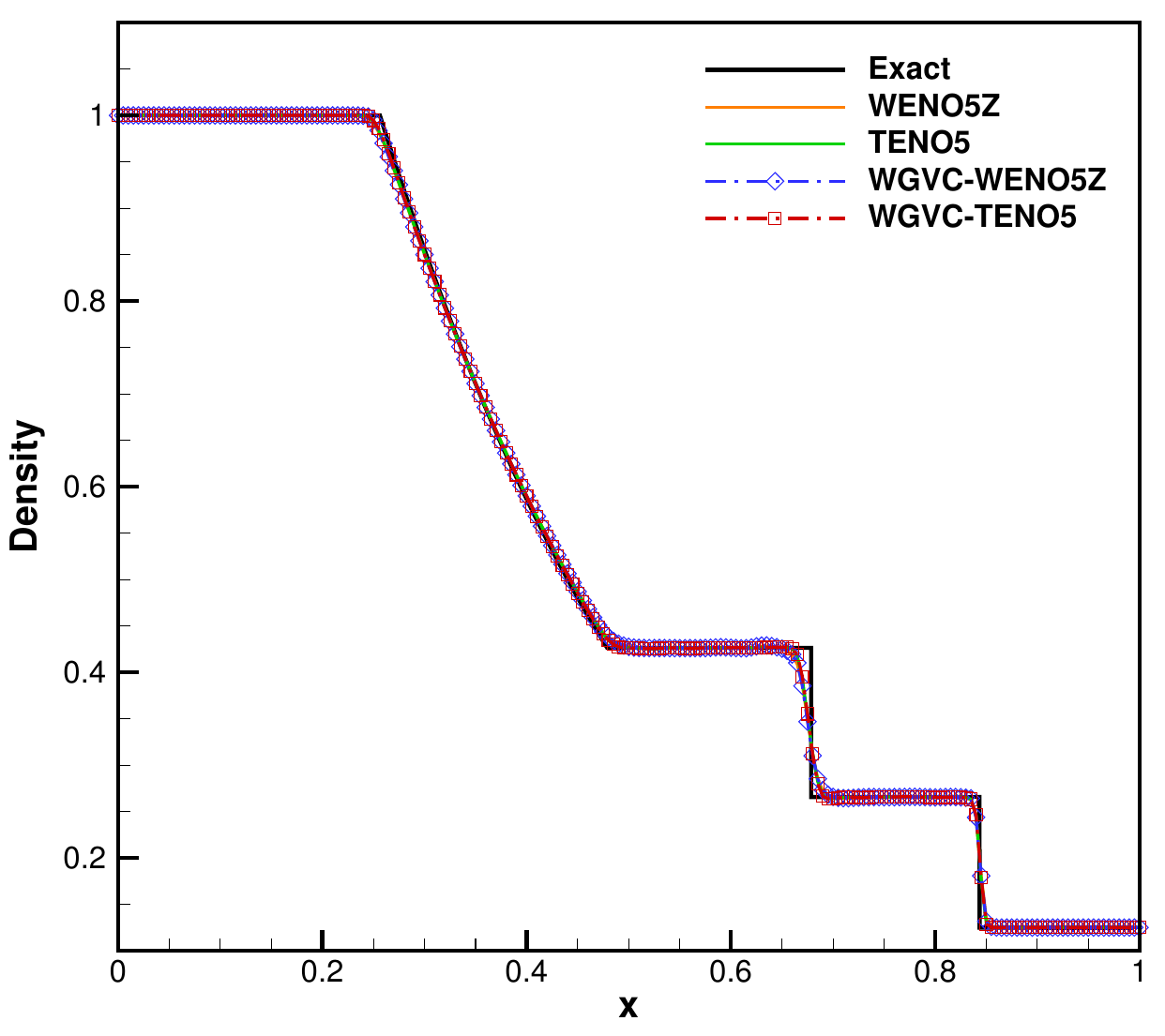}}
\subfigure[Enlarged view]{
\includegraphics[width=8cm]{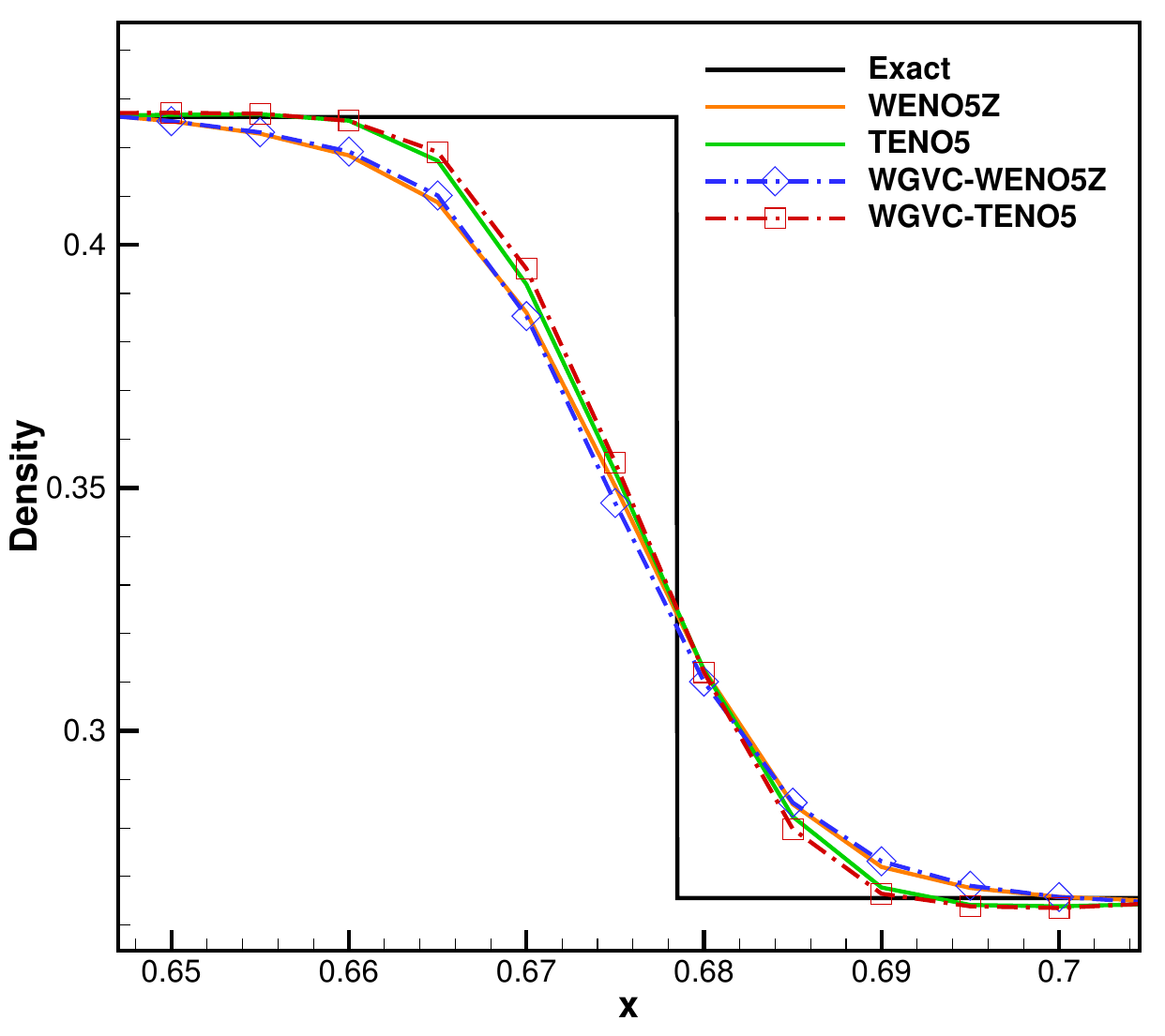}}
\caption{Density profiles of Sod shock-tube problem at $t=0.2$ with 200 grid number.}
\label{fig.6}
\end{figure}

\begin{figure}[H]
\centering
\subfigure[Density profiles of Lax shock-tube problem]{
\includegraphics[width=8cm]{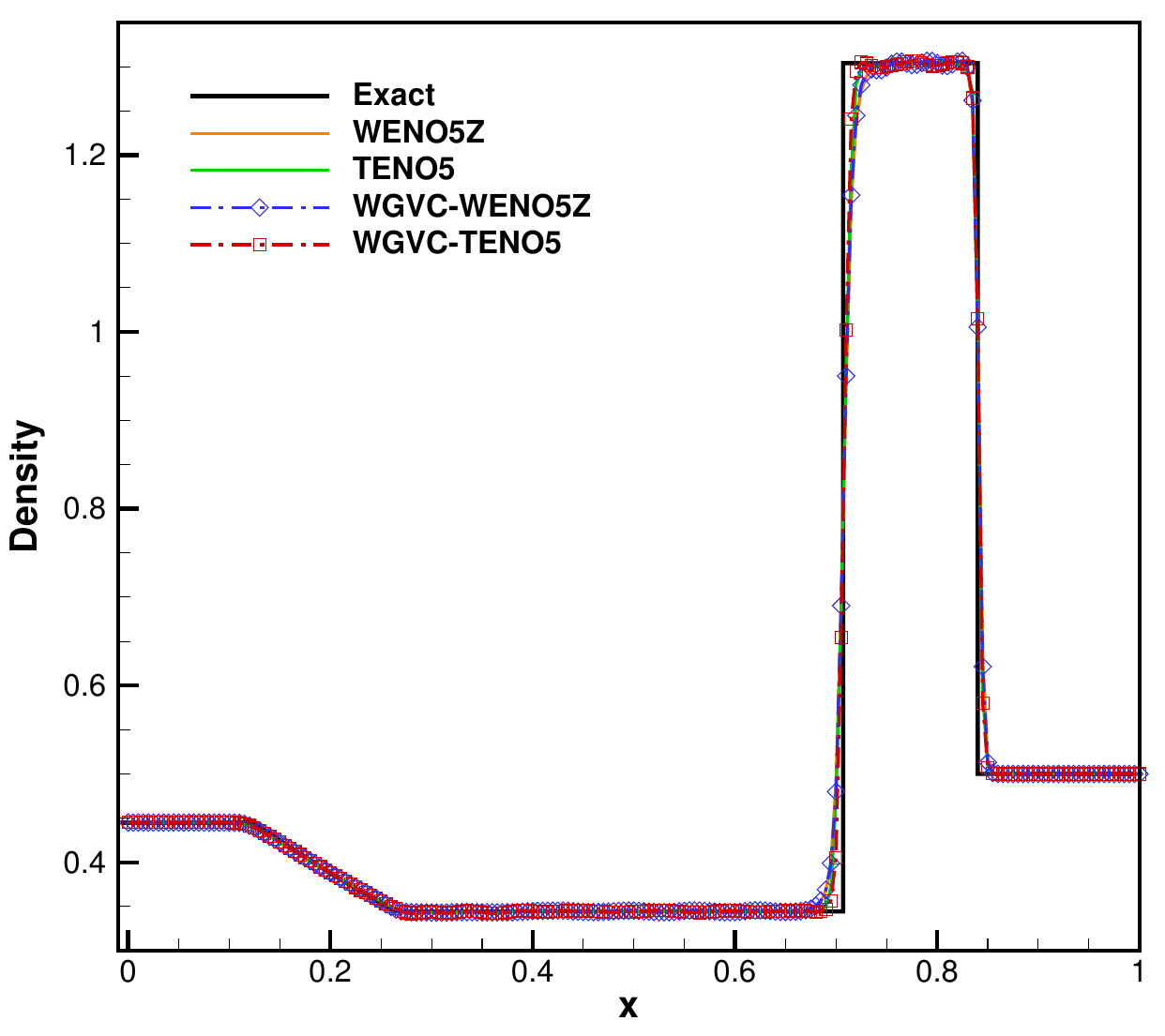}}
\subfigure[Enlarged view]{
\includegraphics[width=8cm]{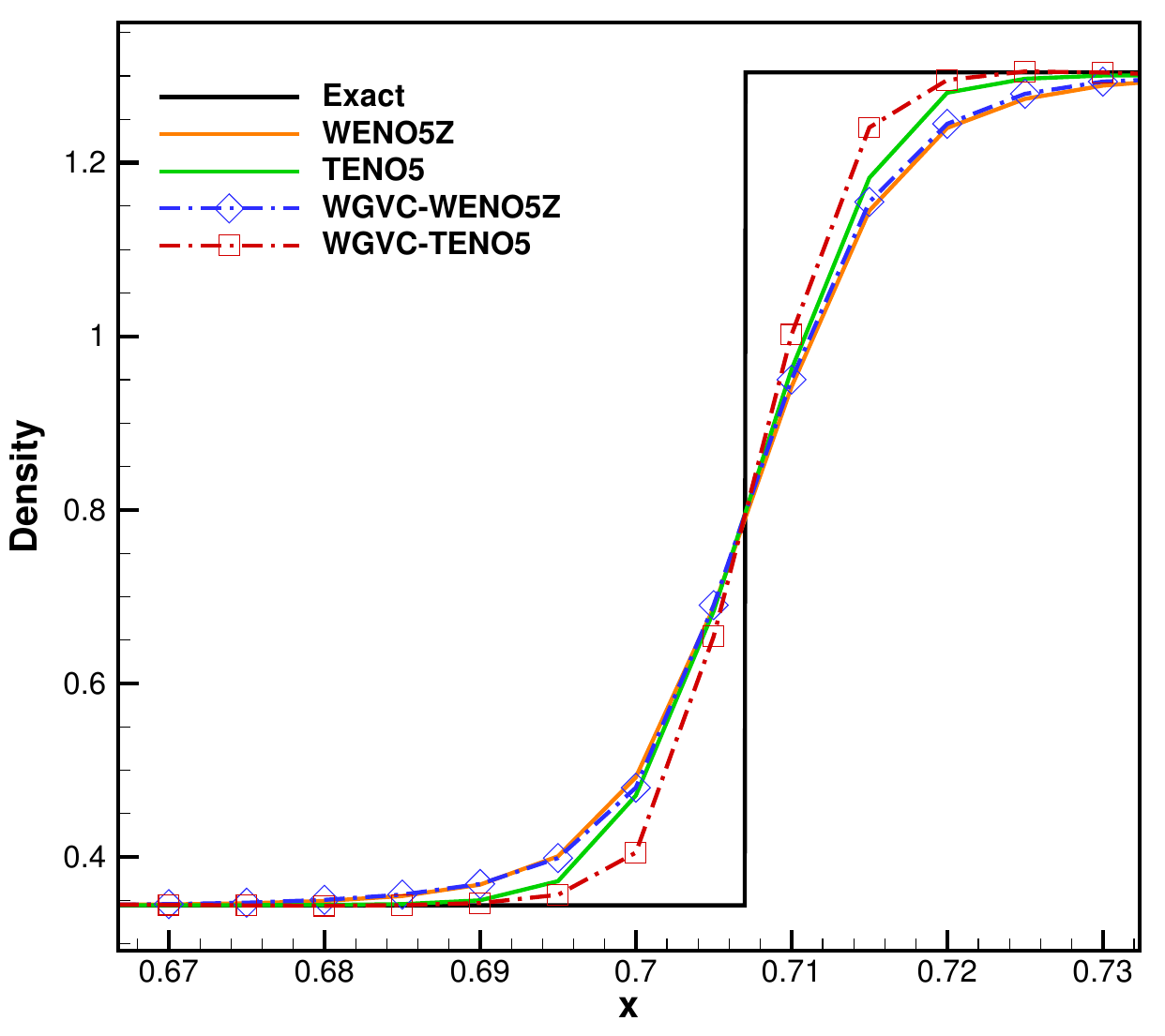}}
\caption{Density profiles of Lax shock-tube problem at $t=0.14$ with 200 grid number.}
\label{fig.7}
\end{figure}

Both Case 1 and Case 2 have a computational domain of [0,1], with tight boundary conditions applied on both left and right boundaries. The calculation time for Case 1 is $t=0.2$, and for Case 2, it is 0.14, with a grid number of $N=200$. From the density results shown in Fig.\ref{fig.6} and Fig.\ref{fig.7}, it can be observed that the calculations with WGVC-WENO5Z and WGVC-TENO5 are closer to the reference solution, and they exhibit higher resolution compared to the corresponding WENO5Z and TENO5 results.

\subsubsection{Blast wave problem}
In this case, the blast wave problem \cite{woodward1984} is considered for verifying the performance of WENO5Z, TENO5, WGVC-WENO5Z and WGVC-TENO5 with the initial flow field:
\begin{equation} \label{eq70}
   (\rho,u,p)=\begin{cases}(1,0,1000), &0\leq x<0.1,\\
   (1,0,0.01), &0.1\leq x<0.8,\\
   (1,0,100), &0.8\leq x\leq1.\end{cases}
\end{equation}

\begin{figure}[ht]
\centering
\subfigure[Density profiles of blast wave problem]{
\includegraphics[width=8cm]{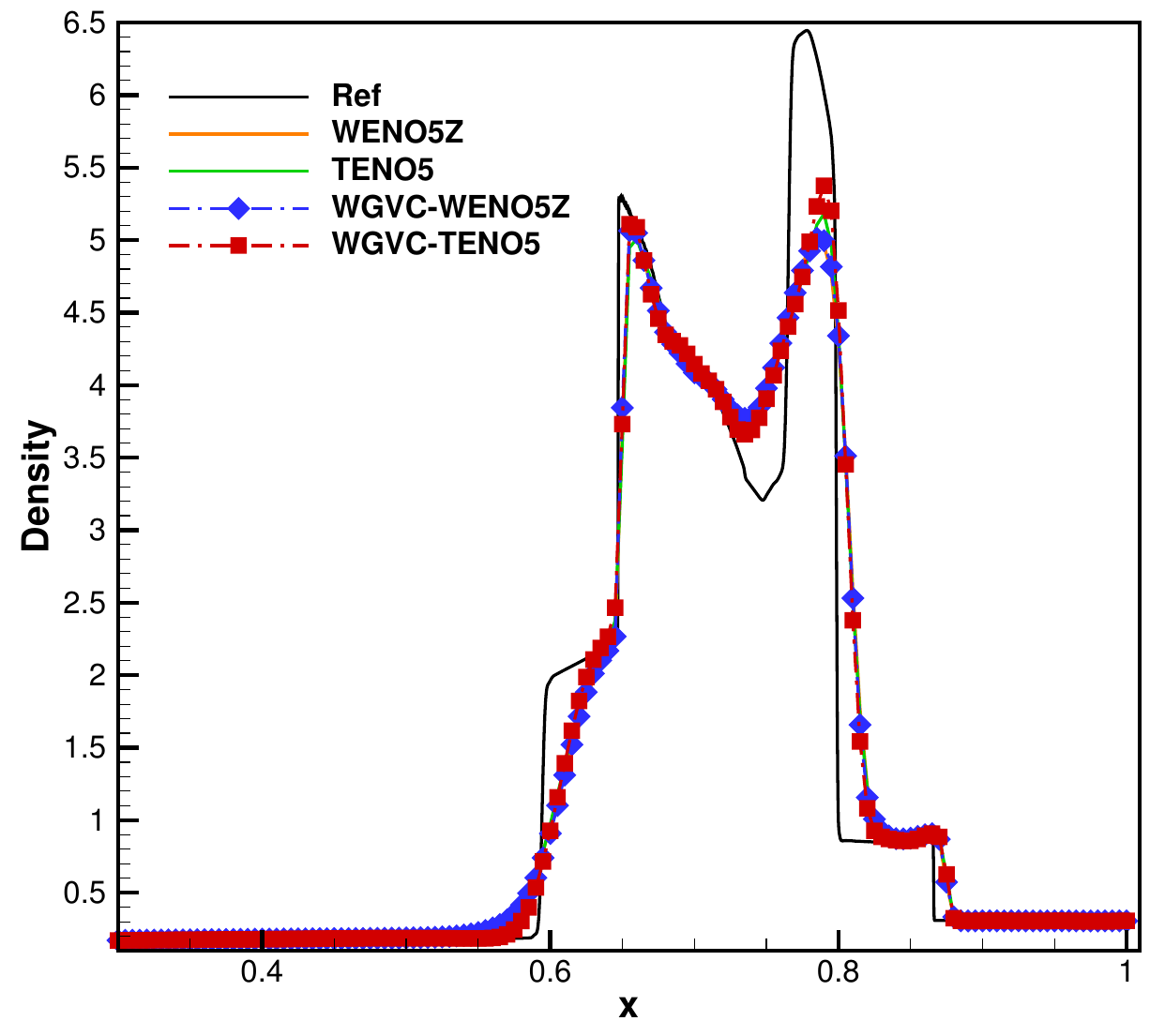}}
\subfigure[Enlarged view]{
\includegraphics[width=8cm]{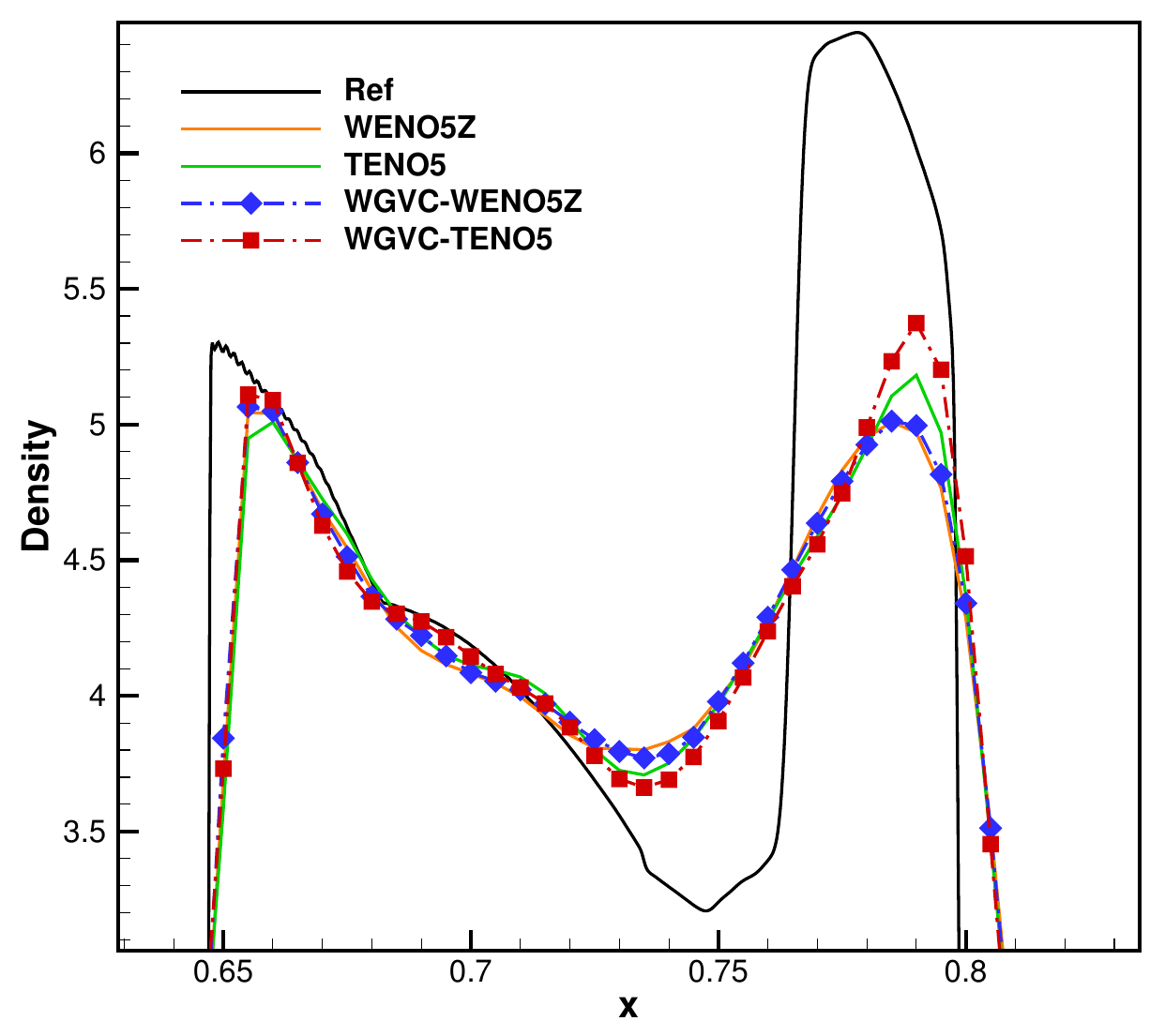}}
\caption{Density profiles of blast wave problem at $t=0.038$ with 200 grid number.}
\label{fig.8}
\end{figure}

The computational domain for this case is defined as $[0,1]$, with reflecting boundary conditions applied on both ends. Fig.\ref{fig.8} illustrates the numerical outcomes with $N=200$ at $t=0.038$. Given the absence of an exact solution for this case, a comparative analysis is conducted using WENO5Z with N=4000. As depicted in Fig.\ref{fig.8}, both WGVC-WENO5Z and WGVC-TENO5 demonstrate enhanced performance compared to WENO5Z and TENO5, particularly in the turning region.

\subsubsection{Shock–density wave interaction}
\begin{itemize}
    \item  Case 1: Shu-Osher problem \cite{shu1989}
    \end{itemize}
    
The Shu-Osher problem is a one-dimensional case involving the interaction of a Mach 3 shock and entropy wave. and it is mainly used to evaluate the resolution of the strong and the small waves. Primarily, it serves as a benchmark to assess the resolution capabilities of numerical methods for both strong and small waves. The initial condition is as follows:
\begin{equation} \label{eq71}
   (\rho,u,p)=\begin{cases}(3.857143,2.629369,10.333333),&0.0\leq x\leq1.0,\\
   (1+0.2\sin(5x),0,1),&1.0<x\leq10.0.\end{cases}
\end{equation}
        
\begin{figure}[H]
\centering
\subfigure[Density profiles of Shu-Osher problem]{
\includegraphics[width=8cm]{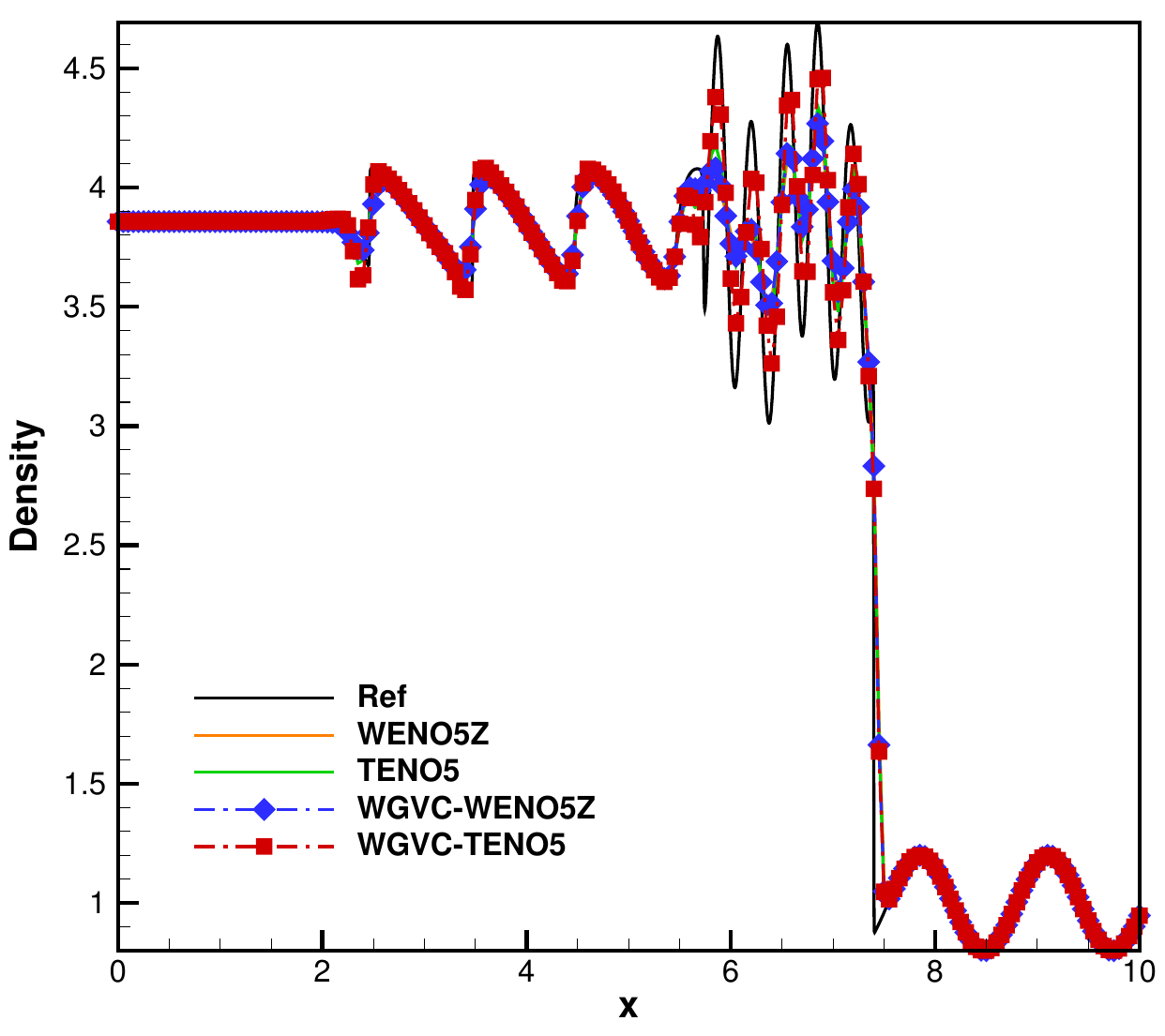}}
\subfigure[Enlarged view]{
\includegraphics[width=8cm]{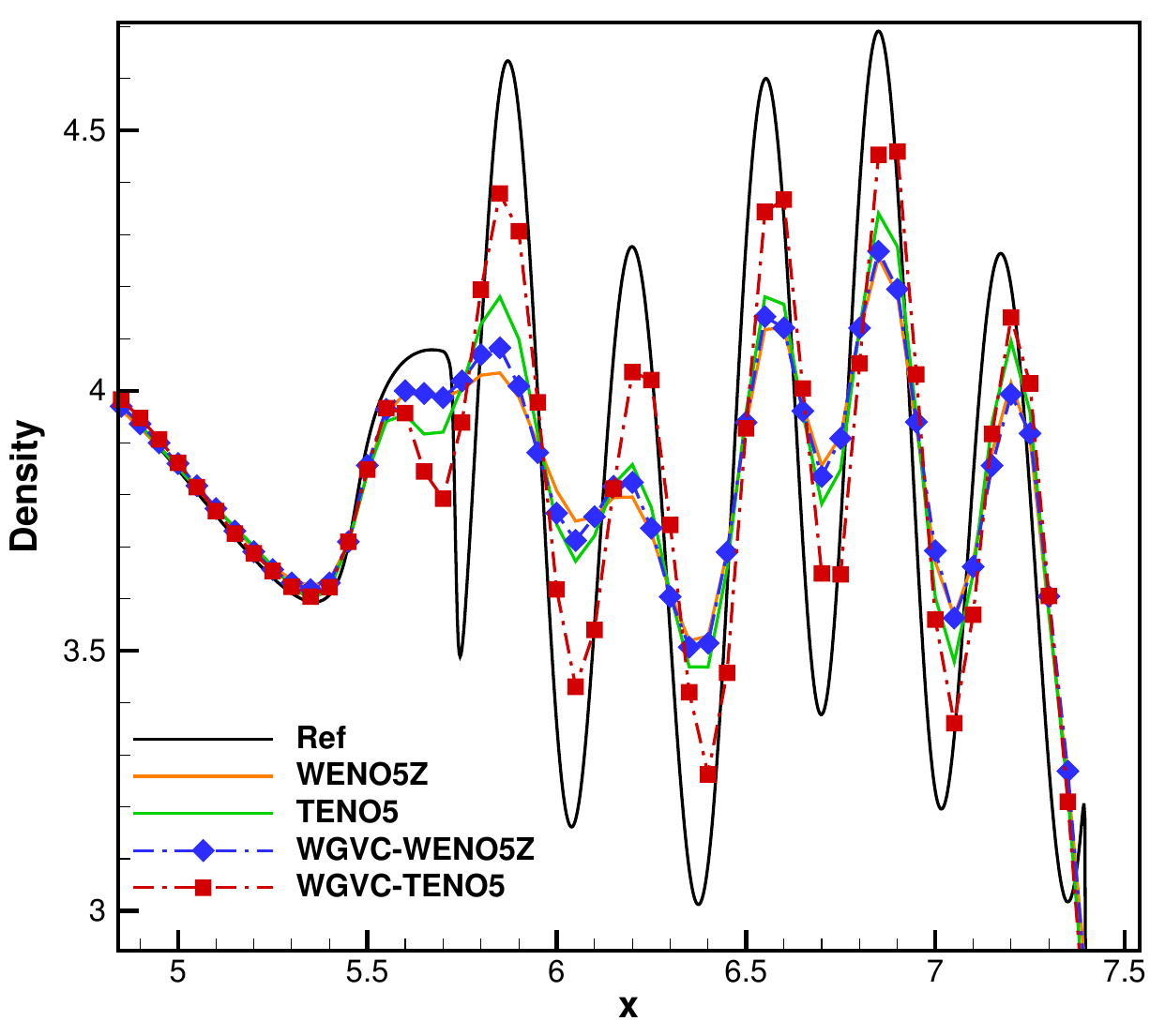}}
\caption{Density profiles of Shu-Osher problem at $t=1.8$ with 200 grid number.}
\label{fig.9}
\end{figure}

The computational time for this case is denoted as $t=1.8$, and 200 grid points were employed to address the problem. Furthermore, the numerical solution with $N=4000$ using WENO5Z is regarded as the reference solution, given the absence of an analytic solution. As depicted in Fig.\ref{fig.9}, both WGVC-WENO5Z and WGVC-TENO5 exhibit significantly improved results compared to WENO5Z and TENO5, respectively, under the same grid resolution.

\begin{itemize}
    \item Case 2: Titarev-Toro problem \cite{titarev2004}
\end{itemize}

The Titarev-Toro problem is an extension of the Shu-Osher problem, introducing increased complexity arising from higher frequency entropy waves. The initial condition is as follows:
\begin{equation} \label{eq72}
   (\rho,u,p)=\begin{cases}(1.515695,0.523346,1.805),&-5\leq x\leq-4.5,\\
   (1+0.1\sin(20\pi x),0,1),&-4.5\leq x\leq5.\end{cases}
\end{equation}

\begin{figure}[ht]
\centering
\subfigure[Density profiles of Titarev-Toro problem]{
\includegraphics[width=8cm]{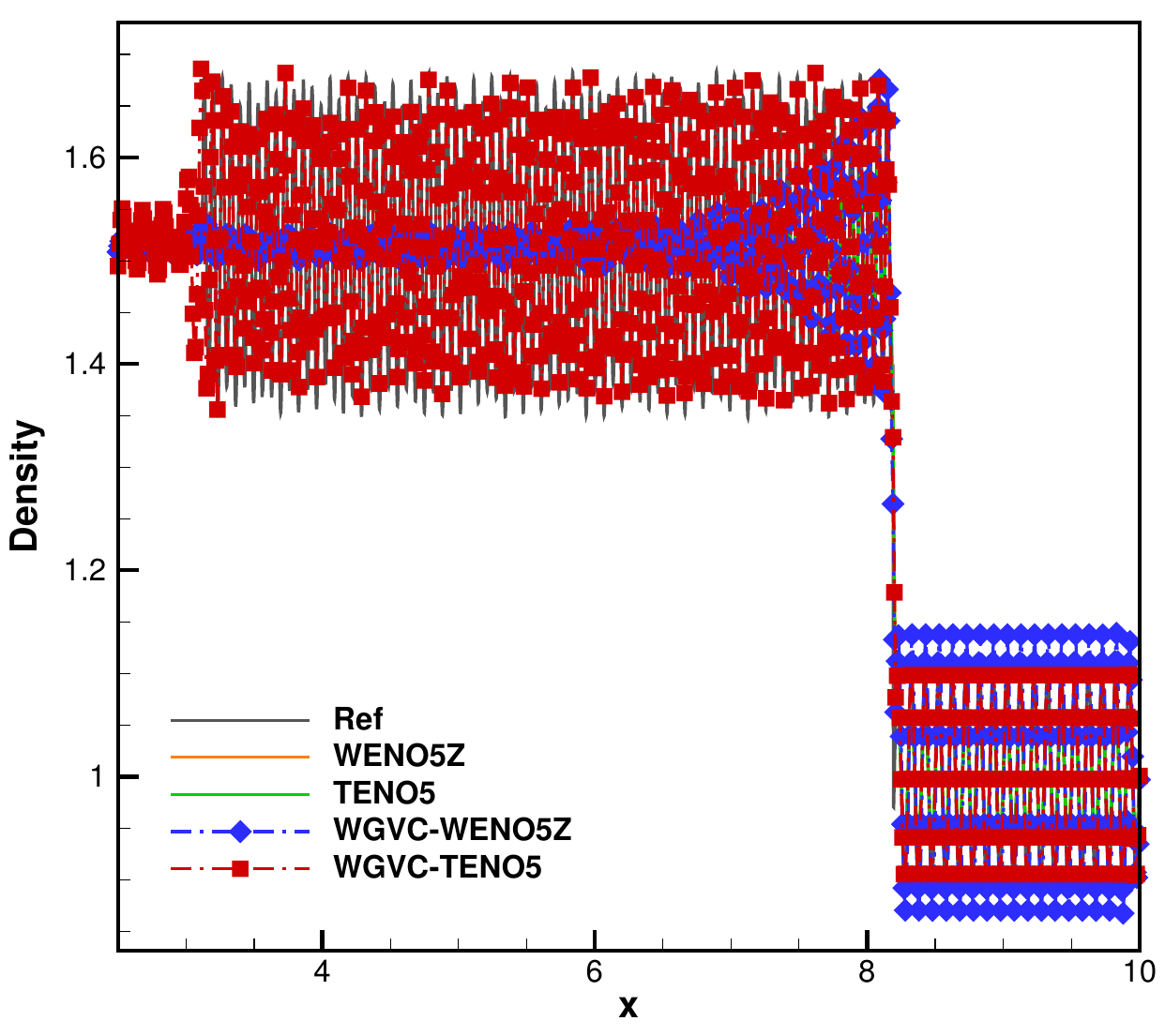}}
\subfigure[Enlarged view]{
\includegraphics[width=8cm]{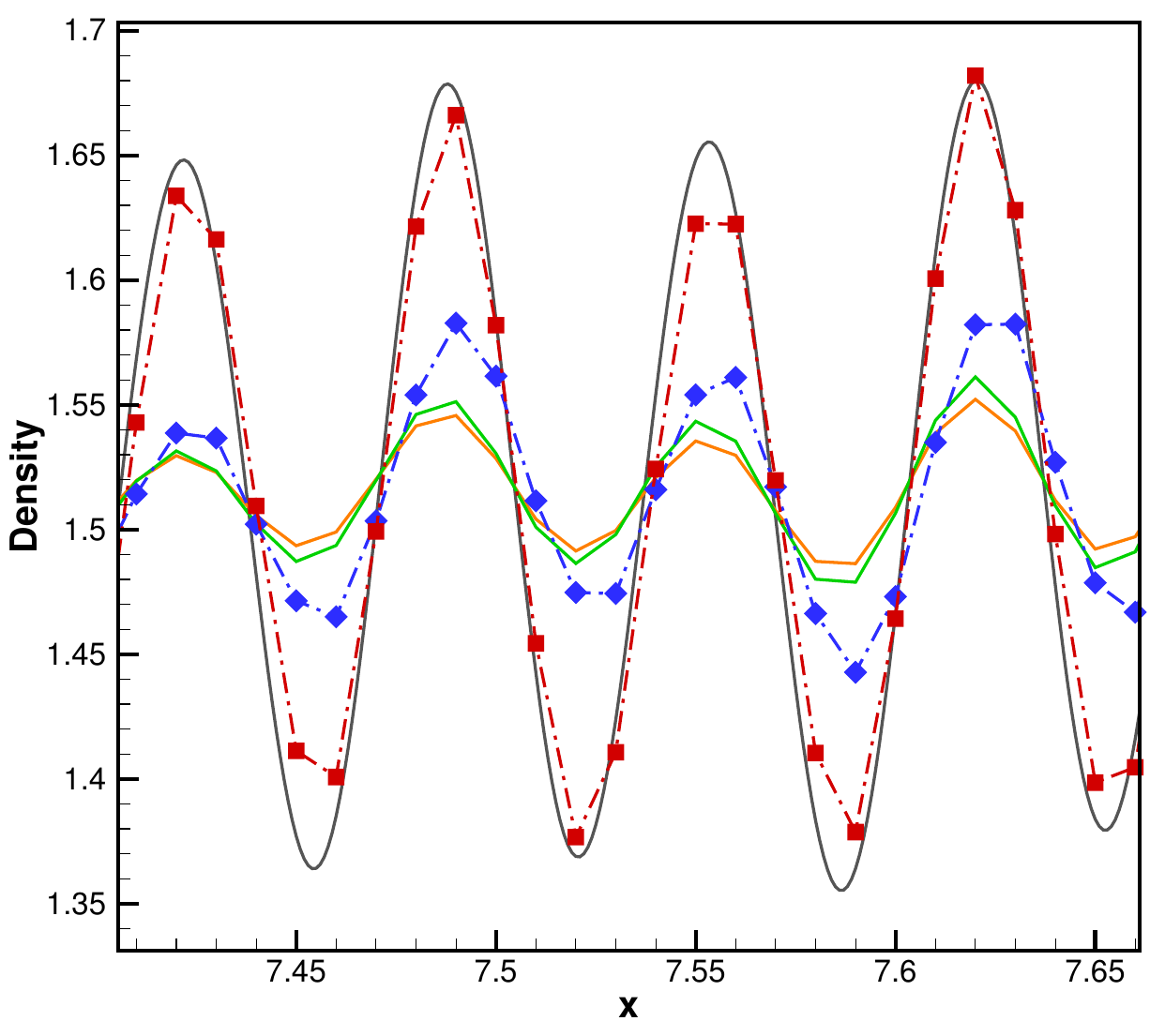}}
\caption{Density profiles of Titarev-Toro problem at $t=5$ with 1000 grid number.}
\label{fig.10}
\end{figure}

The Titarev-Toro problem is addressed using WENO5Z, TENO5, WGVC-WENO5Z, and WGVC-TENO5 schemes with a grid resolution of 1000. The computational time is set at $t=5.0$. Additionally, the numerical solution obtained with WENO5Z at $N = 10000$ is employed as the reference solution. Fig.\ref{fig.10} displays the density profiles of the Titarev-Toro problem, illustrating the decay of high-frequency waves after passing the shock wave. Notably, the numerical solutions of WGVC-TENO5 and WGVC-WENO5Z exhibit sharper profiles compared to TENO5 and WENO5Z at most peaks. This suggests that WGVC-TENO5 and WGVC-WENO5Z demonstrate superior resolution in regions characterized by rapidly changing waves.

\subsection{Two-dimensional cases}
Furthermore, the 2D Euler equations with the following form are solved:
 \begin{equation} \label{eq73}
   \frac{\partial\mathbf{U}}{\partial t}+\frac{\partial\mathbf{F}(\mathbf{U})}{\partial x}+\frac{\partial\mathbf{G}(\mathbf{U})}{\partial y}=0,
\end{equation}
where:
\begin{equation} \label{eq74}
   \begin{aligned}
        &\mathbf{U}=\left(\rho,\rho u,\rho v,E\right)^{T}, \\
        &\mathbf{F}(\mathbf{U})=\left(\rho u,\rho u^{2}+p,\rho uv,u(E+p)\right)^{T}, \\
        &\mathbf{G}(\mathbf{U})=\left(\rho v,\rho uv,\rho v^{2}+p,v(E+p)\right)^{T}.
\end{aligned}
\end{equation}

The 2D equations can be obtained by solving Eq.(\ref{eq67}) through a dimension by dimension manner \cite{jiang1996}. Benchmark cases for 2D Euler equations include the double-Mach reflection problem, Rayleigh-Taylor instability problem, and 2D Riemann problems.

\subsubsection{Double Mach Reflection Problem}
The double-Mach reflection problem \cite{woodward1984} presents a two-dimensional case featuring a right-moving Mach 10 shock wave situated at $x=1/6$, $y=0$, forming a 60-degree angle with the increasing direction. The parameters preceding the shock wave are denoted as $\rho=1.4$, $p=1.0$, $\gamma=1.4$. The exact post-shock solutions are employed for the bottom boundary when $0\leq x\leq 1/6$, and reflecting boundary conditions are applied for other intervals. The upper boundary solutions are imposed to accurately depict the motion of the Mach 10 shock wave. Inflow and outflow boundary conditions are implemented for the left and right boundaries, respectively. The initial condition is as follows:
\begin{equation} \label{eq75}
   (\rho,u,v,p)=\begin{cases}(8,7.145,-4.125,116.5),&y>\sqrt{3}\left(x-1/6\right),\\ 
                                 (1.4,0,0,1),&y\leq\sqrt{3}\left(x-1/6\right).\end{cases}
\end{equation}

\begin{figure}[H]
\centering
\subfigure[WENO5Z]{
\includegraphics[width=8cm]{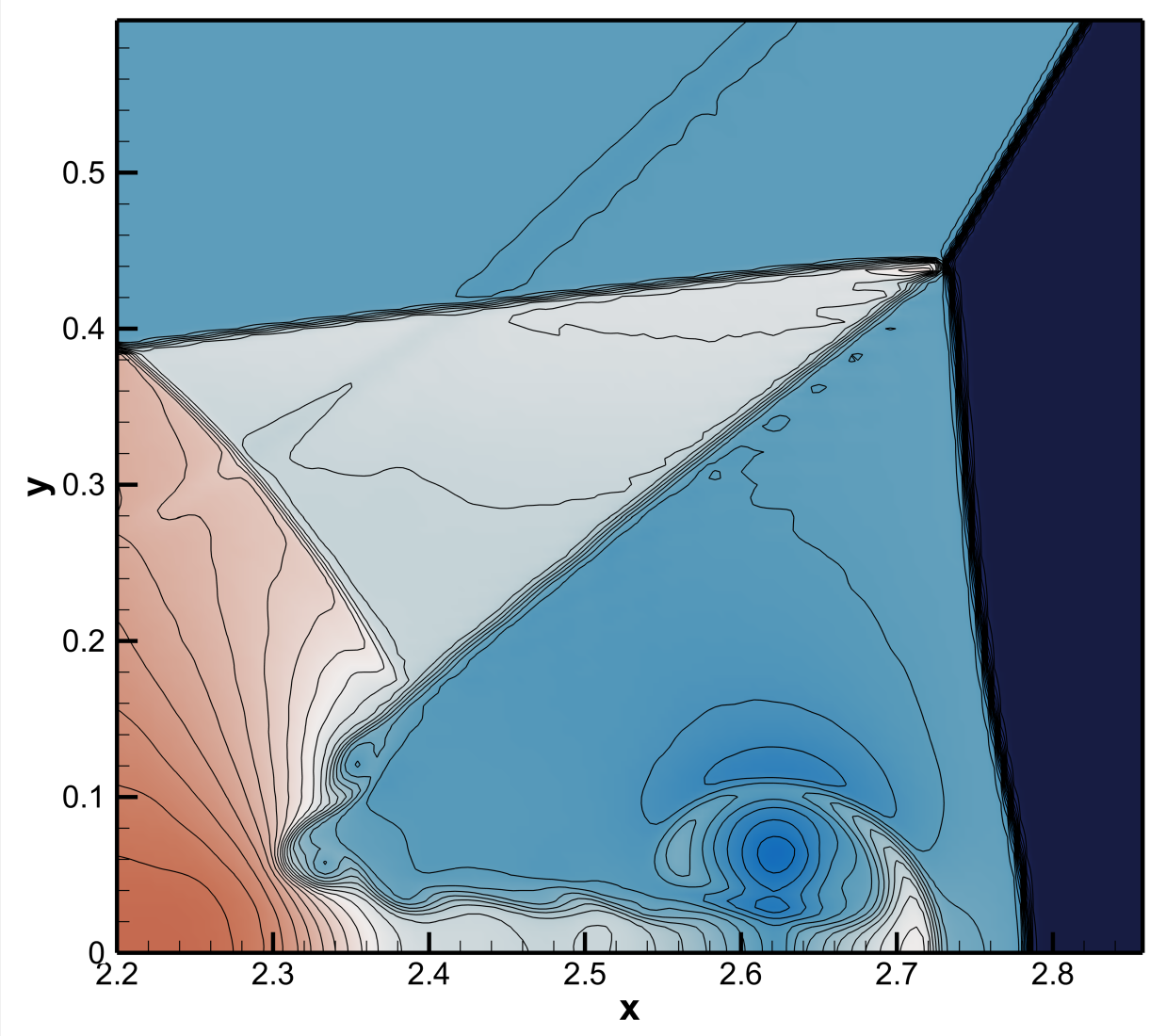}}
\subfigure[WGVC-WENO5Z]{
\includegraphics[width=8cm]{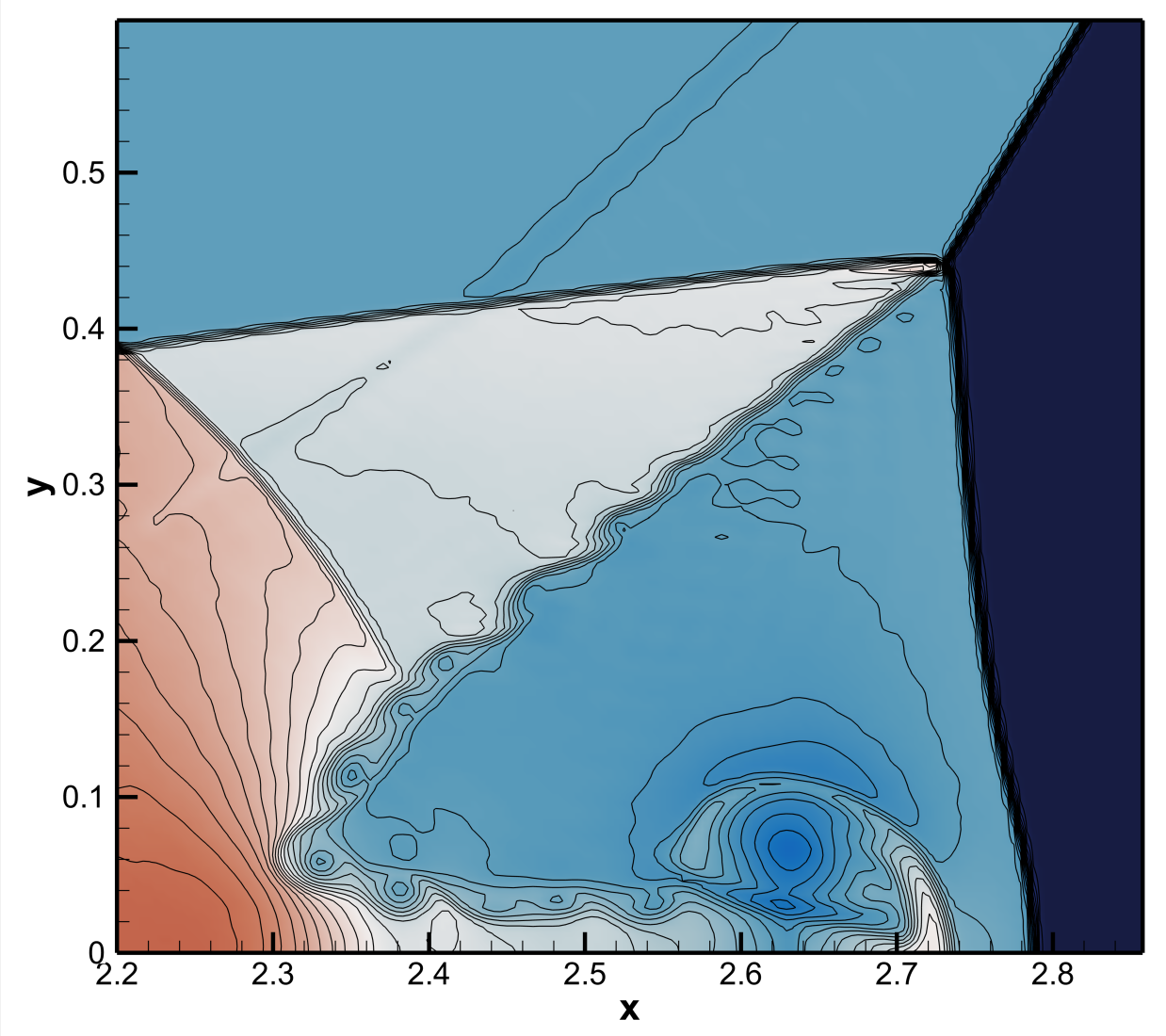}}
\subfigure[TENO5]{
\includegraphics[width=8cm]{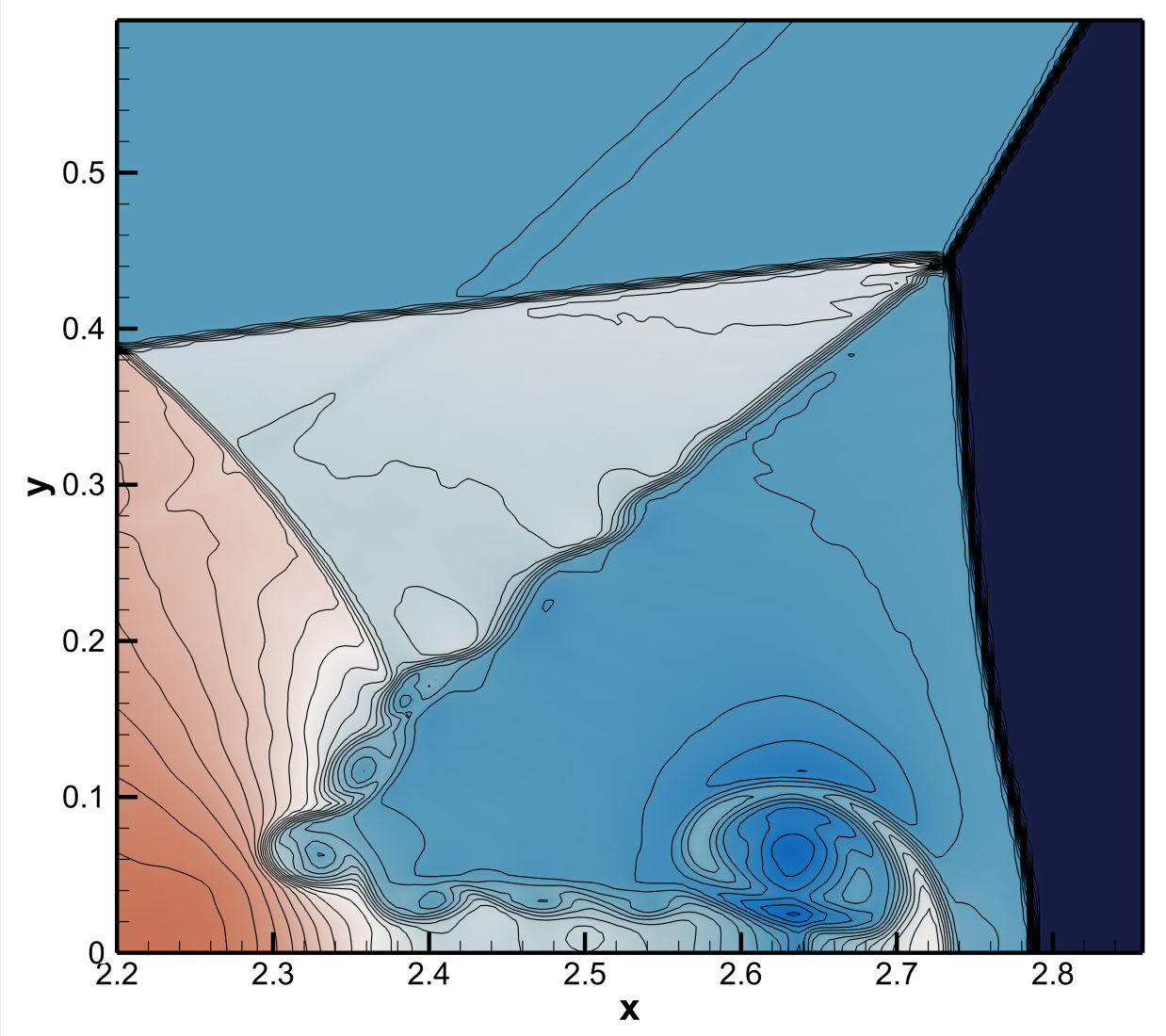}}
\subfigure[WGVC-TENO5]{
\includegraphics[width=8cm]{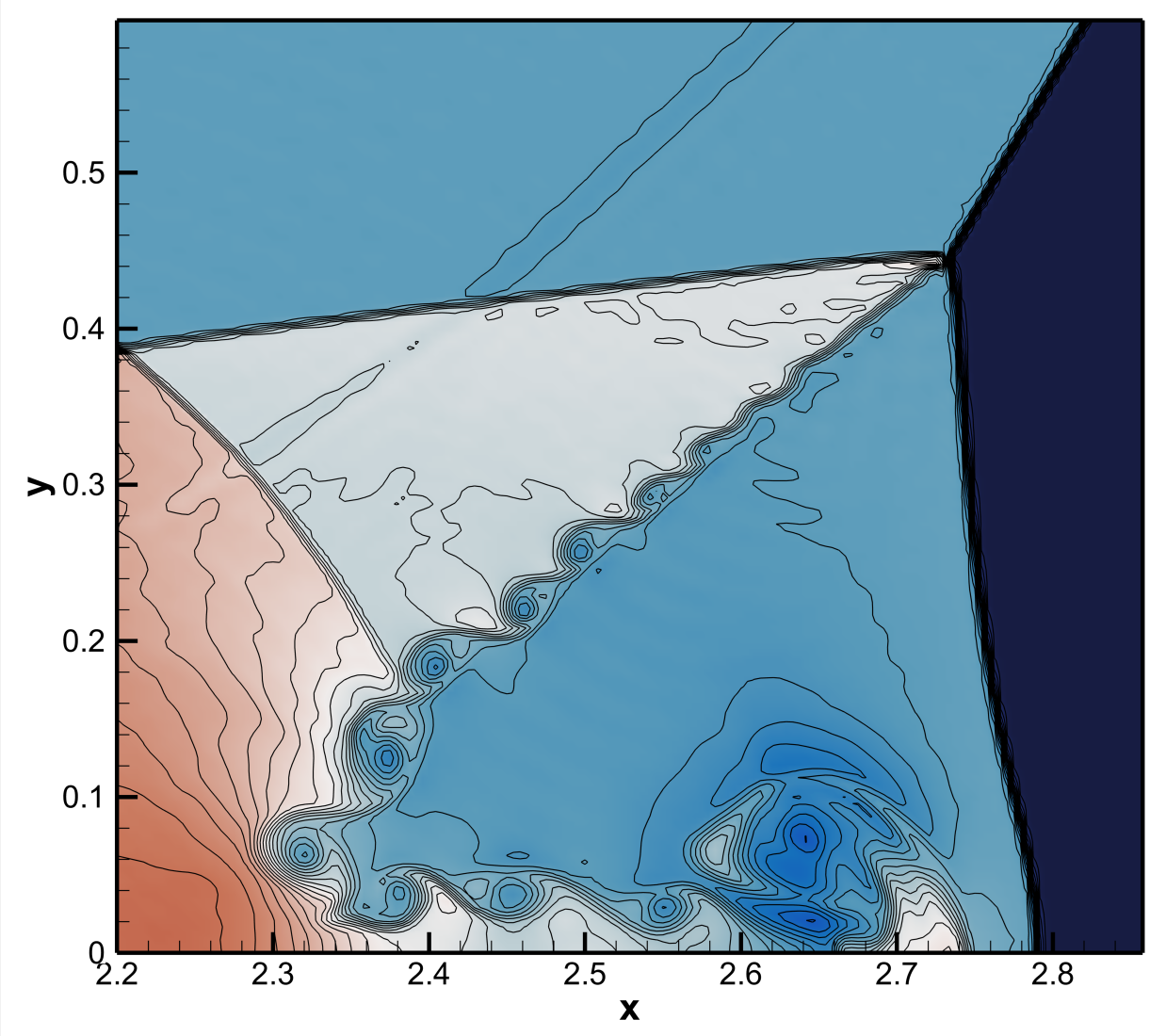}}
\caption{Density profiles of double Mach reflection problem at $t=0.2$ with $960\times240$ grid number; 40 equally spaced contour lines from $\rho=1.77$ to $\rho=21.8$.}
\label{fig.11}
\end{figure}

The computational domain is defined as $[0,4]\times[0,1]$, with a computational time set at $t = 0.2$ and a grid resolution of $960\times240$. Given the extremely high Mach number, the computational stability is crucial to prevent divergence. Flow characteristics, including Mach stems, slip lines, and the incident shock, are evident in the density contours depicted in Fig.\ref{fig.11}. In comparison to alternative schemes, WGVC-WENO5Z and WGVC-TENO5 exhibit superior performance in capturing tightly rolled-up small vortices and intricate shear surfaces along the inclined Mach stems. This observation indicates that WGVC-WENO5Z and WGVC-TENO5 exhibit reduced dissipation compared to the WENO5Z and TENO5 schemes.

\subsubsection{RT instability problem}
The Rayleigh-Taylor (RT) instability \cite{shi2003} typically occurs under the influence of gravity when a heavy fluid flows into a light fluid. During this process, bubbles generated by the light fluid rise into the heavy fluid, while the peaks of the heavy fluid descend into the light fluid. This process gives rise to structures like mushroom vortices, making it a valuable test case for evaluating numerical schemes. The initial conditions are as follows:
\begin{equation} \label{eq76}
   (\rho,u,v,p)=\begin{cases}(2,0,-0.025\sqrt{\gamma p/\rho}\cdot\cos(8\pi x),2y+1),&0\le y<0.5,\\(1,0,-0.025\sqrt{\gamma p/\rho}\cdot\cos(8\pi x),y+1.5),&0.5\le y\le1.\end{cases}
\end{equation}

\begin{figure}[H]
\centering
\subfigure[WENO5Z]{
\includegraphics[width=0.24\linewidth]{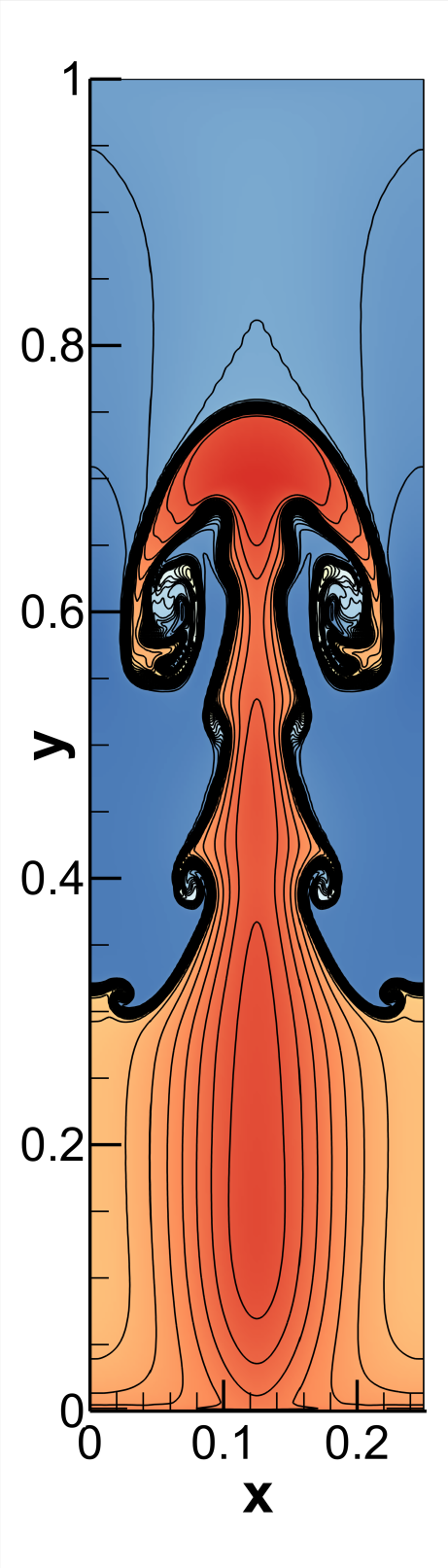}}
\subfigure[WGVC-WENO5Z]{
\includegraphics[width=0.24\linewidth]{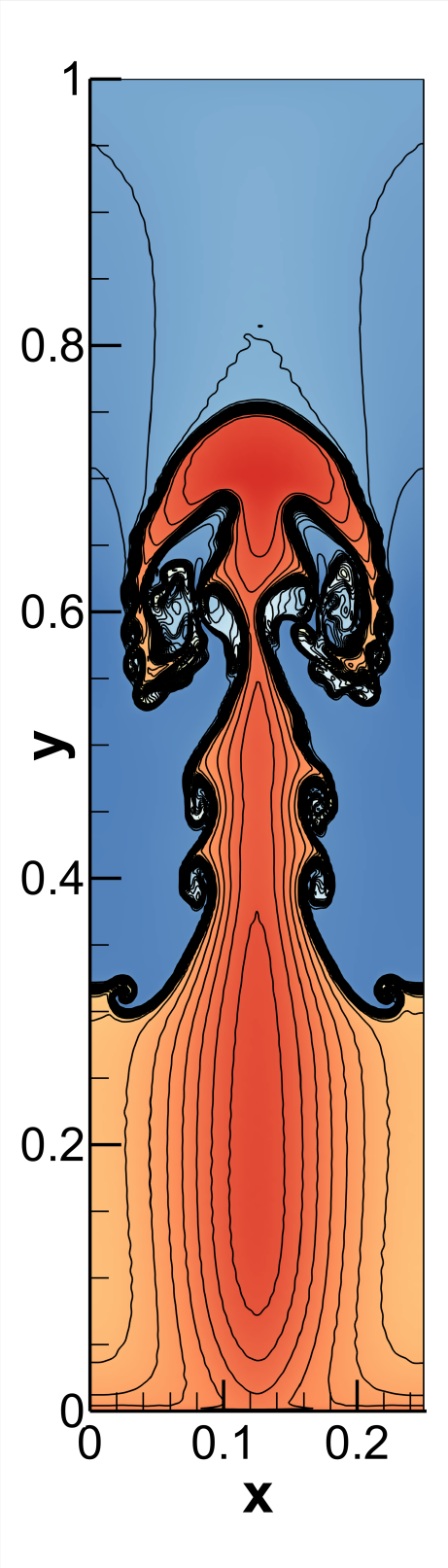}}
\subfigure[TENO5]{
\includegraphics[width=0.24\linewidth]{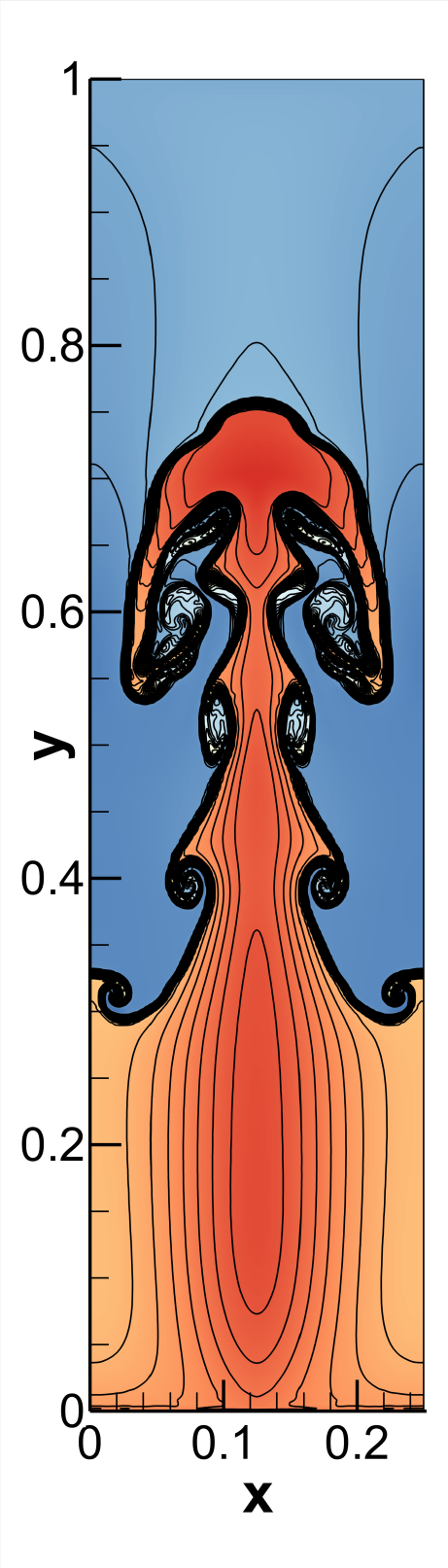}}
\subfigure[WGVC-TENO5]{
\includegraphics[width=0.24\linewidth]{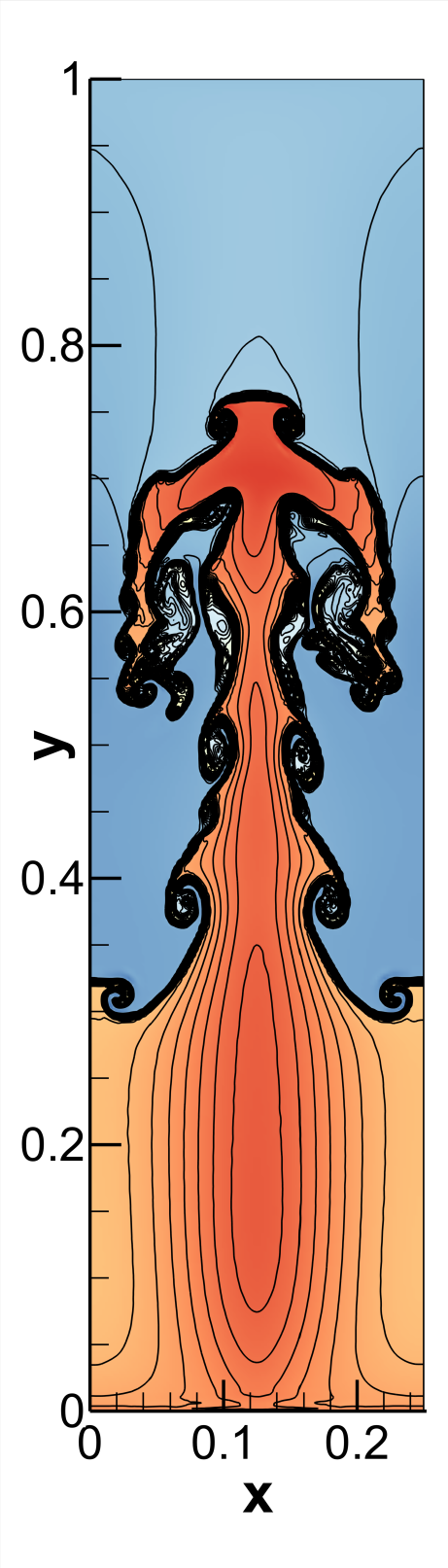}}
\caption{Density profiles of RT instability problem at $t=1.95$ with $120\times480$ grid number; 30 equally spaced contour lines from $\rho=0.95$ to $\rho=2.15$.}
\label{fig.12}
\end{figure}

The computational domain is $[0,0.25]\times[0,1]$, with a calculation time of $t=1.95$. Reflective boundary conditions are applied to the left and right boundaries, while the top and bottom boundaries are subjected to Dirichlet boundary conditions given by:
\begin{equation} \label{eq77}
   (\rho,u,v,p)=\begin{cases}(2,0,0,1),&y=0,\\(1,0,0,2.5),&y=1.\end{cases}
\end{equation}

\begin{figure}[H]
\centering
\subfigure[WENO5Z]{
\includegraphics[width=0.2\linewidth]{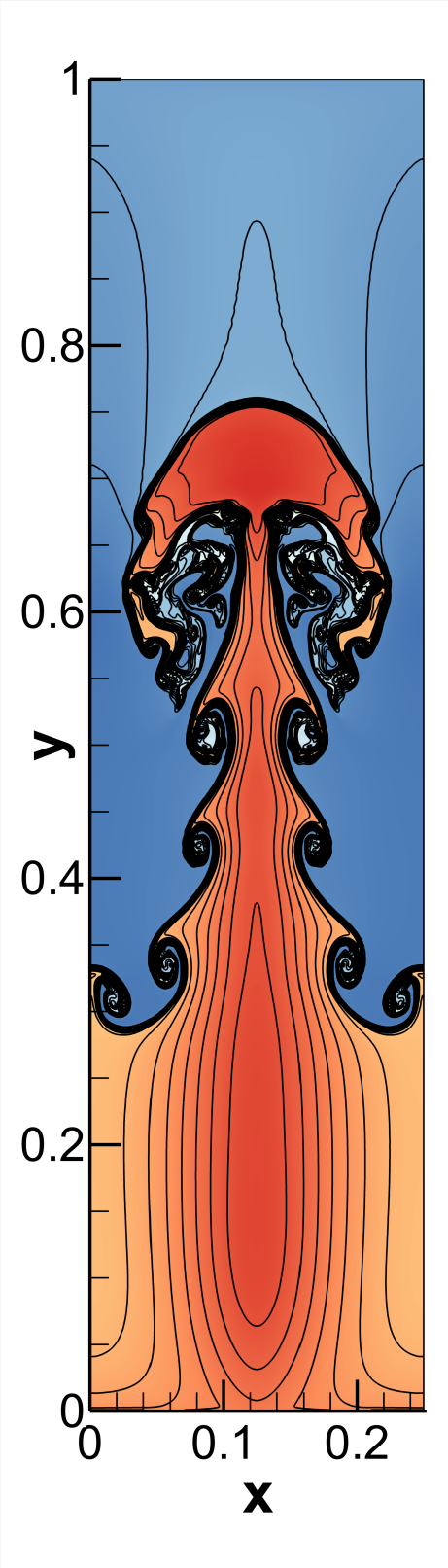}}
\subfigure[WGVC-WENO5Z]{
\includegraphics[width=0.2\linewidth]{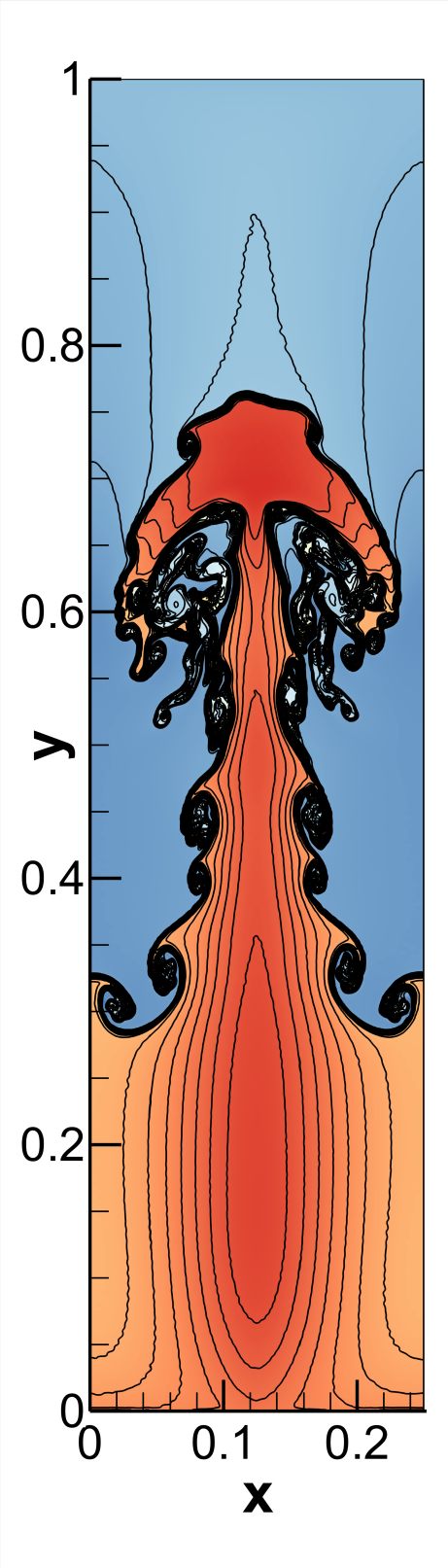}}
\subfigure[TENO5]{
\includegraphics[width=0.2\linewidth]{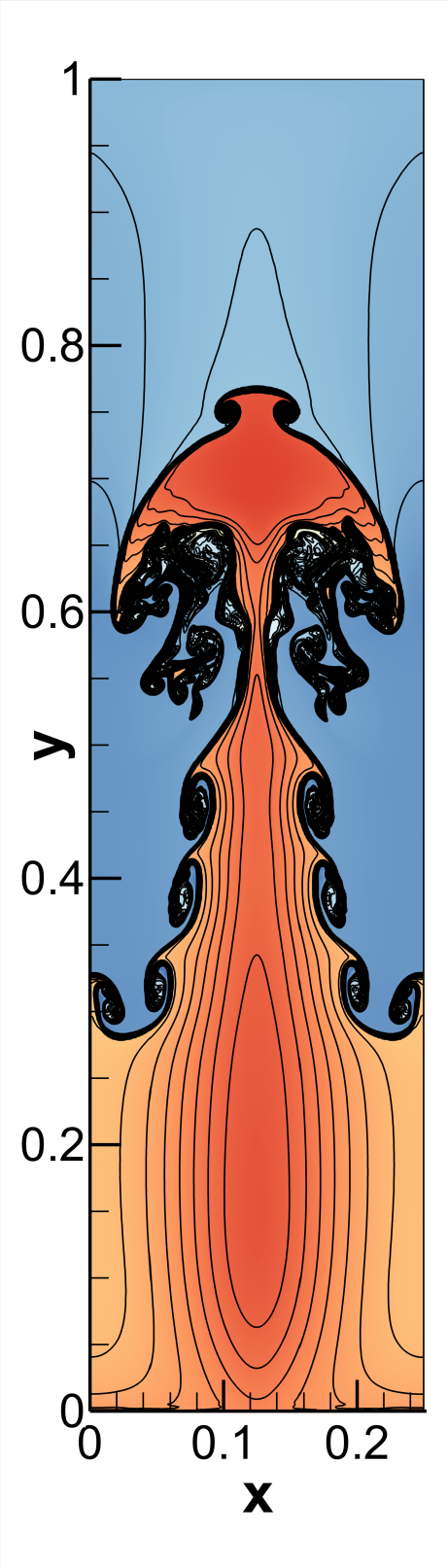}}
\subfigure[WGVC-TENO5]{
\includegraphics[width=0.2\linewidth]{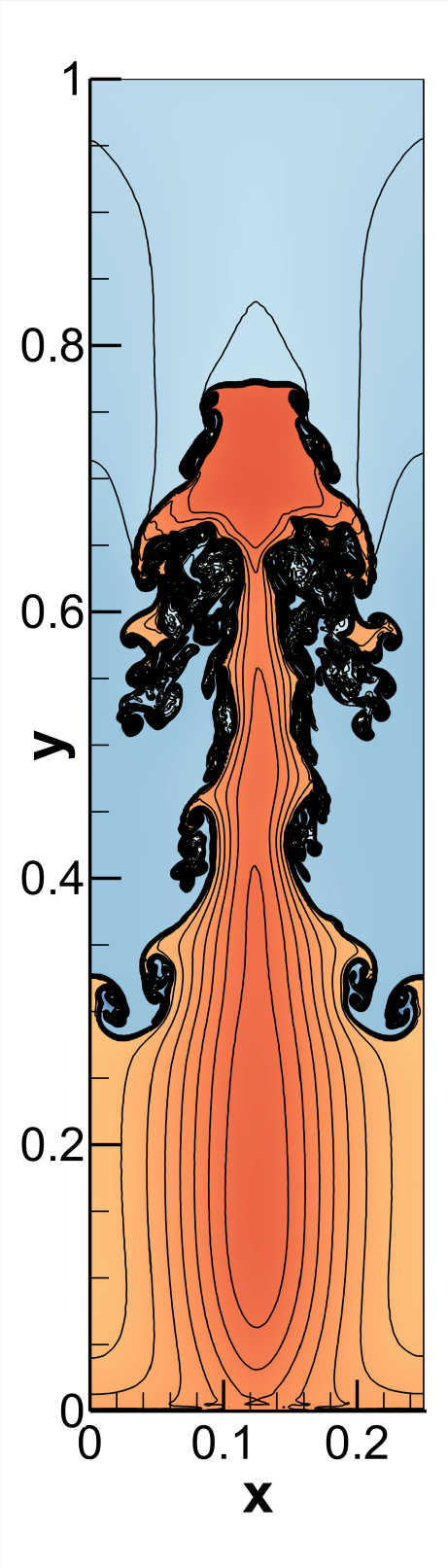}}
\caption{Density profiles of RT instability problem at $t=1.95$ with $240\times960$ grid number; 30 equally spaced contour lines from $\rho=0.95$ to $\rho=2.15$.}
\label{fig.13}
\end{figure}

Additionally, a source term $(0,0,\rho,\rho v)^{T}$ needs to be included, and typically, $\gamma$ is set to 5/3. The density results obtained from calculations using WENO5Z, WGVC-WENO5Z, TENO5, and WGVC-TENO5 with $120\times480$ grid number are shown in Fig.\ref{fig.12} and the results with a refined grid number $240\times960$ are shown in Fig.\ref{fig.13}. From Fig.\ref{fig.12} and Fig.\ref{fig.13}, it can be observed that WGVC-WENO5Z scheme captures a more intricate vortex structure compared to WENO5Z scheme. Similarly, WGVC-TENO5 scheme exhibits superior resolution compared to TENO5 scheme. This further underscores that the embedded method effectively enhances the resolution of shock-capturing schemes, reducing numerical dissipation.

\subsubsection{2D Riemann problems}
In this subsection, two classical 2D Riemann problems \cite{schulz1993} are resolved by different schemes.

\begin{itemize}
    \item Case 1: The initial condition is:
    \begin{equation} \label{eq78}
       (\rho,u,v,p)=\begin{cases}
       (1.0,0.75,-0.5,1.0),&1.0\leq x\leq2.0,1.0\leq y\leq2.0,\\
       (2.0,0.75,0.5,1.0),&0.0\leq x<1.0,1.0\leq y\leq2.0,\\
       (1.0,-0.75,0.5,1.0),&0.0\leq x<1.0,0.0\leq y<1.0,\\
       (3.0,-0.75,-0.5,1.0),&1.0\leq x\leq2.0,0.0\leq y<1.0.\end{cases}
    \end{equation}
\end{itemize}

\begin{itemize}
    \item  Case 2: The initial condition is:
    \begin{equation} \label{eq79}
        (\rho,u,\nu,p)=\begin{cases}(0.5313,0.0,0.0,0.4),&0.5\leq x\leq1.0,0.5\leq y\leq1.0,\\ (1.0,0.7276,0.0,1.0),&0.0\leq x<0.5,0.5\leq y\leq1.0,\\(0.8,0.0,0.0,1.0),&0.0\leq x<0.5,0.0\leq y<0.5,\\(1.0,0.0,0.7276,1.0),&0.5\leq x\leq1.0.0.0\leq y<0.5.\end{cases}
   \end{equation}
\end{itemize}

\begin{figure}[H]
\centering
\subfigure[WENO5Z]{
\includegraphics[width=8cm]{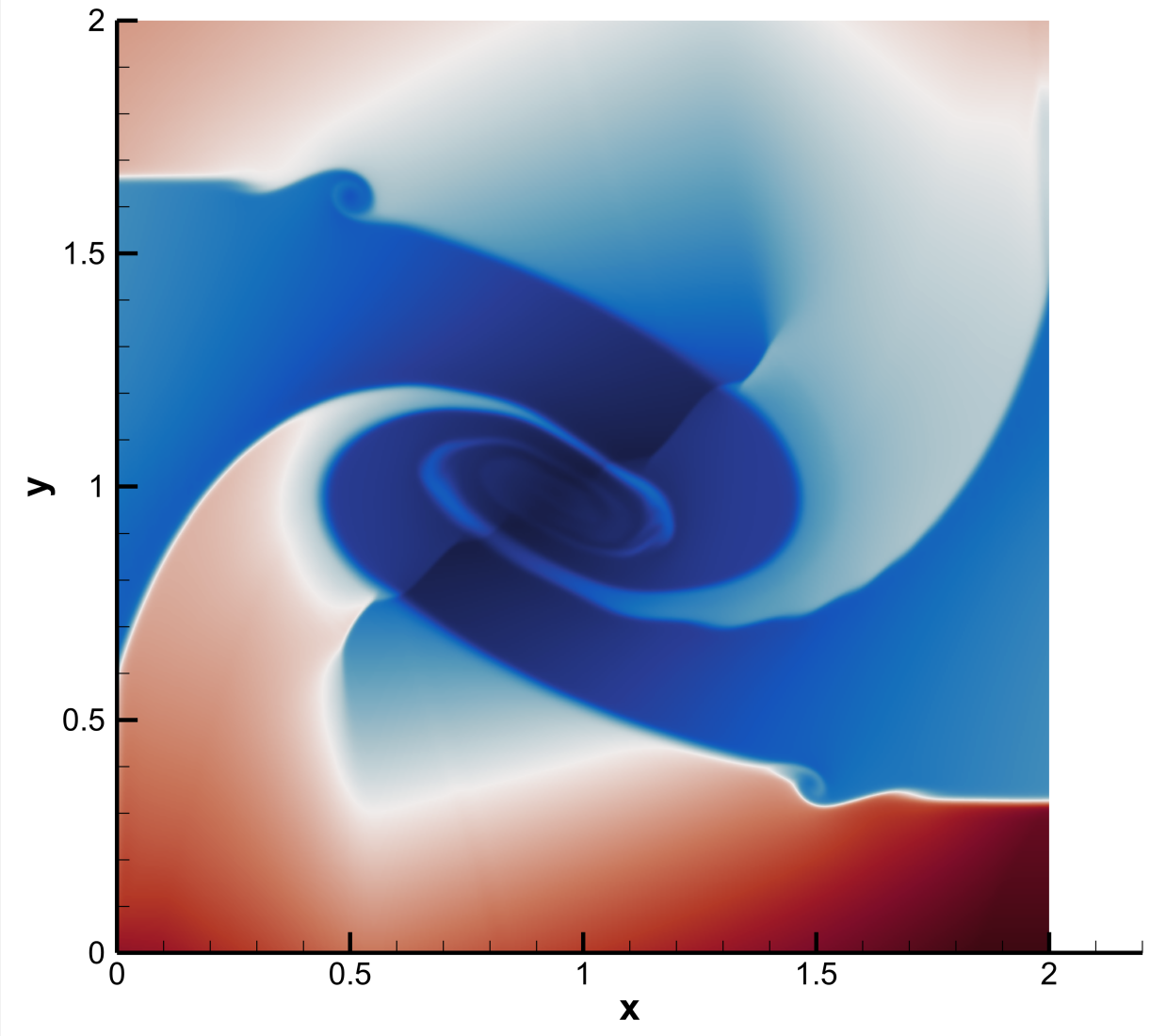}}
\subfigure[WGVC-WENO5Z]{
\includegraphics[width=8cm]{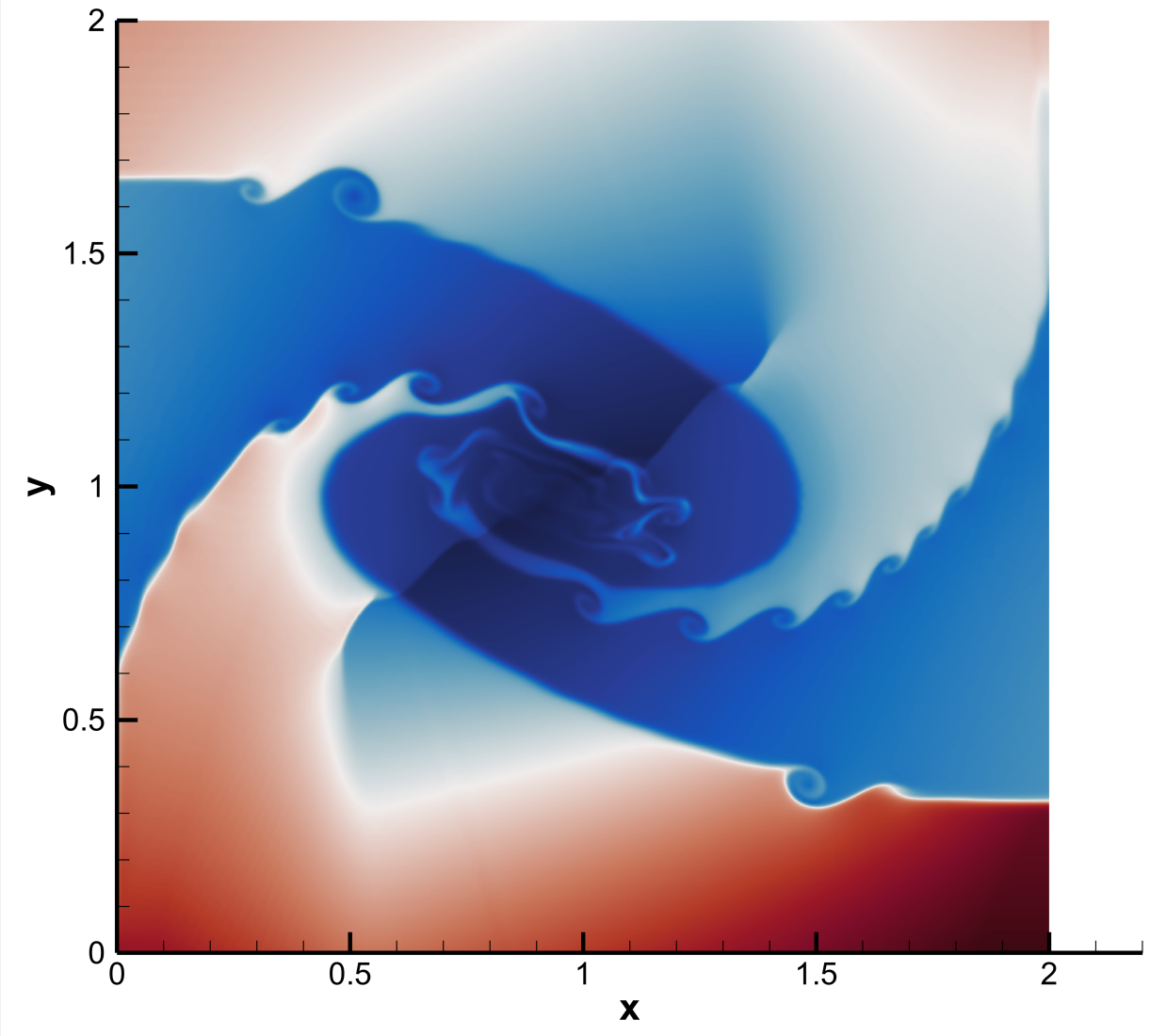}}
\subfigure[TENO5]{
\includegraphics[width=8cm]{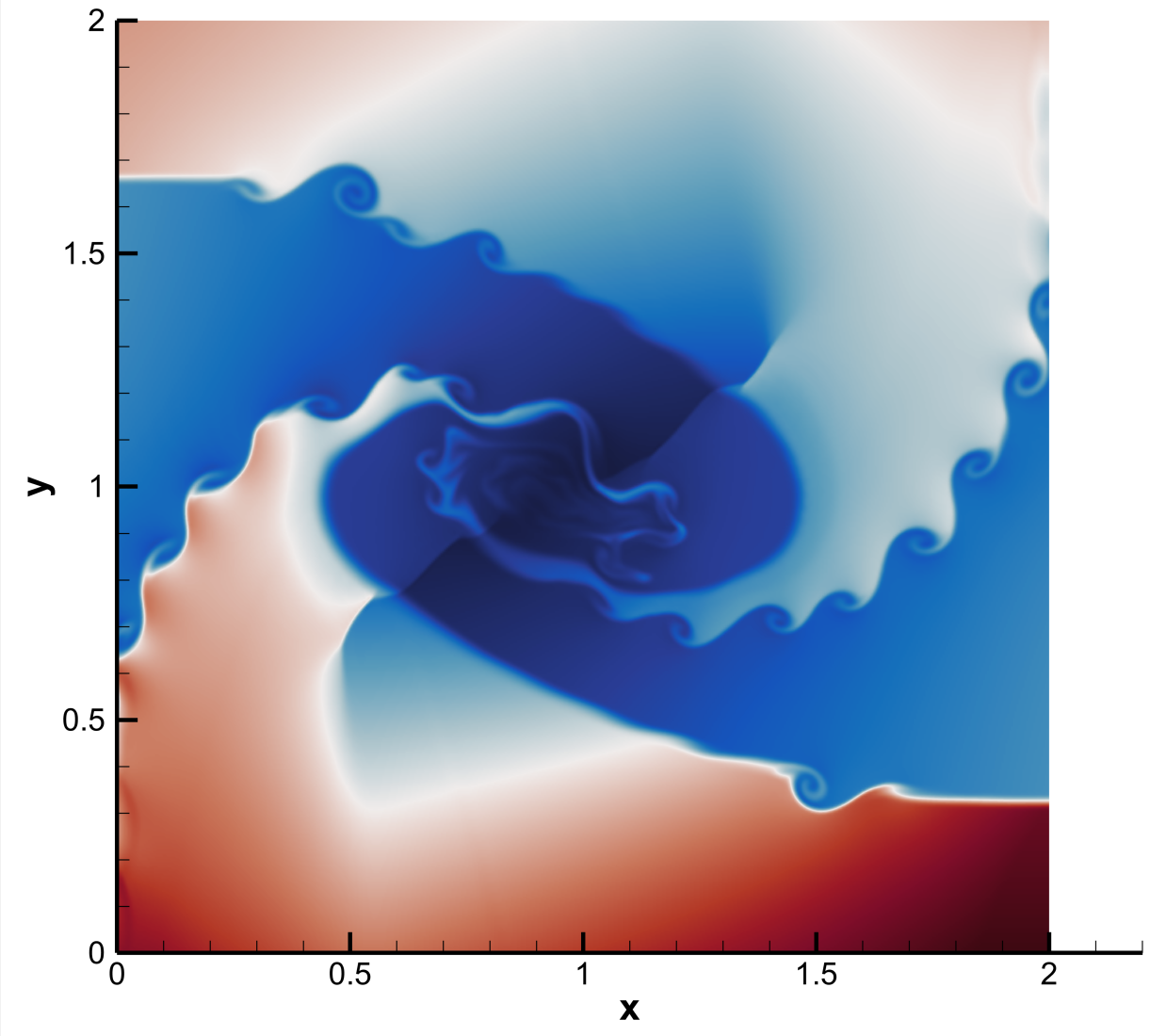}}
\subfigure[WGVC-TENO5]{
\includegraphics[width=8cm]{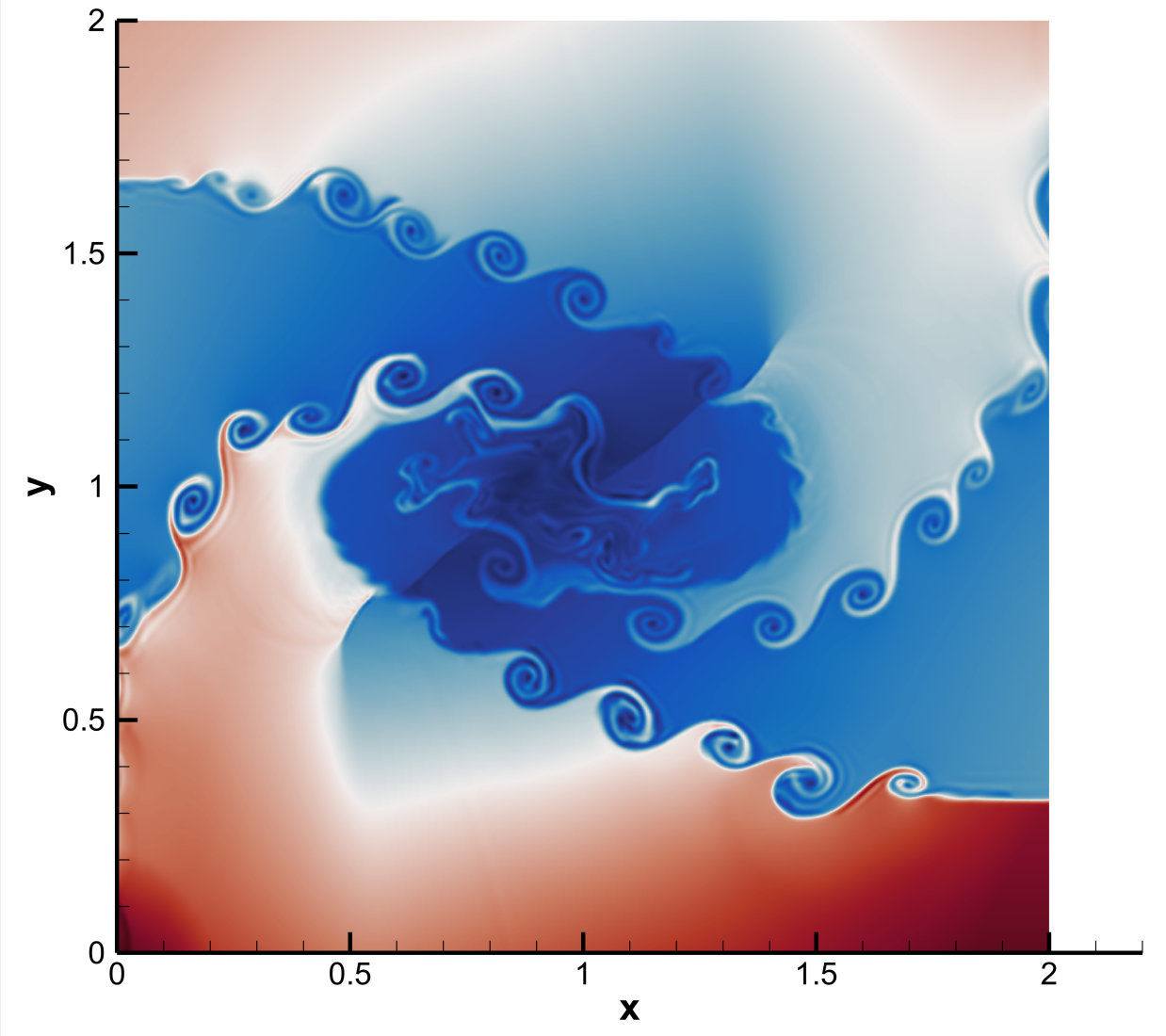}}
\caption{Density profiles of 2D Riemann problem (Case 1) at $t=1.6$ with $500\times500$ grid number.}
\label{fig.14}
\end{figure}

\begin{figure}[H]
\centering
\subfigure[WENO5Z]{
\includegraphics[width=8cm]{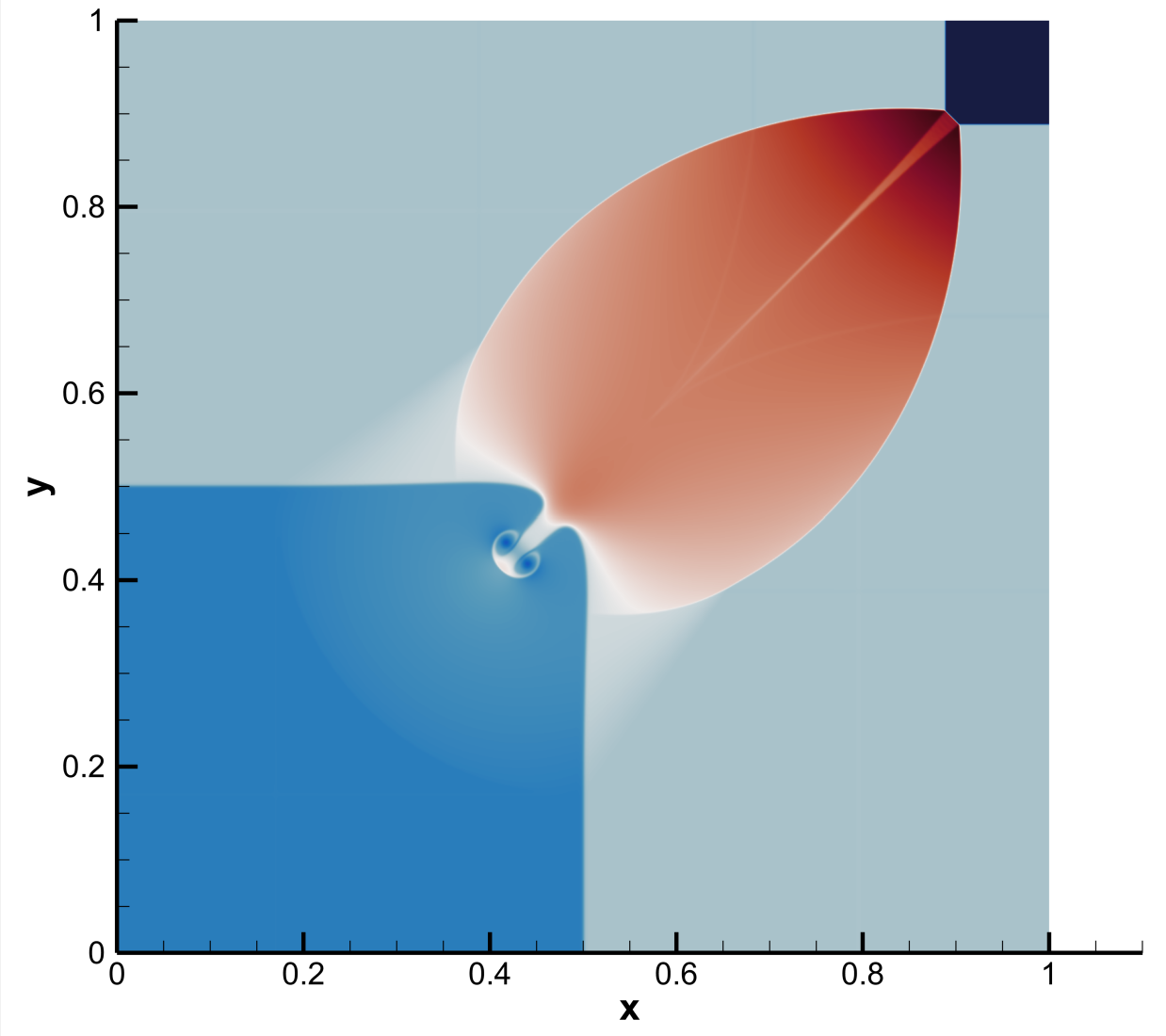}}
\subfigure[WGVC-WENO5Z]{
\includegraphics[width=8cm]{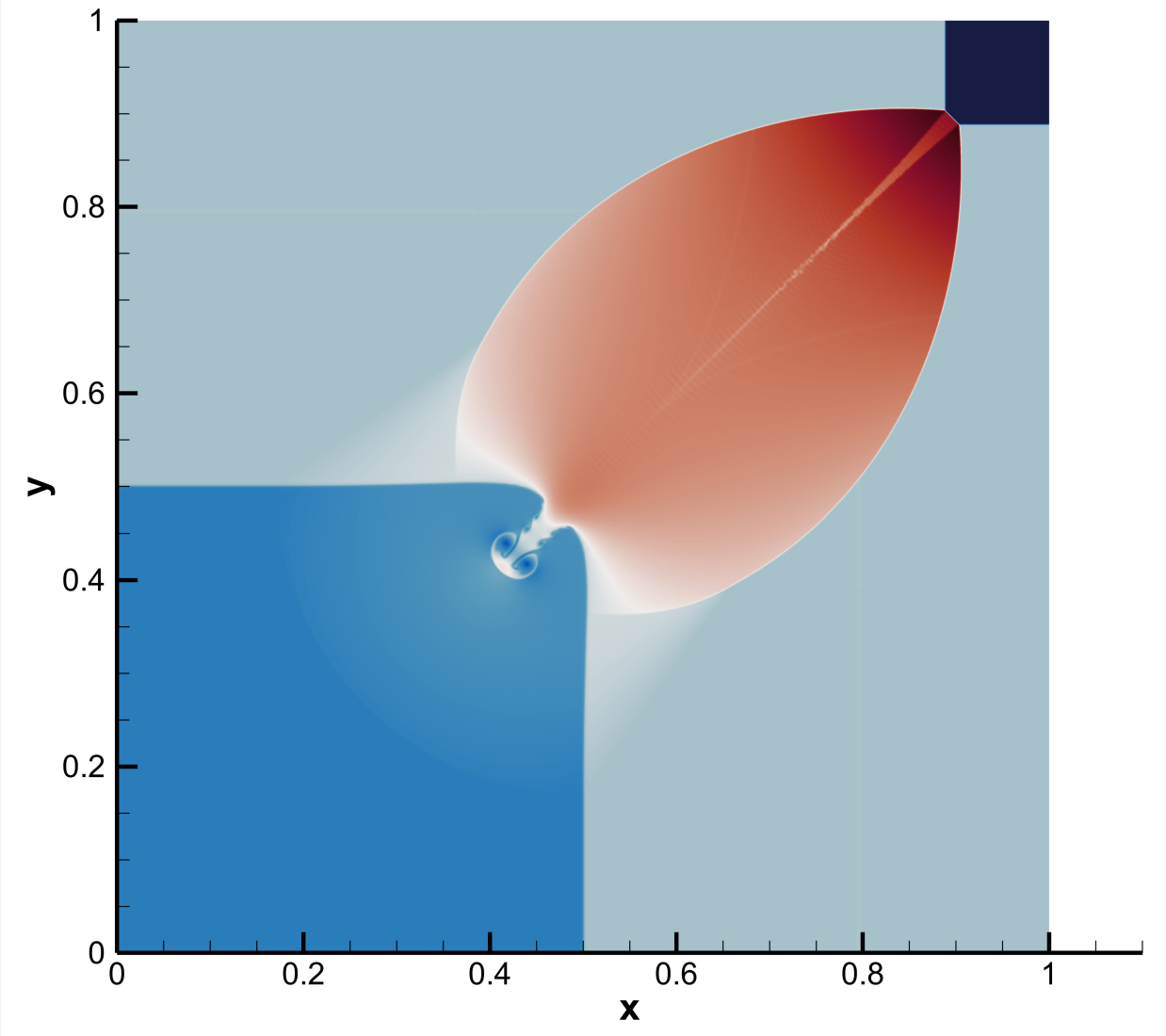}}
\subfigure[TENO5]{
\includegraphics[width=8cm]{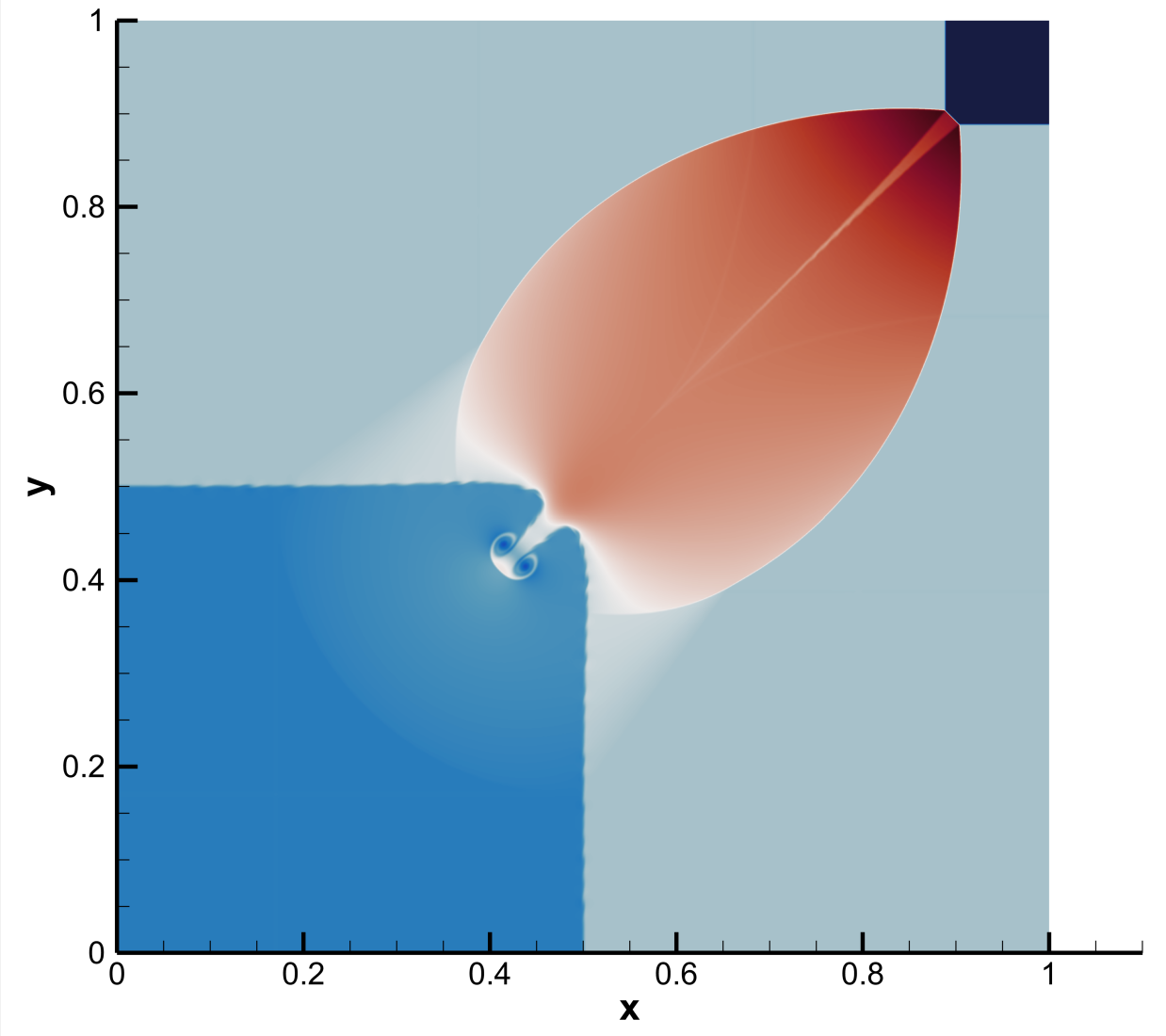}}
\subfigure[WGVC-TENO5]{
\includegraphics[width=8cm]{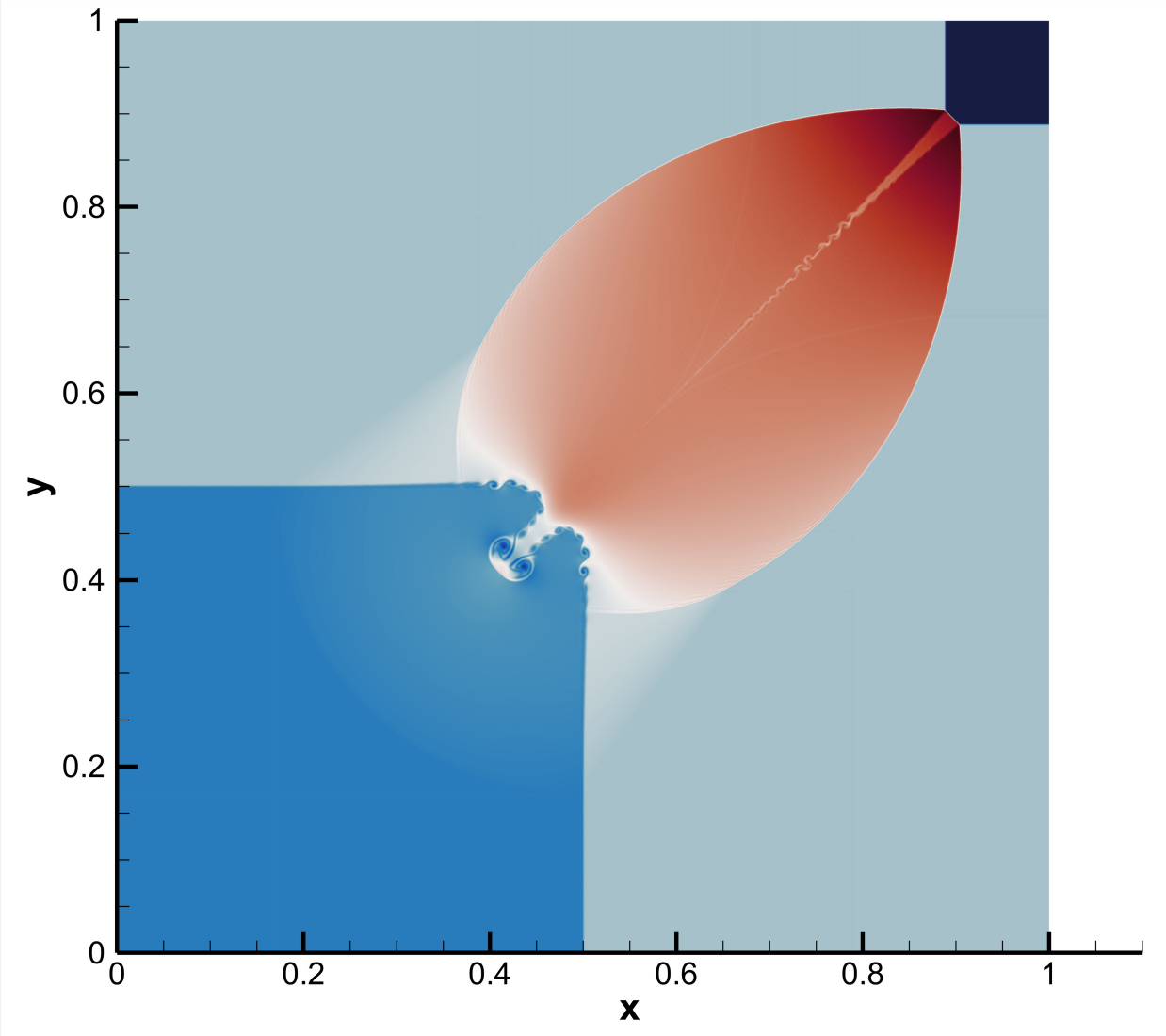}}
\caption{Density profiles of 2D Riemann problem (Case 2) at $t=0.25$ with $1400\times1400$ grid number.}
\label{fig.15}
\end{figure}

\begin{figure}[H]
\centering
\subfigure[WENO5Z]{
\includegraphics[width=8cm]{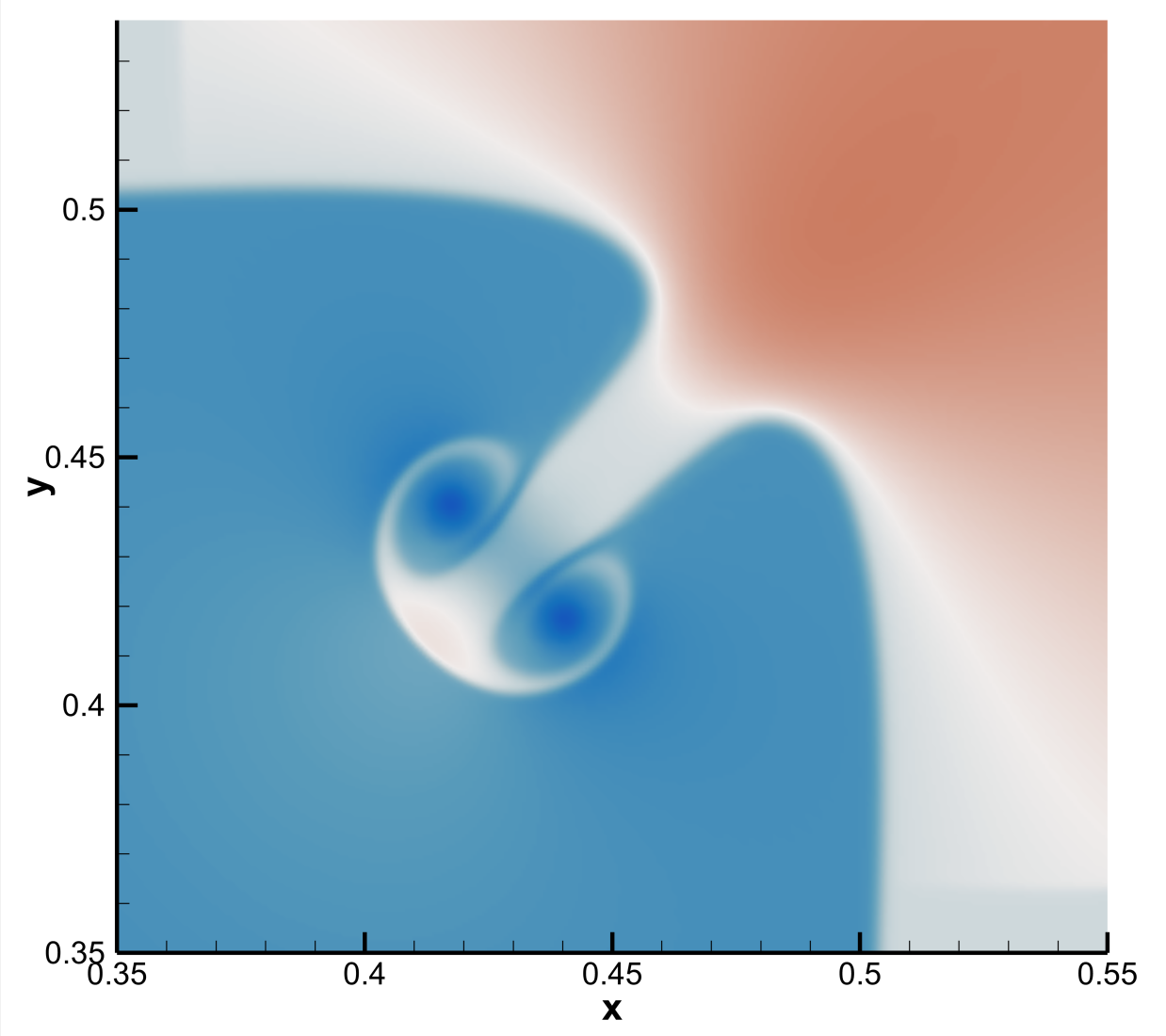}}
\subfigure[WGVC-WENO5Z]{
\includegraphics[width=8cm]{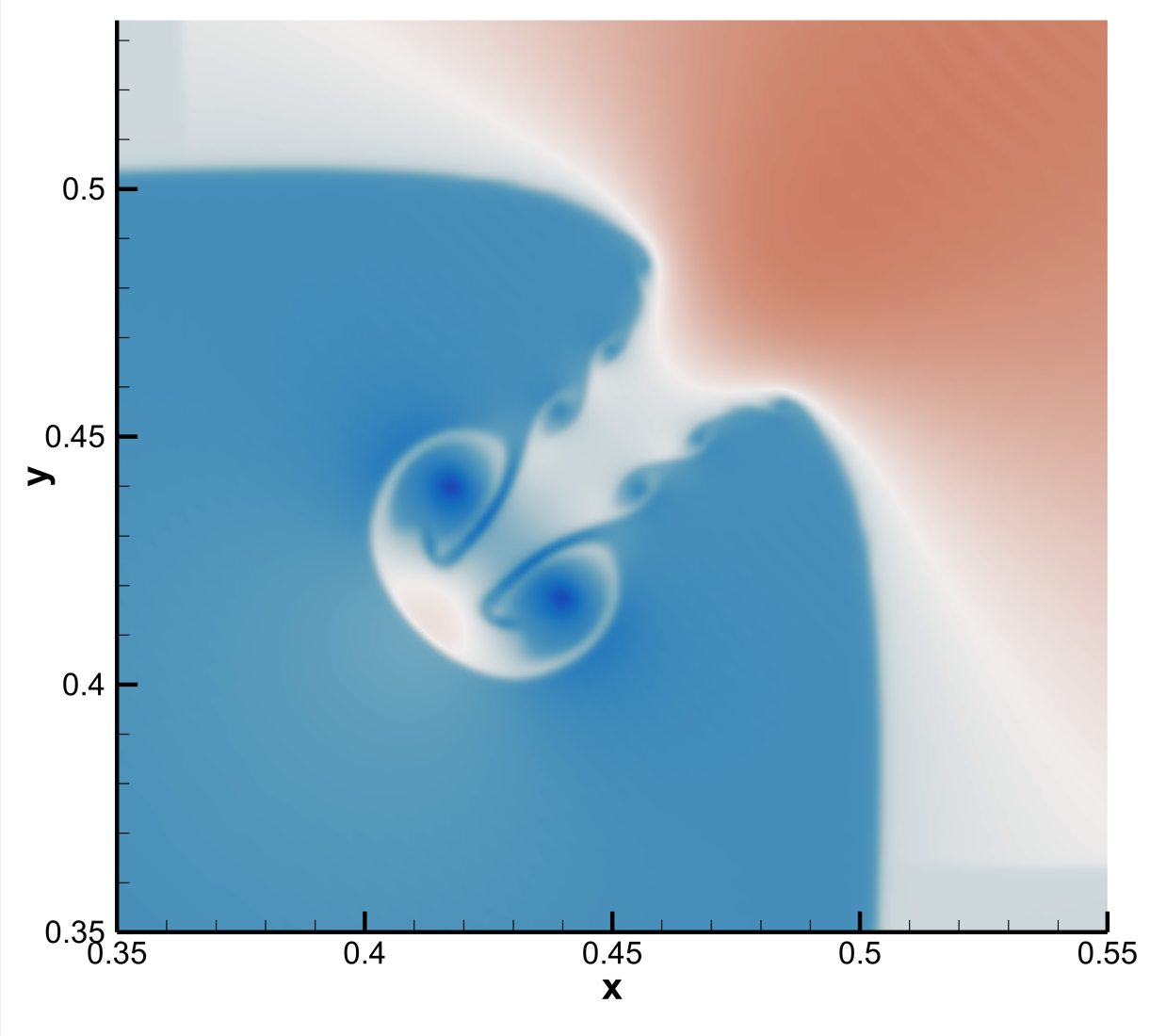}}
\subfigure[TENO5]{
\includegraphics[width=8cm]{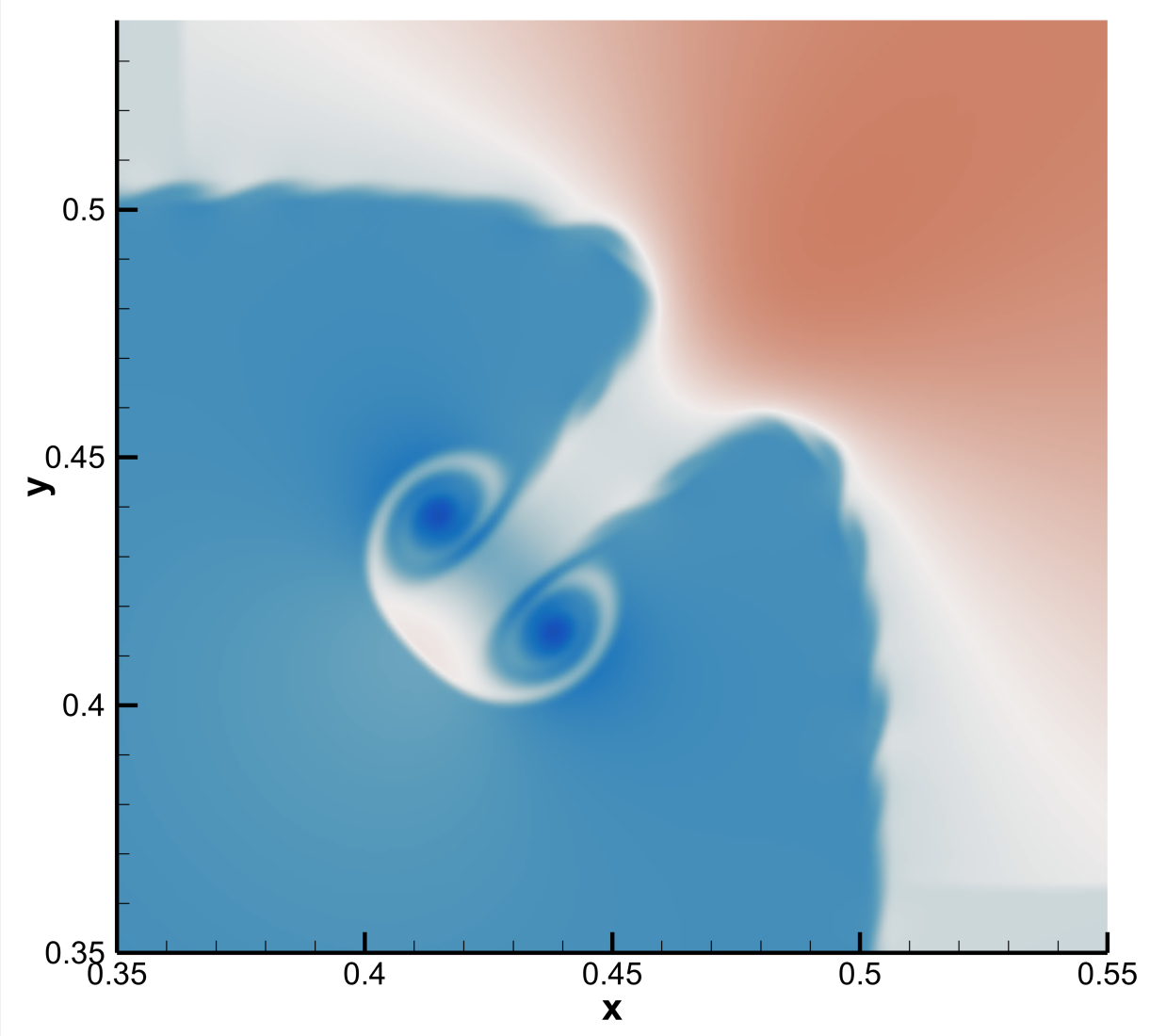}}
\subfigure[WGVC-TENO5]{
\includegraphics[width=8cm]{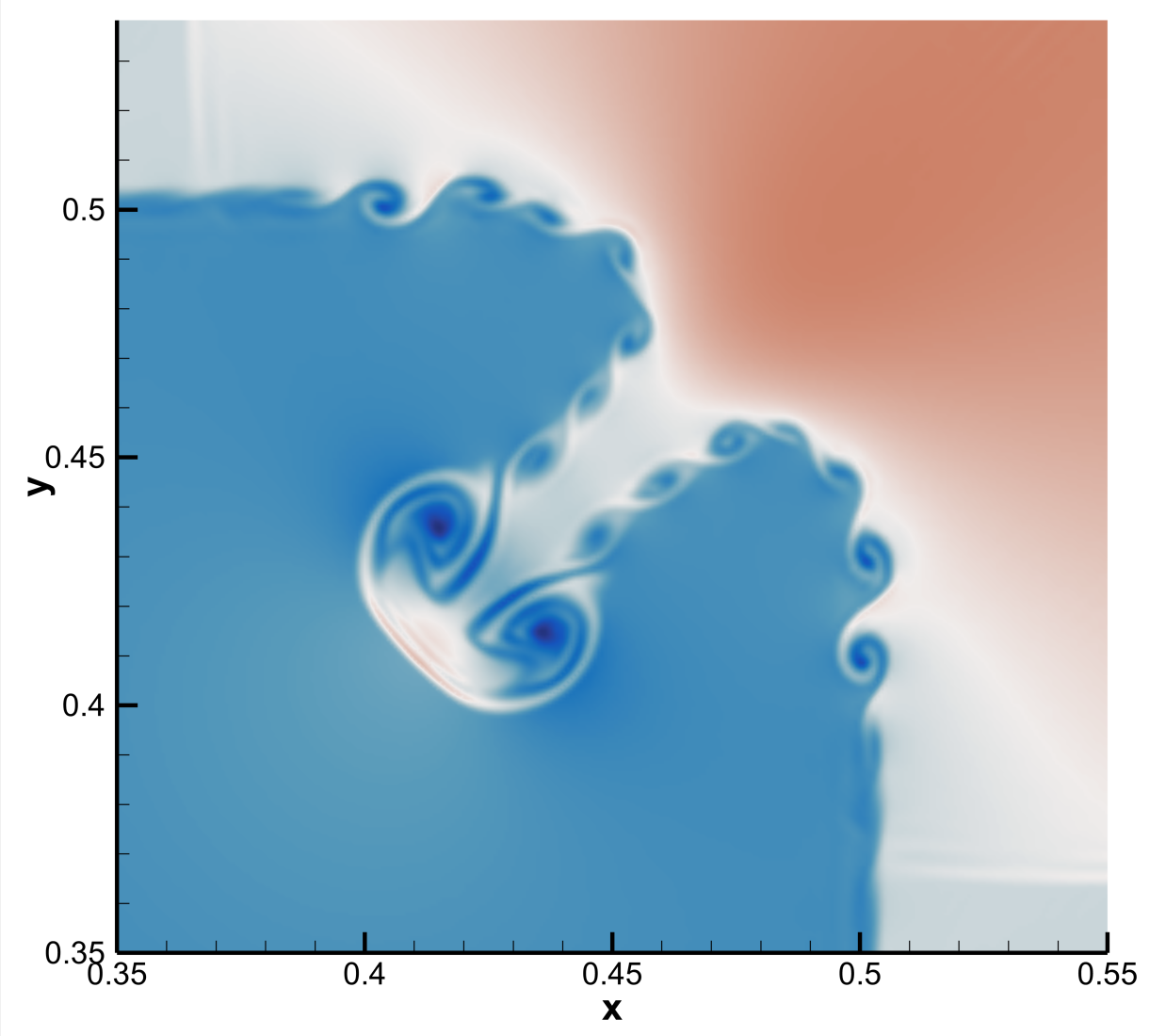}}
\caption{Enlarged view of Fig.\ref{fig.15}}
\label{fig.16}
\end{figure}

Case 1 covers a computational domain extending from $[0,2]\times[0,2]$, and the simulation runs for a duration of $t=1.6$, employing a grid size of $500\times500$. In Case 2, the computational domain is confined to $[0,1]\times[0,1]$, with a simulation time of $t=0.25$ and a grid resolution of $1400\times1400$. Tight boundary conditions are consistently enforced along all four boundaries. 

Fig.\ref{fig.14} illustrates the density profiles for Case 1, while Fig.\ref{fig.15} and Fig.\ref{fig.16} provide contour plots of density for Case 2. Fig.\ref{fig.16} offers an enlarged view of a specific region within Fig.\ref{fig.15}. Both Case 1 and Case 2 initially exhibit contact discontinuities, which progressively evolve into intricate vortical structures over time. From Fig.\ref{fig.14}-\ref{fig.16}, it is evident that WGVC-TENO5 and WGVC-WENO5Z schemes, when compared to TENO5 and WENO5Z scheme, yield more intricate vortical structures due to lower numerical dissipation. This enhancement signifies that embedding the WGVC scheme into WENO/TENO schemes strengthens the numerical methods' ability to resolve small-scale structures, thus aiding in capturing finer flow details.

\section{Conclusion}

In this paper, we proposed an order-preserving and spectral property optimization scheme, named WGVC scheme, with the aim of optimizing spectral properties of finite difference methods and enhancing the resolution of multiscale structures such as turbulence. This scheme is centered around the concept of group velocity and guided by the group velocity control theory. By designing smoothness indicators and employing a nonlinear weighting approach for wave packets, it combines two schemes with different group velocity characteristics ($SLW$ scheme and $MXD$ scheme) through weighting. This approach achieves order control in the low-wavenumber range, group velocity control in the mid-wavenumber range, and significantly improves the spectral properties of the difference scheme. To handle discontinuous structures like shock waves, the proposed WGVC scheme is further embedded into shock-capturing schemes such as WENO and TENO, resulting in the novel development of WGVC-WENO and WGVC-TENO schemes. Numerical results indicate that WGVC-WENO and WGVC-TENO schemes, while preserving accuracy, possess both the spectral properties of the WGVC scheme in the medium to low-wavenumber range and the shock-capturing capabilities of WENO/TENO schemes. These schemes are highly suitable for the numerical simulation of multiscale complex flow problems with discontinuities, such as shock-turbulence boundary layer interaction.

\section*{Acknowledgment}
This work was supported by the Strategic Priority Research Program of Chinese Academy of Science (XDB0500301), NSFC Projects (12372285, 12232018, 12072349, 12202457), and the National Key Research and Development Program of China (2019YFA0405300). The authors thank National Supercomputer Center in Tianjin (NSCC-TJ), and National Supercomputer Center in Guangzhou (NSCC-GZ) for providing computer time.






\bibliographystyle{elsarticle-num-names}
\bibliography{sample.bib}







\end{document}